\documentclass[a4paper]{article}

\providecommand{\esymbol}{e}
\providecommand{\esymbolv}{\ev}

\providecommand{\cext}{x}
\providecommand{\cint}{c}
\providecommand{\cextv}{\xv}
\providecommand{\cintv}{\cv}

\usepackage{setspace} 
\usepackage{xr}
\usepackage{wrapfig}
\usepackage{enumitem}
\usepackage{color}
\usepackage{amsmath,amsfonts,amssymb}
\usepackage{epsfig}
\usepackage{graphicx}
\usepackage{colortbl}
\usepackage{url}
\usepackage{lineno}

\definecolor{darkblue}{rgb}{0.05,0.,0.65}
\definecolor{grey}{rgb}{0.9, 0.9, 0.9}

\newcommand{\myparagraph}[1]{\vspace{-3mm}\paragraph{#1}}

\newcommand{\coout}[1]{[{\color{magenta} #1}]}
\newcommand{\co}   [1]{[{\it \color{red} #1}]}
\newcommand{\todo} [1]{[{\color{blue} #1}]}

\usepackage{amsmath,amsfonts,amssymb}
\usepackage{relsize}
\usepackage{ marvosym }

\newcommand{\const}{\mbox{\rm const.}}

\newcommand{\inv}{^{-1}}

\newcommand{\md}{{\rm d}}

\newcommand{\diag}{\mbox{\rm Dg}}

\newcommand{\e}   {\mbox{\rm e}}

\newcommand{\cv}{{\bf c}}
\newcommand{\vv}{{\bf v}}

\newcommand{\kv}{{\bf k}}

\newcommand{\zv}{{\bf z}}

\newcommand{\ev}{{\bf e}}
\newcommand{\xv}{{\bf x}}

\newcommand{\tauv}{{\boldsymbol \tau}}

\newcommand{\kappav}{{\boldsymbol \kappa}}

\newcommand{\Nmat}{\mathbf{N}}
\newcommand{\Imat}{\mathbf{I}}
\newcommand{\Cmat}{\mathbf{C}}
\newcommand{\Mmat}{\mathbf{M}}

\newcommand{\Gmat}{\mathbf{G}}

\newcommand{\Lmat}{\mathbf{L}}
\newcommand{\Kmat}{\mathbf{K}}
\newcommand{\Rmat}{{\bf R}}

\newcommand{\trans}{^{\top}}

\DeclareMathAlphabet{\mathpzc}{OT1}{pzc}{m}{it}
\DeclareMathAlphabet{\mathcalligra}{T1}{calligra}{m}{n}

\newcommand{\oslashs}{/}

\newcommand{\deltapar} {{\delta_\mathsmaller{\parallel}}}

\providecommand{\esymbol}{e}

\newcommand{\steady}{^{\rm st}}
\newcommand{\vvsteady}   {\vv\steady}
\newcommand{\vsteady}    {v\steady}
\newcommand{\cvsteady}   {\cv\steady}

\newcommand{\csteady}    {c\steady}

\newcommand{\modevector}{\kappav}

\newcommand{\kcatl}{k_{\rm cat,l}}

\newcommand{\Euns}    {E}
\newcommand{\Eun}     {\mathbf{\Euns}}
\newcommand{\Eunc}    {\Eun_{\rm c}}
\newcommand{\Eunint}  {\Eun_{\rm c}} 
\newcommand{\Eunu}    {\Eun_{\rm \esymbol}}
\newcommand{\Eune}    {\Eun_{\rm e}}

\newcommand{\Eunul}   {{\Euns^{\rm v_{l}}_{\rm \esymbol_{l}}}}
\newcommand{\Eunvlci} {{\Euns^{\rm v_{l}}_{{\rm c}_{i}}}} 
\newcommand{\Eunextlj}{{\Euns^{\rm v_{l}}_{{\rm \cextj}}}}
\newcommand{\Eunvlpn} {{\Euns^{\rm v_{l}}_{{\rm p}_{n}}}}
\newcommand{\Eunvlpr} {{\Euns^{\rm v_{l}}_{{\rm p}_{r}}}}

\newcommand{\Escs}    {\hat{E}}
\newcommand{\Esc}     {\mathbf{\Escs}}

\newcommand{\Escint}  {{\Esc^{\rm v}_{\rm c}}} 
\newcommand{\Escu}    {{\Esc^{\rm v}_{\rm \esymbol}}}

\newcommand{\Escvlci} {\Escs^{\rm v_{l}}_{{\rm c}_{i}}} 
\newcommand{\Escvlp} {\Escs^{\rm v_{l}}_{{\rm p}}}
\newcommand{\Escvlpn} {\Escs^{\rm v_{l}}_{{\rm p}_{n}}}
\newcommand{\Escvlpr} {\Escs^{\rm v_{l}}_{{\rm p}_{r}}}

\newcommand{\Ccmat} {\Cmat^{\rm S}}
\newcommand{\Cc}    {C^{\rm S}}

\newcommand{\Cvmat} {\Cmat^{\rm V}}
\newcommand{\Cv}    {C^{\rm V}}

\newcommand{\CJ}{C^{\rm J}}

\newcommand{\CJmat}{{\bf C}^{\rm J}}

\newcommand{\Rc}    {R^{\rm C}}
\newcommand{\Rv}    {R^{\rm V}}

\newcommand{\Rcmat} {\Rmat^{\rm C}}
\newcommand{\Rvmat} {\Rmat^{\rm V}}

\usepackage{etoolbox}

\newtoggle{bookversion}      \togglefalse{bookversion}
\newtoggle{shortbookversion} \togglefalse{shortbookversion}
\newtoggle{anhang}           \togglefalse{anhang}
\newtoggle{hidecomments}      \togglefalse{hidecomments}

\newcommand{\bookco}[1]{}

\definecolor{brown}{rgb}{0.9,0.69,0.34}
\definecolor{samoabrownlight}{rgb}{0.89,0.69,0.4}
\definecolor{samoabrowndark} {rgb}{0.5,0.3,0.15}
\definecolor{cbasamoabrown1}{rgb}{0.87,0.6,0.23}
\definecolor{cbasamoabrown2}{rgb}{0.87,0.6,0.23}
\definecolor{cbabrown1}{rgb}{0.87,0.6,0.23}
\definecolor{cbabrown2}{rgb}{0.87,0.6,0.23}
\definecolor{cbabrown3}{rgb}{0.87,0.6,0.23}
\definecolor{cbabrown4}{rgb}{0.87,0.6,0.23}
\definecolor{cbaecoblue1}{rgb}{0.8,0.8, 1.0}
\definecolor{cbaecoblue2}{rgb}{0.7,0.7, 1.0}
\definecolor{cbaecoblue3}{rgb}{0.87,0.6,0.23}
\definecolor{cbaecoblue4}{rgb}{0.87,0.6,0.23}
\definecolor{cbablue2}{rgb}{0.87,0.6,0.23}
\definecolor{cbapink}{rgb}{.99,0.92,0.75}
\definecolor{cbabeige1}{rgb}{0.86, 0.797, 0.625} 
\definecolor{cbabeige2}{rgb}{0.93, 0.812, 0.56}  
\definecolor{cbabeige3}{rgb}{1.0, 0.97, 0.88}  
\definecolor{cbahelleslila}{rgb}{1.0, 0.99, 1.0}  
\definecolor{cbalightgrey}{rgb}{0.95,0.95,0.95}
\definecolor{cbatablecolor1}{rgb}{0.86, 0.797, 0.625} 
\definecolor{cbatablecolor2}{rgb}{1,1,1}         

\newcommand{\myvalue}      {value}

\newcommand{\gain}         {gain}

\newcommand{\stress}       {stress}

\newcommand{\investment}   {investment}

\newcommand{\enzymeinvestment}{{enzyme \investment}}

\newcommand{\price}        {price}

\newcommand{\burden}       {burden}

\newcommand{\enzymeprice}  {enzyme \price}

\newcommand{\fluxburden}   {flux \burden}

\newcommand{{\fluxvalue}}  {flux \myvalue}
\newcommand{{\fluxgain}}   {flux \gain}
\newcommand{\metabolicobjective}{metabolic objective}

\newcommand{\valid}  {valid}
\newcommand{\invalid}  {invalid}

\newcommand{{\valueflow}}    {value flow}
\newcommand{{\Valueflow}}    {Value flow}

\newcommand{ {\flow}}        {flux profile}
\newcommand{ {\Flow}}        {Flux profile}

\newcommand{\point}         {point} 

\newcommand{\benefitshade}  {{\point} benefit}
\newcommand{\costshade}     {{\point} cost}

\newcommand{\fluxbenefit}   {flux {\point} benefit}

\newcommand{\fluxcostshade} {flux {\point} cost}
\newcommand{\Fluxcostshade} {Flux {\point} cost}
\newcommand{\metabolitecost}{metabolite {\point} cost}

\newcommand{\enzymecost}    {enzyme     {\point} cost}

\newcommand{\enzymebenefit} {enzyme     {\point} load}

\newcommand{\complete}   {all-active}

\newcommand{\futile}{futile}
\newcommand{\wasteful}{wasteful}
\newcommand{\nonbeneficial}{non-beneficial}

\newcommand{\fluxbenefitbalance}    {reaction balance}
\newcommand{\Fluxbenefitbalance}    {Reaction balance}
\newcommand{\compoundbenefitbalance}{metabolite balance}
\newcommand{\Compoundbenefitbalance}{Metabolite balance}
\newcommand{\loadbalance}           {reaction-metabolite balance}
\newcommand{\Loadbalance}           {Reaction-metabolite balance}
\newcommand{\summationcondition}    {flux variation rule}
\newcommand{\Summationcondition}    {Flux variation rule}
\newcommand{\connectivitycondition} {metabolite variation rule}
\newcommand{\Connectivitycondition} {Metabolite variation rule}
\newcommand{\summationconnectivitycondition}{variation rules}
\newcommand{\Summationconnectivitycondition}{Variation rules}

\newcommand{\MCA}{MCT}

\newcommand{\pointbenefitform}{value production form}
\newcommand{\valueform}{value form}

\providecommand{\esymbol}   {e}
\providecommand{\esymbolv}  {\mathbf{e}}
\providecommand{\cext}      {c^{\rm ext}}
\providecommand{\cextv}     {\cv^{\rm ext}}
\providecommand{\cint}      {c^{\rm int}}
\providecommand{\cintv}     {\cv^{\rm int}}
\providecommand{\prodrate}  {r}

\newcommand{\enzymev}   {\mathbf{\esymbol}}
\newcommand{\intprod}   {\prodrate^{\rm int}}

\newcommand{\rate}{\nu}

\newcommand{\ratelaw}{k}
\newcommand{\ratelawv}{\kv}
\newcommand{\enzdegrate}{\lambda^{\rm deg}}

\newcommand{\ffit}        {{\mathcal F}} 
\newcommand{\fluxbene}    {b}
\newcommand{\fluxcost}    {a}
\newcommand{\metcost}     {g}
\newcommand{\gplus}       {q}
\newcommand{\hminus}      {h}

\newcommand{\wsymbol}     {w}
\newcommand{\loadsymbol}  {y}

\newcommand{\partialder}{^{\centerdot}} 

\newcommand{\Nint}    {\Nmat^{\rm int}}
\newcommand{\Ntot}    {\Nmat^{\rm tot}}

\newcommand{\Next}    {\Nmat^{\rm x}}
\newcommand{\NR}      {\Nmat^{\rm ind}}
\newcommand{\Nobj}    {{\Nmat^{\rm ben}}}
\newcommand{\Kint}    {{\Kmat}}

\newcommand{\Lmatplus}{{\Lmat^{+}}}

\newcommand{\Deltar}{\square}

\newcommand{\cextj}{\cext_{j}}

\newcommand{\vvt}{{\tilde{\bf v}}}
\newcommand{\cvt}{{\tilde{\bf c}}}

\newcommand{\virtphi}    {{\prodrate}^{\rm int, vrt}}
\newcommand{\virtphiv}   {{\mathbf{\prodrate}}^{\rm int, vrt}}

\newcommand{\virtphiindv}{{\mathbf{\prodrate}}{^{\rm ind, vrt}}}

\newcommand{\virtgamma}  {{c^{\rm vrt}}}

\newcommand{\Ffit}   {{\bf F}}
\newcommand{\Fuu}    {\Ffit_{\rm uu}}

\newcommand{\stresssymbol}     {t}

\newcommand{\fu}     {{\bf \stresssymbol}_{\rm \esymbol}}

\newcommand{\ful}    {\stresssymbol_{\esymbol_l}}
\newcommand{\fuvs}   {\stresssymbol_{\rm v}}

\newcommand{\fes}    {\fes}
\newcommand{\fel}    {\fel}
\newcommand{\fevs}   {\fevs}

\newcommand{\fudotl} {\stresssymbol_{\esymbol_l}\partialder}
\newcommand{\fudots} {\stresssymbol_{\esymbol}\partialder}

\newcommand{\gus}   {\loadsymbol_{\rm \esymbol}}
\newcommand{\gul}   {\loadsymbol_{\esymbol_l}}

\newcommand{\gu}    {\loadvsymbol_{\rm \esymbol}}

\newcommand{\gusdot}   {\loadsymbol_{\rm \esymbol}\partialder}

\newcommand{\husdot}   {\hus\partialder}
\newcommand{\huldot}   {\hul\partialder}

\newcommand{\gc}    {\gplusv_{\rm \cint}}
\newcommand{\gci}   {\gplus_{\rm \cint_i}}

\newcommand{\gcm}   {\gplusv_{\rm cm}}

\newcommand{\gvtot}    {\gvsymbolv_{\rm v}}

\newcommand{\gvtotl}   {\gvsymbol_{v_{l}}}
\newcommand{\gvtots}   {\gvsymbol_{\rm v}}
\newcommand{\gvtoti}   {\gvsymbol_{\rm v_i}}

\newcommand{\gvindl}    {\loadsymbol_{v_l}}

\newcommand{\gtotlshade}{\gsymbol_{v_{l}}\partialder}

\newcommand{\hminusfun} {\hminus}
\newcommand{\hminusv}{{\bf \hminus}}

\renewcommand{\u}     {\esymbol}
\newcommand{\hul}     {\hminus_{\esymbol_l}}

\newcommand{\husweight}{\hat{\hminus}_{\esymbol}}
\newcommand{\huweight}{\hat{\mathbf \hminus}_{\esymbol}}

\newcommand{\hus}     {\hminus_{\rm \esymbol}}
\newcommand{\hu}      {\hminusv_{\rm \esymbol}}

\newcommand{\metcostfun}{\metcost}
\newcommand{\metcosti}  {\metcost_{c_i}}
\newcommand{\metcostc}  {{\bf \metcost}_{\rm c}}

\newcommand{\metcostdotci}   {\metcost_{c_i}\partialder}

\newcommand{\fluxbenev}     {{\bf \fluxbene}}

\newcommand{\bbenefit}  {\fluxbene}

\newcommand{\fluxbenetot}  {\fluxbene}
\newcommand{\fluxbenetotv} {\fluxbenev}
\newcommand{\bvtot}        {\fluxbenetotv_{\rm v}}
\newcommand{\bvtots}        {\fluxbenetot_{\rm v}}

\newcommand{\bvtotl}       {{\fluxbenetot _{v_l}}}

\newcommand{\bvtotj}       {{\fluxbenetot _{v_j}}}

\newcommand{\bvtotweight}  {\hat{\fluxbenev}_{\rm v}}

\newcommand{\bvdir}        {{\fluxbenev_{\rm v}^{\rm int}}}

\newcommand{\bvdirs}       {{\fluxbene_{\rm v}^{\rm int}}}
\newcommand{\bvdirl}       {{\fluxbene_{v_l}^{\rm int}}}

\newcommand{\bextv}        {\fluxbenev_{\rm \prodrate}^{\rm ext}}

\newcommand{\bvl}          { {\fluxbenetot _{v_l}}\partialder}
\newcommand{\bvv}          { {\fluxbenetotv_{\rm v}}\partialder}

\newcommand{\hmet}   {\metcost}
\newcommand{\hmetv}  {\mathbf{\metcost}}

\newcommand{\hc}     {\hmetv_{\rm c}}
\newcommand{\hci}    {\hmet_{c_i}}
\newcommand{\hcidot} {\hmet_{c_i}\partialder}

\newcommand{\hcs}    {\hmet_{c}}

\newcommand{\hctot}  {\hmetv^{\star}_{\rm c}}

\newcommand{\extprodv} {\mathbf{\prodrate}^{\rm ext}}

\newcommand{\bpsi}     {\fluxbenev_{\rm x}}

\newcommand{\bpsij}    {\fluxbene_{\rm x_j}}

\newcommand{\acost}     {\fluxcost}  
\newcommand{\acostenz}  {\fluxcost^{\rm enz}}  
  
\newcommand{\acostv}    {{\bf \acost}_{\rm v}}
\newcommand{\acostvl}   {\acost_{v_l}}

\newcommand{\acostvlweight}{\hat{\acost}_{v_l}}

\newcommand{\apointcostvl}   {\acost_{v_l}\partialder}

\newcommand{\hvv}   {\acostv}

\newcommand{\hvl}   {\acost_{v_{l}}}

\newcommand{\wvsymbol}{{\bf \wsymbol}}

\newcommand{\loadvsymbol}{{\bf \loadsymbol}}

\newcommand{\wtot} {\wvsymbol_{\rm \prodrate}}

\newcommand{\wtoti}{\wsymbol_{\prodrate_i}}
\newcommand{\wtotl}{\wsymbol_{\prodrate_l}}

\newcommand{\wint}  {\wvsymbol_{\rm \prodrate}^{\rm int}}
\newcommand{\wints} {\wsymbol_{\rm \prodrate}^{\rm int}}

\newcommand{\wintl} {\wsymbol_{\rm \prodrate_{l}}^{\rm int}}
\newcommand{\winti} {\wsymbol_{\rm \prodrate_{i}}^{\rm int}}

\newcommand{\wext}   {\wvsymbol_{\rm \prodrate}^{\rm ext}}
\newcommand{\wextv}  {\wvsymbol_{\rm \prodrate}^{\rm ext}}

\newcommand{\wextj}  {\wsymbol_{\prodrate_j}^{\rm ext}}

\newcommand{\wexts}  {\wsymbol_{\rm \prodrate}^{\rm ext}}

\newcommand{\loads}  {\loadsymbol}
\newcommand{\loadv}  {\loadvsymbol}
\newcommand{\loadint}{\loadv_{\rm c}}
\newcommand{\loadinti}{\loadv_{\rm c_i}}

\newcommand{\loadcm} {\loadv_{\rm cm}}
\newcommand{\loadi}  {\loads_{c_{i}}}
\newcommand{\loadeffi}  {\loadi^{\rm eff}}
\newcommand{\loadidot}{\loads_{c_{i}}\partialder}

\newcommand{\loadj}  {\loads_{\cextj}}

\newcommand{\hudot}        {{{\bf \hminus}_{\esymbol}\partialder}}
\newcommand{\hudots}       {{\hminus_{\esymbol}\partialder}}

\newcommand{\hudotl}       {{\hminus_{\esymbol_l}\partialder}}

\newcommand{\hudotlplusone} {\hminus_{\esymbol_{l+1}}\partialder}
\newcommand{\yvcostl}       {\fluxcost_{v_l}\partialder}

\newcommand{\ubenes}   {\loadsymbol_{\rm \esymbol}\partialder}
\newcommand{\ubenel}   {\loadsymbol_{{\esymbol_{l}}}\partialder}

\renewcommand{\modevector}{\kappav}

\newcommand{\fluxenzymecostl}  {\Delta \wsymbol_{\intprod_l:}}

\newcommand{\us}{\esymbol}
\newcommand{\ul}{\esymbol_{l}}

\newcommand{\hel}{\hminus_{u_l}}

\renewcommand{\gc} {\wvsymbol_{\rm c}}
\renewcommand{\gci}{\wsymbol_{\rm c_i}}
\renewcommand{\gcm}{\wvsymbol_{\rm cm}}

\newcommand{\gcint}{\wvsymbol_{\rm c}^{\rm int}}

\renewcommand{\gvtot}  {\wvsymbol_{\rm v}}
\renewcommand{\gvtotl} {\wsymbol_{v_{l}}}
\renewcommand{\gvtots} {\wsymbol_{\rm v}}

\renewcommand{\gvtoti} {\wsymbol_{\rm v_i}}

\renewcommand{\gtotlshade}{\wsymbol_{v_{l}}\partialder}

\newcommand{\gphialli}{\loadsymbol_{\prodrate_i}}

\renewcommand{\CJ}{C^{\rm v}}

\renewcommand{\CJmat}{{\bf C}^{\rm v}}

\renewcommand{\Nmat} {{\bf N}}

\renewcommand{\Lmat} {{\bf L}}
\renewcommand{\Kmat} {{\bf K}}

\renewcommand{\coout}[1]{}

\definecolor{brown}{rgb}{0.9,0.69,0.34}
\definecolor{lightyellow}{rgb}{1,0.99,0.85}

\newcommand{\psfileskinetic}{ps-files}

\renewcommand{\todo}[1]{#1}
\renewcommand{\co}[1]{}
\renewcommand{\myparagraph}[1]{}
 
\begin{document}

\title{Enzyme economy and metabolic control}
\coout{sounds like theory of everything?}

\date{} 

\author{Wolfram Liebermeister\\[3mm] 
Universit\'e Paris-Saclay, INRAE, MaIAGE, 78350 Jouy-en-Josas, France}

\maketitle

\begin{abstract} 
  \coout{m.e. einfuehren und bezug zu kinetischen modellen klarmachen}
  \coout{use standard terms!  make sure keywords are there!  one
    sentence intro, maybe two sentences am ende conclusion and impact}
  The metabolic state of a cell, comprising fluxes, metabolite concentrations
  and enzyme levels, is shaped by a compromise between metabolic
  benefit and enzyme cost. This hypothesis and its consequences can be
  studied by computational models and using a theory of metabolic
  value.  In optimal metabolic states, any increase of an enzyme level
  must improve the metabolic performance to justify its own cost, so
  each active enzyme must contribute to the cell's benefit by
  producing valuable products.  This principle of value production
  leads to {\summationconnectivitycondition} that relate metabolic
  fluxes and reaction elasticities to enzyme costs.  Metabolic value
  theory provides a language to describe this. It postulates a balance
  of local values, which I derive here from concepts of metabolic
  control theory.  Economic state variables, called economic
  potentials and loads, describe how metabolites, reactions, and
  enzymes contribute to metabolic performance.  Economic potentials
  describe the indirect value of metabolite production, while economic
  loads describe the indirect value of metabolite concentrations.
  These economic variables, and others, are linked by local balance
  equations.  These laws for optimal metabolic states define
  conditions for metabolic fluxes that hold for a wide range of rate
  laws.  To produce metabolic value, fluxes run from lower to higher
  economic potentials, must be free of futile cycles, and satisfy a
  principle of minimal weighted fluxes. Given an economical flux mode,
  one can systematically construct kinetic models in which all enzymes
  have positive effects on metabolic performance.
\end{abstract}


\co{JA!}  \co{move some text ABOUT tensions and evolutionary
  optimisation to CBA reg, discussion (optimal choice of enzyme levels
  in evolution)} \co{wort fuer ``load'': sowas wie ``ausstrahlung''?
  ``reach''?}  \co{WICHTIG: verweis auf SI, in dem neue CC definiert
  werden und in dem die beziehungen zwischen verschiedenen CC gezeigt
  und im sinne oekonomischer gesetze interpretiert werden (notfalls
  das noch dort schreiben)} \co{explain early on: MVT is based on
  assessing the fitness effects of state variations; here we descibe
  state variations by perturbation parameters and their control
  coefficients; we define metabolic values by control coefficients,
  derive formulae between these values from CC formulae.}
\co{nachschauen:
  https://en.wikipedia.org/wiki/Maximum\_power\_principle} \co{try to
  move some footnotes to the SI}

\co{WOERTER UND SYMBOLE} \co{neue bezeichnungen br, yr, wr einbauen
  und klar erklaeren; constraints on external prod rates are possible
  (erwaehnen)! prices for external concentrations are possible!
  (erwaehnen)} \co{clarify different meaning of $\bextv$ and the
  IDENTICAL $\wextv$, auch in table} \co{wort uea ``kinetic enzyme
  optimality problem'' hier und CBA lag usw?}  \co{``metabolic
  utility'' statt ``fitness'' uea?}  \co{enzyme-balanced means:
  extremal point; enzyme-optimal means optimal point} \co{ueberall
  point cost (\costshade): cost contribution, point benefit: benefit
  contribution usw?} \co{``production economics'' and
  ``concentration'' economics''} \co{does ``flux mode'' describe only
  the support, or also signs, or also relative magnitudes? stefan
  fragen} \co{in text, use ``enzyme-economical state'' for state in
  which all enzyme loads are positive} \co{general, use ``overhead''
  for ``indirect''; use ``load'' for ``overhead value'' and ``burden''
  for ``overhead price''?}  \co{think about symbols for ec pot and
  load (both INDIRECT); assume that internat production br and
  external concentrations gx can also be scored, and find good symbols
  and words that capture all this!}  \co{JA! machen und erklaeren!
  hier cint: c; cext: s; rint: r; rext:x! note that gx und br werden
  kaum verwendet! gc und bs schon!} \co{darauf achten: usage of
  enzyme-balanced (positive benefit, balanced by specified positive
  cost) vs enzyme-economic UEA (positive benefit). ist schon definiert
  in SI} \co{JA allgemein, in allen artikel oefters auch ``economic
  value'' statt ``economic variable''?}

\co{SCHREIBEN} \co{latex-symbole entruempeln} \co{indizes l weglassen
  wenn nicht direkt noetig? OFT vektoriell schreiben!}  \co{woerter in
  bildern?  // farben wie in letzten bildern // auf anderem bildschirm
  testen // erklaerungsbilder} \co{bilder mit enzymen: immer gleiche
  symbole (braune polygone) // einheitliche farben: enzym braun, immer
  gleich e8aa50; (dunklerer ton: dd8822) ziel als kasten}
\coout{interior optimum ueberall} \coout{VERWENDEN: flux benefit
  balance; {\fluxvalue} balance; compound benefit balance; }
\coout{bilder: EXT RAND DICK UEBERALL!}  \coout{``teleological''
  verwenden?  ueberall? //} \coout{cell doubling time ist nicht
  1/lambda!}  \coout{reconstruction of {\flow}s: von blaettern}
\coout{allg. clarify connection between lagrange and rules, laws etc:
  what are definitions and what are facts // beispiele? //
  terminologie und symbole } \coout{From Ron: - I think I need the
  posssible titles that give me a sense of the overarching story.
  currently it is very general.  - What is new in this paper?  It will
  help to clarify exactly what is new.  Do we have one "punchline" of
  the analysis? } \coout{konkreter titel; worum geht es genau?  was
  ist das ergebnis?  (kein "ing", econ pot benennen und sagen was
  passiert)} \coout{lagrange jetzt in CBA II: ans umschreiben denken}
\coout{in presenting the equations, follow the order from the lagrange
  derivation?}  \coout{feinheiten bei lagrange: use hv oder gv (d.h.,
  y oder nicht?)}  \coout{Use $\Upsilon$} \coout{manchmal: {\flow}:
  pathway flux?  flux vector?}  \coout{alles transponieren?}
\coout{hergo: erst ein konkretes beispiel} \coout{vielleicht solllten
  alle oekonomischen vektoren zeilen sein?}
\coout{Naturwissenschaften; JTB; BioSystems; MSB (Anhang abtrennen?)}
\coout{glycolysis example: make a number of optimal states for
  different {\flow}s: include PPP (for ribose phosphate production);
  anaplerotic and productive TCA cycle.  Growth on ethanol, glucose,
  fermentation, respiration, warburg effect} \coout{Nutzen: oph'eleia}
\coout{dass elastizitaeten durch oekonomie eingeschraenkt werden ist
  gut!}  \coout{Varian, Introduction to Microeconomics}
\coout{Biophysical Journal: limit 10 pages; they convert latex to doc}
\coout{allgemein: schoene referenzen in supplement von molenaar-paper}
\coout{Klamt S, Schuster S and Gilles ED. 2002. Calculability analysis
  in underdetermined metabolic networks illustrated by a model of the
  central metabolism in purple nonsulfur bacteria. Biotechnology \&
  Bioengineering 77 (7): 734-751 }

\co{DENKEN} \coout{nochmal ueber
  nicht-enzymatische reaktionen nachdenken (mit ähnlichem ansatz wie
  multifunktionalle enzyme behandeln?)}  \co{realistic didactic
  example, first solved by actual optimization or ECM // echte
  beispiele!  (hier oder CBA II?)  einfaches beispiel mit allen loads
  and potentials in paper?  //kleines beispiel mit ATP/ADP -
  e-coli-glycolyse + tca; einmal fermentation, einmal respiration
  (zahlen realistisch, zt aus ED-paper \"ubernehmen), enzymkosten
  \"uber enzymmengen absch\"atzen; evtl Noor 2016 in MCA variante?}

\co{PROGRAMMIEREN} \co{Problem in enzym-optimierung: Situation, in der
  enzyme abgeschaltet werden sollten, führt zu kleinen flüssen, die
  das numerische stationaritätskriterium erfüllen, aber
  nicht-stationär sind}

\iftoggle{bookversion}
{\section{Metabolic value theory in kinetic models}}
{\section{Introduction}}

\myparagraph{\ \\Optimal enzyme levels in cells} The metabolic fluxes
in cells are catalysed and steered by enzyme activities.  How should
the cell's enzyme resources be allocated to pathways, to reactions
along a pathway, and between the reactions around a metabolite? How
will enzyme investments in one place change the incentives for
investments elsewhere, given the complex metabolic dynamics and
competition for protein resources?  At what enzyme cost will a pathway
cease to be profitable?  And when an enzyme is inhibited, should it be
overexpressed (to compensate its lower efficiency) or be shut down
together with the rest of the pathway (because the pathway is now
inefficient)?  An optimal allocation of protein resources implies
compromises between {\metabolicobjective}s (resulting from fluxes and
metabolite concentrations) and enzyme cost (arising, e.g.~from a
competition for protein resources with other cell processes).  \co{ref
  rosen's book on biol optimality? heinrich + schuster book} Since
J.~Reichs seminal work on enzyme expression as a cost-benefit problem
\cite{reic:83}, \co{also fruehere ref proposed by herbert sauro (siehe
  book MCA chapter)} many optimality principles for optimal fluxes and
enzyme profiles have been proposed \cite{hesc:98}. In kinetic models,
enzyme levels were chosen to maximise metabolic flux at a given total
enzyme budget \cite{klhe:99} or to minimise enzyme cost
\cite{deal:05}.  Optimality assumptions can be used to predict how
enzyme investments should be distributed along pathways, which
pathways should be used, and how these choices depend on the cell's
life conditions.  Kinetic models in enzyme-optimal states also serve
as starting points for modelling optimal enzyme adaptations
\cite{lksh:04} and metabolic cycles \cite{lieb:14c}.

\myparagraph{Laws for optimal metabolic states} While
optimal enzyme profiles can be found numerically
\cite{klhh:02,zmrb:04}, some  questions remain. Can a given
flux distribution be realised by an enzyme-optimal state, and can we
construct this state and the kinetic model behind it?  And are there
general principles behind optimal metabolic states or, in other words,
economic laws of metabolism?  Intuitively, we may expect that the
``investome'' -- the pattern of enzyme costs spent in reactions or
pathways -- reflects a ``usefulness'' \cite{lnfd:14}, that
is, a benefit these reactions or pathways provide (or in other words,
the fitness loss if the reation or pathway did not exist). A
relationship between enzyme investments and metabolic control
\cite{brow:91} was shown by Klipp and Heinrich \cite{klhe:99}. They
asked how a pathway flux can be maximised at a given total enzyme
amount (and without costs or bounds for metabolite concentrations) and showed
that the enzyme levels in the optimal states must be proportional to
the scaled flux control coefficients \cite{hesc:96,hoco:00}. So in
this case, if enzyme levels are seen as investments, scaled flux
control can be seen as usefulness! This confirms our intuition: in
optimal states, there must be a balance between investments and
usefulness, or between cost and benefit -- i.e.~between the cost of a
virtual extra amount of enzyme and the benefit of the resulting flux
increase. But do such principles hold more generally?  What if
different enzymes are differently costly? And what if metabolite costs
and other constraints are taken into account? Below I derive economic
laws for a wide class of metabolic optimality problems, formulated as
balance equations and resembling Kirchhoff's rules for voltages and
currents in electrical circuits. \co{ref CBA fluxes} In contrast to Kirchhoff's rules,
these balances do not concern our physical variables (such as metabolite
concentrations or fluxes), but economic variables -- the costs and
benefits, defining the value structure of a metabolic state.

\begin{figure*}[t!]
    \begin{center}
      \includegraphics[width=16.5cm]{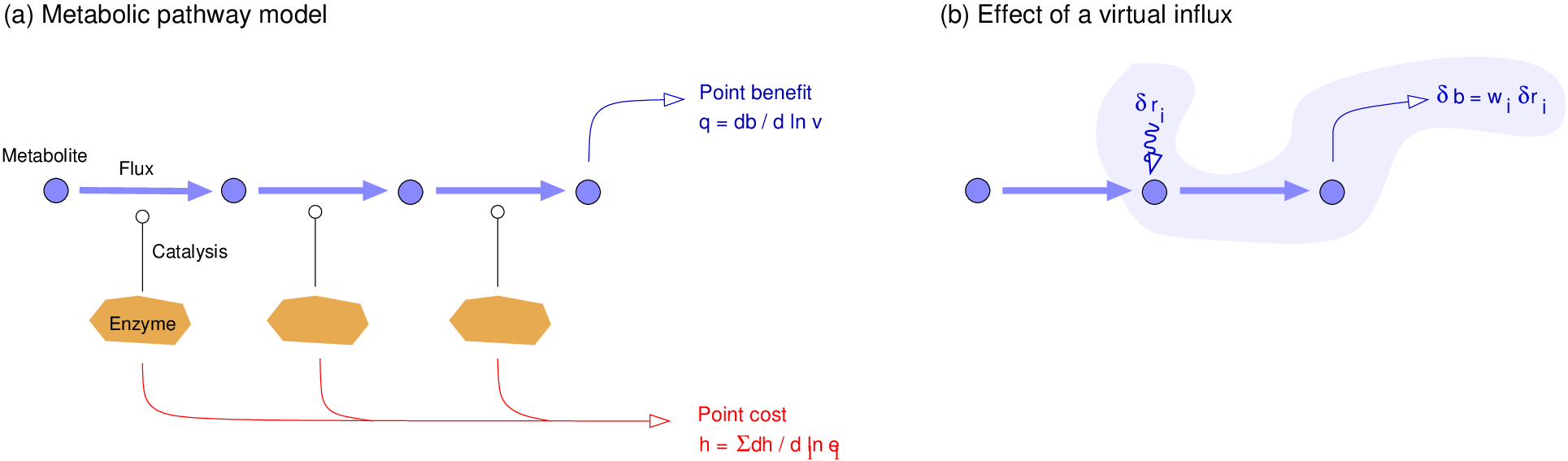}\\
    \end{center}
    \caption{Metabolic cost/benefit problem. (a) {\metabolicobjective} and
      enzyme cost. In the pathway, our running example, metabolic
      production is scored by a benefit $\bbenefit(v_{\rm prod})$
      while enzyme levels $\esymbol_{l}$ are scored by a cost
      $\hminus(\esymbolv)$. The cost function may describe opportunity
      costs in a cell with a limited protein budget, limited space,
      and limited material and energy. The model may describe a
      biosynthesis pathway (a series of metabolic reactions) or the
      cell as a whole (with reactions describing nutrient uptake,
      metabolism, and macromolecule production, and a flux
      proportional to the cell growth rate \cite{lieb:18theory}. In
      each possible state, we can define a benefit
      $b=\sum_{l} \partial b_{v_{l}}\, v_{l}$ and a cost
      $h= \sum_{l} \hul\, \esymbol_{l}\, \esymbol_{l}$ (where $\hul$
      may stand, for example, for enzyme molecular mass).
      (b) Economic
      potentials.  We imagine a virtual influx $\delta \prodrate_{i}$
      of metabolite $i$, leading to a change of steady-state fluxes
      and concentrations and to a benefit change $\delta
      \bbenefit$. In a linear approximation
      $\delta \bbenefit = \winti\,\delta \prodrate_{i}$, the
      prefactors $\winti$ are called economic potentials and describe
      the metabolites' ``use values''. In an optimal state, the
      potentials increase along the pathway and the equation
      $v_{l}\,\Delta \wtotl = \hul\,\esymbol_{l}$ must hold in every
      reaction.}
 \label{fig:simpleExample} 
\end{figure*}

\myparagraph{Optimal state of a linear metabolic pathway} As a running example,
we consider a chain of reactions, e.g.~a metabolic pathway or a peptide
synthetase assembly line consisting of polymerisation reactions (see
Figure \ref{fig:simpleExample}).  What are the optimal enzyme levels in this pathway?
To specify  the problem, we describe the pathway by a kinetic model and
define a flux benefit function $\fluxbenetot(\vv)$ that saturates at
high fluxes. The enzyme levels $\esymbol_l$ are scored by a linear
cost function $\hminus = \sum_l \hul'\,\esymbol_l$ and our aim is to
find the enzyme profile that maximises the benefit-cost difference. We can
do that by numerical optimisation, with two possible outcomes:
if the enzymes are too costly, all reactions will be switched off;
otherwise, we obtain an optimal enzyme profile sustaining a positive steady flux. A closer
look at this state yields some curious observations: by multiplying
the flux $v$ with the benefit derivative $\md b/\md v$, we obtain
 the total enzyme cost.  More surprisingly, a similar
balance holds for each single reaction. To see this, we  define
the notion of economic potentials. If we increase the production of
metabolite $i$ by an additional ``free'' influx $\delta r_i$, this
``gift'' will improve the overall benefit, and if we write the
increase, in a linear approximation, as
$\delta b \approx \winti\,\delta r_i$, the coefficient $\winti$ is
called economic potential. For external metabolites on the pathway
boundaries, a different definition of economic potentials is used: the pathway
substrate, which does not appear in the benefit function, has a
potential of 0, while  the pathway product has a potential of
$\md b/\md v$.  If we now compute an optimal state, the economic
potentials will always increase along the chain: in every reaction
$l$, the difference $\Deltar \wtotl$ is positive. Moreover, in each
reaction the difference $\Deltar \wtotl$, multiplied by the flux
yields exactly the enzyme cost of the reaction!  Since all reactions
share the same flux, the differences $\Deltar \wtotl$ and enzyme costs
are proportional.

\myparagraph{Generality of the findings} Importantly, all these
findings hold for models with \emph{any} reversible rate laws and any
choice of model parameters. For example, if we increase all enzyme
cost weights $\hul'$ by a constant factor and optimise again, the
optimal flux decreases but we obtain the same relationships as before
(unless the cost weights become too large; then the enzymes are shut
off and the flux will vanish).  Even more surprisingly, we can find
similar relationships not only for linear pathways, but for metabolic
networks of any structure or size, and even for other optimality
problems, for example, with regulation arrows or with objectives that
penalise high metabolite concentrations. Of course, all these problems
could be treated numerically, but since the relations are so general,
we may study them as general laws in their own right.

\coout{MP: here I have a question: do we need general principles
 (/laws) to answer the questions above? or are you switching now to
 actually other questions? If the former, do we need general
 principles because the questions are general (i.e.~aim at
 generalizations about many cells / pathways / etc.)? or do we need
 general principles also to answer specific questions (e.g.~because
 we need general criteria in order to be able to do some decisions
 even in specific models)? I imagine: the former; then the
 latter. But I think your being a bit too cursory in both moving away
 from kinetic models and in speaking of "general laws" generates some
 confusion here, or causes overlaps of different sets of questions. i
 think part of the confusion is also generated by "in general" (p.2,
 1.10). Because this may mean / refer to various things at various
 levels of analysis. Maybe a way to frame my confusion is: what does
 it mean to "study enzyme usage in general"? E.g., "study in general"
 or "usage in general" etc. Is the fact of generalizing (vs. the
 case-specific scope of existing approaches) your main contribution
 here? This part of the text suggests this (to me), but I don't think
 this captures your point fully. I wonder whether the point would be
 more explicit if you said (along the lines of bc. this is a bit
 clumsy) that what we still don't have is an analytical tool /
 approach / framework that enables us to focus specifically on enzyme
 usage but as it occurs across cases; that is, on relationships
 holding for enzymes as such insofar as they involved in a specific
 type of process. To me, what is new is the type of process on which
 you decide to focus (or the frame through which you look at a type
 of process, or the question you ask of the process), because it is
 this that then enables you to, as it were, add another sense to the
 definition of "enzyme usage". In this sense there is a "redundancy"
 to general laws (if this is a possible analogy: they add sinn to
 many cases but don't define for any case a new
 bedeutung?). Basically I think you need to make clear that you're
 generalizing precisely by focusing on a specific question, or
 whatever the nature of your generalizing is. If it is by focusing,
 then say on what question. }

\coout{MP: [this is my brief an die grossmutter (/mp) version] Insofar they
 only model enzyme usage given some specific metabolic objective,
 these approaches do not enable us to answer the questions raised
 above. This is the case because those questions investigate enzyme
 usage given any (i.e.~independently of any specific) metabolic
 objective. My questions leave specific metabolic objectives out of
 the picture in order to focus on the following. One, the ways in
 which any instance of enzyme usage is related to other instances of
 enzyme usage in the system considered. In other words, aspects of
 enzyme usage that ensue from its being part of a set of economic
 processes. Two, properties that can be attributed to enzymes once
 their nature as economic units in such processes is appreciated
 [some other verb; or ?: once they are seen in the context of
  economic processes]. }

\begin{figure*}[t!]
\begin{center}
  \includegraphics[width=16.5cm]{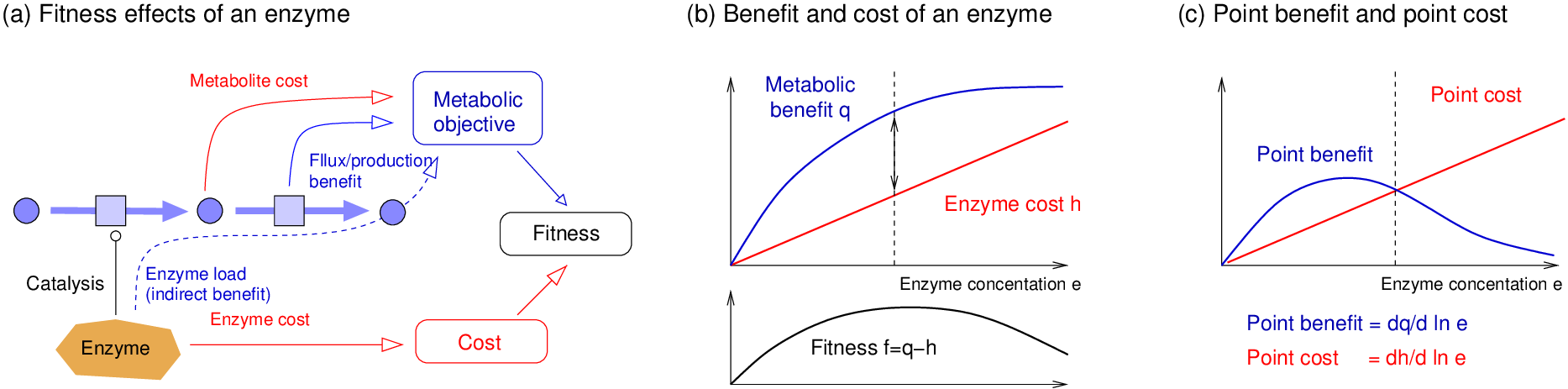}\\
\end{center}

\hspace{7cm}
\parbox{4cm}{Value/price balance\\[2mm]
  $\underbrace{\frac{\partial \gplus}{\partial \us}}_{\mbox{value}\,  \gus}=
  \underbrace{\frac{\partial \hminus}{\partial \us}}_{\mbox{price}\,\hus} \stackrel{!}{>}0$}
\parbox{5.5cm}{Benefit/cost balance\\[2mm]
  $\underbrace{\frac{\partial \gplus}{\partial \us}\,\us}_{\mbox{point benefit}\,\gusdot} =
  \underbrace{\frac{\partial \hminus}{\partial \us}\,\us}_{\mbox{point cost}\,\husdot}\stackrel{!}{>}0$}
\caption{Enzyme-optimal states and economic values. (a) Optimal enzyme
  levels in a kinetic metabolic model. An enzyme (with level
  $\esymbol$) catalyses a reaction with an indirect effect on fluxes
  $v$ and concentrations $c$ everywhere in the network. A varying
  enzyme level affects fitness in two ways, via cost and benefit. The
  {\metabolicobjective}
  $\gplus(\esymbol)=\bbenefit(\vsteady(\esymbol)) -
  \metcostfun(\csteady(\esymbol))$ consists of flux benefit
  $\bbenefit(v)$ and metabolite cost $\metcostfun(c)$, and a cost
  function $\hminusfun(\esymbol)$ penalises high enzyme levels.  The
  fitness function is defined as the difference
  $\ffit\steady(\esymbol)=\gplus(\esymbol)-\hminusfun(\esymbol)$.  (b)
  {\metabolicobjective} $\gplus$ and enzyme cost $\hminus$ as functions of
  one of the enzyme levels. In the example, benefit and cost depend,
  respectively, nonlinearly or linearly on enzyme levels. In the
  optimal point the benefit-cost difference (i.e.~fitness), must be
  maximal, so both curves must have the same slopes.  If we call the
  two slopes ``{\myvalue}'' (slope of benefit curve) and ``{\price}''
  (slope of cost curve), then {\myvalue} and {\price} must be equal in
  optimal states. Since all enzyme prices are positive, active enzymes
  must have a positive (indirect) influence on the {\metabolicobjective}.
  (c) The derivatives of cost and benefit function with respect to
  logarithmic enzyme levels are called {\costshade} and
  {\benefitshade}.  Again, in optimal states {\costshade} and
  {\benefitshade} must be equal. }
 \label{fig:backpropagation} 
\end{figure*}

\myparagraph{Metabolic value theory} Metabolic value theory (MVT)
\cite{lieb:18theory,lieb:18lagrange} defines economic variables that
complement the physical variables in a model, describe their fitness
values, and are subject to balance equations (Figure
\ref{fig:backpropagation}). Economic values can be defined for a wide
range of models by shadow values obtained from optimality problems
\cite{lieb:18lagrange}: to this aim, optimality problems must be
written in an ``expanded form'', in which all physical laws are
formulated as explicit constraints. Each active constraint defines a
shadow value. The shadow values arising from mass-balance constraints
define values of individual metabolites, called economic potentials
\cite{lieb:18lagrange, lieb:14b}.  Economic variables for a variety of
(kinetic and constraint-based) metabolic models can be defined
similarly. Economic values describe the ``use value'' of network
elements, i.e.~the effect of small variations of physical variables on
fitness as defined in our model.  In optimal states, these use values
equal to ``embodied values'' that result from enzyme
investments. Formally, the economic laws for metabolic states resemble
laws of thermodynamics.  Thermodynamic laws relate metabolite
concentrations to chemical potentials, thermodynamic forces and flux
directions, and must hold in any metabolic system, with any reaction
kinetics. Similarly, economic balance equations represent optimal
enzyme usage, independent of the details of enzyme kinetics.  The laws
can also be seen as conservation laws for economic value, describing
conserved value flows. \co{REF physical field theories:
  \cite{lieb:cbafield}} In this picture of metabolism, value flows
into the system in the form of substrate and enzyme investments,
accumulates, and leaves the system in the form of metabolic
benefit. 

\myparagraph{A derivation from Metabolic Control Theory} Economic
variables describe the value of physical variables, that is, the
fitness effects of small variations. The variations need not occur in
reality, but are used for mathematical arguments.  Optimal states can
be characterised by a simple condition: as shown in
\cite{lieb:18lagrange}, no legal (that is, constraint-respecting)
variation of the state can improve fitness. In the same paper,
economic variables were defined through Lagrange multipliers. In a
metabolic optimality problem, after expressing all dependencies
between model variables by explicit constraints, we obtain Lagrange
multipliers associated with these constraints, which can be
interpreted as economic variables.  Alternatively we can describe
constraint-violating variations by perturbation variables: for
example, violations of a metabolite's mass-balance can be described by
a virtual influx of the metabolite. In kinetic models, the effects of
these perturbations on steady states can be captured by metabolic
control or response coefficients, a concept from Metabolic Control
Theory ({\MCA}) \cite{hesc:96,hofm:01}. \co{kurze erklaerung von MCA
  fuer leute, die es nicht kennen (grundbegriffe!); und warum es hier
  relevant ist} The economic variable associated with a constraint can
be defined by the response coefficient between a virtual variable
(perturbing the constraint) and the system's objective function. Here
I will use this idea for an alternative derivation of metabolic value
theory: I consider kinetic metabolic models with enzyme levels as
control variables and derive the economic laws from the summation and
connectivity theorems of {\MCA}.  Economic values are defined by
metabolic response coefficients, matching the existing definition by
shadow values (see SI section
\ref{sec:ProofsPotentialsAreIdentical}). While the previous definition
is more general (and also applicable to constraint-based models), the
link between metabolic values and metabolic control provides
interesting insights and makes a direct connection to Klipp and
Heinrich's results \cite{klhe:99}.

\myparagraph{Overview of the article} In this article, we consider
kinetic metabolic models whose states are scored by a fitness
function, a function of fluxes, metabolite concentrations, and enzyme
levels.  For steady states with fluxes $\vvsteady$ and internal
concentrations $\cvsteady$, we obtain the steady-state fitness
$\ffit\steady(\enzymev,\cextv)
=\ffit(\vvsteady(\enzymev,\cextv),\cvsteady(\enzymev,\cextv),\enzymev)$.
Optimal states must be enzyme-balanced, that is, the condition
$\partial \ffit\steady/ \partial \enzymev=0$ must hold for all active
(i.e.~expressed) enzymes.  As a consequence, each active enzyme must
have a positive benefit derivative to balance its cost derivative\co{:give math symbols for clarity?}: in
the language of {\MCA}, the response coefficient between enzyme level
and benefit function must be positive. Using the theorems of {\MCA}, I
show that this implies a principle of local value production: enzymes
must produce valuable metabolites from less valuable ones (unless the
catalysed flux has a direct benefit). Therefore, fluxes must run from
low to high economic potentials, which excludes futile submodes just
like thermodynamic constraints would exclude certain flux cycles. Such
fluxes are called economical, and only economical fluxes are
compatiable with an optimal choice of enzyme levels. Next, I define
economic values for metabolite concentrations and metabolic
production, derived from the global benefit function.  Economic rules
and balance equations connect these economic values between
metabolites, reactions, and enzymes in the network.  Based on these
laws, we can construct kinetic models in enzyme-balanced states with
predefined fluxes. Such models are useful for studying enzyme
adaptation in changing environments \cite{lksh:04} or optimal enzyme
regulation by effector molecules \cite{uhle:09}. The theory holds not
only for simple examples -- as shown in this paper -- but also for
large metabolic or non-metabolic systems (e.g.~including protein
biosynthesis).

\coout{MP: Then I would continue with the rest of your intro: My approach
 to this [/the] question is as follows. - whether we can require
 that what holds globally also holds locally; this entails an analogy
 with labor value in economics; thus: what needs to be done to
 transfer it here - what is shown here, what shape is given to the
 theory (metabolic control theory rather than Lagrange multipliers,
 etc.) (current p. 3), analogies with other types of potential and
 why you call your concept potential; also here most of the current
 last paragraph of the introduction?]; also here the third [ ] on
p. 4 - what the theory allows us to do at the theoretical level - what
the theory allows us to do when modeling - why these advances are
important to progress The paper is organized as follows. Etc. }

\section{Kinetic models with cost and benefit terms} 
\label{sec:economickinetic}

\coout{ beispiel mit numerischer loesung; lineare kette? bsp:
 gekruemmte nutzenkurve z(vprod(product)), gerade kostenkurve sum hat
 hul ul? }

\myparagraph{\ \\Enzyme optimisation in kinetic models} To study
enzyme-optimal states, here we consider kinetic metabolic models and
score their metabolic states by a fitness function (Figure
\ref{fig:backpropagation} (a)), given by a difference of flux benefit,
metabolite cost, and enzyme cost \cite{reic:83,lksh:04,deal:05}
(Figure \ref{fig:backpropagation} (b)).  In theory, optimal enzyme
profiles can be computed numerically, but here we are not interested
in in numerical results, but in general laws. To obtain such laws, all
network elements (reactions, metabolites, and enzymes) are
characterised by economic variables, describing costs and benefits
associated with these elements. An economic variable describes how
virtual changes in a physical variable affect the overall fitness,
either directly or indirectly. While such variables can be defined by
Lagrange multipliers, I present here an alternative definition based
on methods from {\MCA}: we consider a violation of mass balances by
virtual supply fluxes, and study their effects on stationary
concentrations and fluxes.  Mathematical definitions and proofs can be
found in the Supplementary Information (SI). Metabolic value theory
introduces new terminology and mathematical symbols: for an overview,
see tables \ref{tab:listofsymbols} and \ref{tab:listofsymbols2} in the
SI. \co{note: math details and proofs are in separate supplements!}
MATLAB code is available on github \cite{github:meteco}.  For more
information about metabolic value theory, see
www.metabolic-economics.de.

\myparagraph{Kinetic model and metabolic {\flow}s} Kinetic metabolic
models describe the dynamics of metabolite concentrations $\cint_{i}$
and chemical reaction rates $v_l$. Aside from internal metabolites,
there are external metabolites with fixed concentrations $\cext_{j}$,
treated as model parameters. The reaction rates are determined by rate
laws\footnote{For simplicity, we assume that each reaction is
  catalysed by a single specific enzyme. Generalisations will be
  discussed below.}
$\rate_{l}(\enzymev,\cintv,\cextv) = \esymbol_{l} \,
\ratelaw_{l}(\cintv,\cextv)$ with enzyme levels\footnote{For
  simplicity, we assume that enzyme concentrations (or ``enzyme
  levels'') directly determine enzyme activities.  In reality, enzyme
  activities can be modulated by posttranslational modification
  (e.g.~phosphorylation).} $\esymbol_l$, internal metabolite
concentrations $\cint_{i}$, and external metabolite concentrations
$\cext_{j}$. A flux distribution $\vv$ is called stationary or steady
(or a {\flow}) if inflows and outflows of internal metabolites are
balanced: internal metabolites do not accumulate nor deplete. \co{If N
  contains dependent rows (ie if there exists a matrix $\Gmat$ such as
  $\Gmat\,\Nmat=0$), then some linear combination of th einternal
  metabolite concentrations (``conserved moiety concentrations'') will
  be constant in time, that is,
  $\cv_{\rm cm}(t)=\Gmat\,\cv_{\rm int}(t)=\const$} If all reactions
in a {\flow} carry non-zero fluxes, the {\flow} is called
\emph{\complete}.

\myparagraph{Fitness function, {\metabolicobjective}, and enzyme cost}
To develop a metabolic value theory for kinetic models, we treat
enzyme levels in a metabolic pathway or network as control variables
that determine a steady state\footnote{For simplicity, we assume that
  given enzyme levels (and conserved moiety concentrations, determined
  by initial conditions for $\cv$) lead to a unique metabolic steady
  state. If multiple steady states exist, we consider only one of
  them. The theory does not apply at bifurcation points, where steady
  states appear or disappear.}. To describe the effects of enzyme
levels $\enzymev$, metabolite concentrations $\cintv$, and fluxes $\vv$ on
cell fitness, we assume an effective fitness objective
$\ffit(\vv,\cintv,\enzymev)$.  Enzymes in cells are costly: even
beneficial pathway fluxes may not be profitable if a pathway requires
excessive amounts of enzyme \cite{brow:91,klhe:99}. If an enzyme does
not contribute to {\metabolicobjective}, it should be repressed to save
costs. Moreover, the higher a pathway's enzyme cost, the higher the
benefit the pathway needs to provide to balance this cost.  To capture
these trade-offs, we consider a metabolic pathway or network with
variables $\vv$, $\cv$, and $\esymbolv$, and score the possible states
by a fitness function \co{lieber einfaches symbol f!}
\begin{eqnarray}
 \label{eq:fitnessFunctionFormula}
 \ffit(\vv,\cintv,\enzymev) = \bbenefit(\vv)-\metcostfun(\cintv)-\hminusfun(\enzymev)
\end{eqnarray}
comprising a flux benefit $\bbenefit(\vv)$, a metabolite cost
$\metcostfun(\cintv)$, and an enzyme cost $\hminusfun(\enzymev)$
\cite{lieb:18theory}.  For convenience, we sometimes combine the
objective terms and define the {\metabolicobjective}
$\gplus(\vv,\cintv) = \bbenefit(\vv)-\metcostfun(\cintv)$ or the
kinetic cost
$\metcostfun^{\rm kin}(\cintv,\enzymev) =
\metcostfun(\cintv)+\hminusfun(\enzymev)$. The flux benefit function
$\bbenefit(\vv)$ may score metabolic production or conversions,
cofactor conversion, or biomass production. The cost terms
$\metcost(\cintv)$ and $\hminus(\enzymev)$ penalise high metabolite or
enzyme levels \cite{sche:91,tnah:13}: they describe costly effects
(e.g.~of occupying space) that arise outside our pathway model
\cite{lksh:04}. \co{:oben einmal sagen, was das heisst! (auch
  boundary)} In some cases, the function $\metcostfun(\cintv)$ may
also penalise low metabolite concentrations (e.g.~to account for a
metabolite's concentration benefits outside the model pathway).

\myparagraph{Choice of fitness functions} A fitness function describes
what a cell, according to the modeller, strives to maximise to obtain
a selection advantage in the growth condition considered. How should
we choose it?\co{refer to CBA opt, where all this should be clearly
  explained // verbindung zu dahinterliegenden kosten auch besprechen
  in CBA local / CBA growing cells / CBA labour / CBA opt} Fitness
functions of the form (\ref{eq:fitnessFunctionFormula}) -- a
benefit-cost difference -- do not follow from deeper biological
principles, but are used for mathematical convenience\footnote{A
  benefit/cost ratio may be even more plausible than a benefit-cost
  difference. For example, the biomass/catalytic rate, defined as
  ``biomass production rate $v_{\rm BM}$ per total amount of metabolic
  enzyme $\esymbol_{\rm met}$'', can be treated as a proxy for cell
  growth \cite{sgmz:10, wnfb:18, lieb:18fcm}. By taking logarithms,
  this ratio can be converted into a difference
  $\ffit=\ln v_{\rm BM} - \ln \esymbol_{\rm met}$.}.  \co{FN! lieber
  nur CBA lag und hier darauf verweisen? A fitness function of this
  form may also reflect a Pareto-optimal state with 3 objectives, the
  correct condition for function (\ref{eq:fitnessFunctionFormula}) would not be not a
  maximality, but an extremality condition (saddle point).  Depending
  on the curvature of the Pareto front, there may be cases in which an
  ``economically unstable'' state still corrresponds to a Pareto
  optimaum between flux benefit, metabolite cost, and enzyme
  cost. HIER (UND IN CBA OPT) sagen: wir fordern hier trotzdem
  maximalitaet, aber bestehen meistens sowieso nicht darauf.} To
obtain fitness functions for pathways, we may start from a cell
fitness function $F$ (e.g.~the cell growth rate) and define an
``optimistic'' pathway objective $\ffit(\vv,\cintv,\enzymev)$ as the
maximal possible value of $F$ given our pathway variables
$\vv,\cintv,\enzymev$.  We can define this function as
$\ffit(\vv,\cintv,\enzymev)=\max_{\zv} F(\zv|\vv,\cv,\enzymev)$ where
$\zv$ denotes cell variables outside the pathway, to be optimised at
given $\vv, \cv,$ and $\enzymev$ and under the constraints of the cell
model\footnote{Constraints in whole-cell models may define bounds and
  dependencies for the variables in a pathway of interest. Here we
  ignore such dependencies except for direct dependencies through
  kinetic rate laws within the pathway).}.  Functions
$\ffit(\vv,\cintv,\enzymev)$ defined in this way may be complicated
and possibly not differentiable. For convenience, we replace or
approximate them by the simple function
Eq.~(\ref{eq:fitnessFunctionFormula}) and assume that all three terms
are differentiable\footnote{Below we mostly usually consider fitness
  derivatives derivatives, so instead of a difference
  Eq.~(\ref{eq:fitnessFunctionFormula}), we may also use a general
  function $\ffit(\vv,\cv,\esymbolv)$, as long as it is differentiable
  in the state in question, and replace, below,
  $\partial \bbenefit/\partial \vv \rightarrow \partial \ffit/\partial
  \vv, \partial \metcostfun/\partial \cv \rightarrow -\partial
  \ffit/\partial \cv, \partial \hminusfun/\partial \esymbolv
  \rightarrow \partial \ffit/\partial \esymbolv$.}.

\myparagraph{Metabolic objectives, flux {\gain}s, and metabolite
  {\price}s} Below we will usually not consider the entire function
$\ffit$, but its derivatives (which represent ``values'').  Typically,
the terms $\bbenefit$ and $\metcostfun$ in our fitness function
(\ref{eq:fitnessFunctionFormula}) score only a small number of model
variables: these are the variables with direct fitness effects.
Direct (i.e.~partial) derivatives of benefit and cost functions,
called {\gain}s and {\price}s, describe how small variations of fluxes
$v_{l}$ or concentrations $\cint_{i}$ would directly affect the
{\metabolicobjective}, and how small variations of $\esymbol_{l}$ would
affect enzyme cost.  The flux benefit function $\bbenefit(\vv)$ yields
the {\fluxgain}s $\bvtotl = \partial \bbenefit/\partial v_{l}$. For
simplicity, we assume flux benefit functions of the form
$\bbenefit^{\star}(\vv) = \bbenefit^{\rm dir}(\vv) + \bbenefit^{\rm
  ext}(\extprodv(\vv))$, with a direct term for fluxes and a term for
the external metabolite rates. With
$\bvdir = \partial \bbenefit^{\rm dir}/\partial \vv$ and
$\bpsi = \partial \bbenefit^{\rm ext}/\partial \extprodv$, the
{\fluxgain} vector reads
\begin{eqnarray}
 \label{eq:splitmarginalfluxbenefit} 
\bvtot = \bvdir + {\Next}\trans\, \bpsi,
\end{eqnarray} 
where $\Next$ is the stoichiometric matrix for external metabolites.
The production {\gain}s $\bpsij$ score the production or consumption
of external metabolites, while the flux {\gain}s $\bvdirl$ score
fluxes directly\footnote{For example, if heat production provides
  benefits, this can be described by an extra term $\bvdirl$.}. The
splitting of $\bvtot$ into flux {\gain}s and production {\gain}s is
not unique \coout{or, one could say, it is a matter of gauging} and
can be chosen by the modeller\footnote{On the one hand, we may set
  $\bpsi=0$, and the {\fluxgain}s $\bvtotl$ are given by direct
  {\fluxgain}s $\bvdirl$. On the other hand, we may formally set
  all~direct {\fluxgain}s $\bvdirl$ to zero and express all
  {\fluxgain}s by production {\gain}s $\bpsij$ of virtual external
  metabolites (which are introduced just for this purpose). For a
  standard convention for splitting the {\fluxgain}s, we may minimise
  $||\bvdir||$ under the constraint $\bvdir + \Next \wext = \bvtot$,
  where $||\cdot||$ can be the Euclidean norm or the 1-norm.}. The
derivatives of the cost terms are called metabolite {\price}s
$\hci = \partial \metcostfun/\partial \cint_{i}$ and {{\enzymeprice}}
$\hul = \partial \hminusfun/\partial \esymbol_{l}$.  Enzyme {\price}s
are positive, and an enzyme cost function
Eq.~(\ref{eq:investmentfunction}) yields
$\hul=h'_l (\lambda+\enzdegrate_{l})\,L_{l}$.  If higher metabolite
concentrations provide an advantage outside the model pathway,
metabolite {\price}s can also be negative. If the metabolic objective
depends only on fluxes (\emph{flux objective} $\bbenefit(\vv)$), and
not on metabolite levels, the concentration {\price}s vanish. A
metabolic objective with a {\fluxgain} $\bvtot = {\Next}\trans \bpsi$
and $\hc=0$ (i.e.~without flux gains or metabolite prices) is called a
\emph{production objective}. Besides fitness effects, the {\gain}s and
{\price}s can also reflect the effects of constraints. For example, if
a reaction rate must be kept above some minimum value, we can describe
this by a flux bound. In this case, our metabolic objective may favour
a low flux, but the constraint will prevent this: if the flux hits its
lower bound, the ``force'' that prevents a further decrease is
described by a shadow value (Lagrange multiplier) which adds to the
{\fluxgain} for this reaction (see appendix on ``constraints on state
variables''). With an an upper flux bound, the effective {\fluxgain}
is negative, and with inactive bounds it is zero\footnote{Shadow
  values differ from state to state.  Bounds may concern single
  metabolites or enzymes or sums of compounds and can ensure positive
  enzyme levels. Similar to bounds, we may also fix external
  metabolite concentrations and conserved moiety constraints. All such
  constraints lead to terms in the metabolite and enzyme {\price}s.}.
Similarly, bounds on metabolite concentrations lead to positive
{\price}s (for upper bounds)\footnote{Inactive enzymes are described
  in a similar way: a lower bound, preventing negative concentrations,
  leads to a negative shadow {\price} that cancels the enzyme
  {\price}.} or negative {\price}s (for lower bounds). As a rule of
thumb, active lower bounds act like benefits, while active upper
bounds act like costs \cite{lieb:18lagrange}.

\myparagraph{Enzyme cost function and enzyme price} Enzyme cost
functions $\hminusfun(\enzymev)$ for growing microbes have been
defined operationally by measuring growth defects caused by an
expression of idle proteins. In the cell, these impairments are
mediated by complicated processes and compromises (involving enzyme
production and maintenance, ribosome production, and limited space due
to crowding)\footnote{Such protein costs increase with protein levels,
  and measurements suggest that they are linear \cite{stdd:08,sgmz:10}
  or positively curved \cite{deal:05,szad:10}. Enzyme cost can be
  attributed to various cell processes: according to \cite{stdd:08},
  protein cost arises mainly in protein synthesis, not in the
  synthesis of amino acids.  The cost of the lac transporter in
  \emph{E.~coli} is mostly due to enzymatic side effects
  \cite{eako:12}.}. Since these effects are not captured by our
metabolic model, they are represented by an enzyme cost function
$\hminus(\esymbolv)$. The cost function represents fitness losses due
to protein expression that are not included in the {\metabolicobjective}
function and is typically linear in $\esymbolv$\footnote{In  pathway
  models, it is convenient to assume that cost and benefit of the
  pathway vanish if the pathway is not expressed. However, a constant
  offset of cost or benefit function will not change the optimal states.}
\footnote{Simple linear cost functions can be obtained from total protein mass
or total protein translation rate:
\begin{eqnarray} 
 \label{eq:investmentfunction} 
 \hminusfun(\enzymev)= \sum_{l} \acute{h}_{l}\, [\lambda+\enzdegrate_{l}]\,L_{l}\,\esymbol_{l}.
\end{eqnarray}
In this formula, the translation rate of an enzyme $l$ is proportional
to enzyme level $\esymbol_{l}$, protein chain length $L_{l}$ (number
of amino acids), and effective degradation rate
$(\lambda+ \enzdegrate_{l})$, where $\lambda$ is the cell growth rate
and $\enzdegrate_{l}$ is a protein-specific degradation rate
constant. By summing over all enzymes, we obtain the total translation
rate $\sum_{l}
(\lambda+\enzdegrate_{l})\,L_{l}\,\esymbol_{l}$. \coout{To account for
  the additional ribosome demand caused by additional proteins, the
  protein translation rate may be multiplied by a \emph{ribosome
    overhead factor} $\rho(\lambda)$, which increases with the growth
  rate $\lambda$, but is identical for all proteins (see SI
  \ref{sec:proofribosomeproteinratio}). For slow growth, this factor
  is close to 1.} A linear relationship
$[\mbox{cost}] = \acute{h}_{l} \cdot [\mbox{translation rate}]$, where
cost describes, e.g., growth defects, yields cost functions of the
form (\ref{eq:investmentfunction}).} The derivative
$\hul=\partial \hminus / \partial \esymbol_{l}$ is called enzyme
price, and the derivative
$\hudotl=\partial \hminus / \partial \ln \esymbol_{l}=\hul\,\ul$ = is
called enzyme investment. With a linear enzyme cost function
$h = \sum_{l} h_{l}=\sum_{l} h_{l}'\,\esymbol_{l}$, the prices
$\hul=h_{l}'$ are constant and the total enzyme investment is given by
the enzyme cost\footnote{A function that satisfies
  $h(\sigma\,\esymbolv) = \sigma^{\kappa}\,h(\esymbolv)$ for all
  positive $\sigma$ is called A positive homogeneous function with
  degree $\kappa$. For cost functions with this property, Euler's
  theorem yields the equality $\husdot = \sum_{l} \hudotl=\kappa\,h$,
  with the degree $\kappa$ as a prefactor. \co{(kommt noch am ende in
    diskussion)} \co{(for more details, see CBA labour
    \cite{lieb:cbalabour})}} $\sum_{l} \hudotl=h$. For a (nonlinear)
convex cost function, the prices increase with the enzyme levels, so
we obtain a bound $\hul\ge \hul^{\rm min}$ on each price, where
$\hul^{\rm min}$ is the price of enzyme $\e_{l}$ at zero expression.

\section{Conditions for enzyme-optimal states}

\myparagraph{\ \\Optimal states: local and global effects} Our fitness
function Eq.~(\ref{eq:fitnessFunctionFormula}) implies that the flux
benefit should be high and metabolite and enzyme costs should be
low. The benefits and costs depend on specific network variables,
which gives these variables ``direct values'' (flux gains for fluxes
and metabolite or enzyme prices for metabolites or enzymes). If a flux
has a positive gain, it has a positive direct value and the flux
should tend to be high. If an enzyme has a positive price, its
concentration has a negative direct valueand its concentration should
tend to be low. However, we know that our state variables cannot be
chosen independently: in steady states, they are coupled. If one of
them changes, the others will change as well. therefore, a variable
has also indirect fitness effects through all other variables. For
instance, if an enzyme (while being costly) catalyses a reaction that
contributes to the flux benefit, the enzyme becomes beneficial itself.
To find the right enzyme level, we need to balance this benefit with
the cost. The principle same applies for all physical variables in the
system: each variable needs to be chosen such that its costs and
benefits are in balance (which includes indirect costs and benefits,
and shadow values due to constraints). To get an impression of the
resulting states, we now consider fitness maximisation in the entire
coupled system.

\myparagraph{Optimality principle for enzyme levels} To define
 enzyme-optimal states in metabolis,, we score the
enzyme profile by a fitness function 
\begin{eqnarray}
 \label{eq:fitness} 
 \ffit(\enzymev,\cextv) &=& \gplus(\enzymev,\cextv) - \hminusfun(\enzymev).
\end{eqnarray}
with a {\metabolicobjective}
$\gplus(\enzymev,\cextv) = \bbenefit(\vvsteady(\enzymev,\cextv)) -
\metcostfun(\cvsteady(\enzymev,\cextv))$ and an enzyme cost
$\hminus(\enzymev)$.  The vectors $\vvsteady$ and $\cvsteady$ describe
steady-state fluxes and concentrations.  We now search for enzyme
levels $\esymbol_l$ that maximise fitness\footnote{The focus on
  \emph{local} fitness maxima is not only for biological reasons, but
  also because the metabolic value theory is mostly about first-order,
  necessary optimality conditions.}.  Variants of this optimality
problem -- models with multi-functional enzymes or non-enzymatic
reactions, other constraints, or multi-objective optimisation -- are
discussed in appendix \ref{sec:ConditionsForOptimalStates} and
\ref{sec:MethodsExtending}.

\coout{notwendigkeit von rev reakt: problem liegt darin, dass nicht alle
 enzymprofile moeglich sind, weil fluss zusammenbricht. wenn beliebig
 hohe konz erlaubt sind: konvexes polytop. rand ist verboten. daher
 gibt es ein supremum, aber kein maximum im erlaubten
 gebiet. innerhalb des gebietes sind die profile der oekonoischen
 potentiale komisch. das auch kurz beachten in fussnote zur existenz
 von loesungen}

 \coout{s und j doch klein?} \coout{USE words "scaled" and
  "unscaled"? hu * u usw "scaled form"? welches wort hatte ich
  dafuer bisher? hu "marginal {{\investment}}" ; hu * u "scaled marginal
  {{\investment}}" (ueberall!)} \coout{FROM FIG 3: $\ubenes = \partial
  \gplus/\partial \ln u$ $\hudots = \partial \hminusfun/\partial \ln u$
  $\fudots = \partial \ffit/\partial \ln u$ $\gplus^{\rm u} =
  \partial \gplus/\partial u$ $\hminus^{\rm u} = \partial
  \hminus/\partial u$ $\ffit^{\rm u} = \partial \ffit/\partial u$
  $\gvtots = (\partial \gplus/\partial \ln u)/v$ $\acosts = (\partial
  \hminus/\partial \ln u)/v$ $\fuvs = (\partial \ffit/\partial \ln
  u)/v$}

\myparagraph{In optimal states, enzyme value and enzyme {\price} must
  be balanced} What can we know about the pattern of enzyme
investments and fluxes in enzyme-optimal states? To see this, we
consider a local optimum of $\ffit(\enzymev)$ and have a look at the
optimality conditions. \co{allg. sortieren, bzgl interior optimum and
  boundary optimum} In an interior optimum state, in which all enzymes
are active, the difference
$t_{e_{l}} = \partial \ffit/\partial \esymbol_{l}=\gul-\hul$ is called
total enzyme value (or enzyme stress). In an optimal state, it must
vanish, $\partial \ffit/\partial \esymbol_{l}=0$, which implies an
equality\footnote{A weighted fitness function
  $\ffit=\alpha\,\gplus - \beta \,\hminus$ would yield the condition
  $\alpha\,\gplus - \beta \,\hminus$.  Since our objective functions
  can be scaled, there is no need for such prefactors in our
  theory. Trade-offs between {\metabolicobjective} $\gplus$ and enzyme
  cost $\hminus$ can be modelled in different ways (maximising the
  {\metabolicobjective} at a given enzyme cost; minimising enzyme cost
  at a given {\metabolicobjective}; or maximising a weighted difference
  of the two). In general, {\metabolicobjective} and enzyme cost are
  measured on different scales (and using different physical units),
  and we obtain optimality conditions of the form $\gu=\tau \,\hu$,
  with some constant ``economic temperature'' $\tau$. In general, in
  optimal states this number must be equal for all subsystems (see
  appendix\co{?}). However, with the right scaling of cost and benefit
  functions we can assume a (non-weighted) fitness
  $\ffit=\gplus-\hminus$, and obtain optimality conditions of the form
  $\gu=\hu$ (with $\tau=1$), as considered here. This justifies our
  additive fitness Eq.~(\ref{eq:fitness}).}  (see Figure
\ref{fig:backpropagation} b),
\begin{eqnarray}
 \label{eq:valuebalanceeq} 
  \gul = \hul
\end{eqnarray}
between enzyme {\myvalue}
$\gul=\frac{\partial
  \gplus(\vvsteady(\esymbolv),\cvsteady(\esymbolv))}{\partial
  \esymbol_{l}}$ and enzyme {\price}
$\hul= \frac{\partial \hminus(\esymbolv)}{\partial \esymbol_{l}}$.  If
the equality holds, then not only enzyme prices, but also enzyme
values must be positive, i.e.~all enzymes must have a positive
influcence on the metabolic objective.  If a metabolic state satisfies
the ``value-price balance'' Eq.~(\ref{eq:valuebalanceeq}) in all
active enzymatic reactions, it is called enzyme-balanced, and all
models in enzyme-optimal states must in fact be
enzyme-balanced\footnote{Similarly, a {\flow} $\vv$ that satisfies
  Eq.~(\ref{eq:valuebalanceeq}) in at least one kinetic model is
  called \emph{enzyme-balanced}, and if all reactions are active it is
  called \emph{strictly enzyme-balanced}.}.  \co{sort rest of
  paragraph; erst boundary, dann non-optimal!} If the equality
(\ref{eq:valuebalanceeq}) does not hold, this indicates a non-optimal
state, and the enzyme stresses describe ``incentives'' to change and
improve the enzyme levels $\esymbol_{l}$.

What about optimal states in which variables hit bounds? All enzyme
levels are bounded from below ($\esymbol_{l} \ge 0$ because they
cannot be negative), and for inactive enzymes ($\esymbol_{l} = 0$) the
bound leads to a negative shadow price that balances out the enzyme
price. Likewise, active upper bounds on enzyme levels lead to shadow
prices that act like cost terms and add to the regular enzyme
price. In a pathway model, enzyme levels are not bounded from above
but penalised by a cost, and we may think of this cost as an
opportunity cost in a larger cell model with a fixed protein budgte,
where an enzyme increase in a pathway leaves less protein for other
pathways, thus reducing their benefit.  If enzymes in the optimal
state remain inactive, the enzyme profile is a boundary optimum and we
obtain optimality conditions $\esymbol_{l}>0$ and $v_{l}\ne 0$ for
active enzymes and $\esymbol_{l}=0$ and $v_{l}=0$ for inactive
ones\footnote{An expressed enzyme ($\esymbol_{l}>0$) with zero
  catalytic rate \co{uea statt efficiency}
  $\ratelaw_{l}=v_{l}/\esymbol_{l}=0$ (due to thermodynamic
  equilibrium or enzyme inhibition), and therefore $v_{l} = 0$, would
  incur a cost without benefit. Fitness maximisiation as postualted
  here implies that such enzymes should not be expressed (``principle
  of dispensable enzyme'').  \co{REF to SI?  to CBA
    optimality \cite{lieb:18theory}?}}.  An active reactions must
satisfy Eq.~(\ref{eq:valuebalanceeq}), while in inactive reactions the
enzyme levels vanish and the optimality condition is an
inequality\footnote{This inequality can also be seen as an equality
  with a shadow value: in the optimality problem, each constraint
  $\esymbol_l \ge 0$ is associated with a Lagrange multiplier
  $\alpha_l$, and the optimality condition reads
  $\partial \ffit/\partial \esymbol_{l} + \alpha_l =0$. In inactive
  reactions, the Lagrange multiplier yields a positive shadow value,
  in line with the inequality
  $\partial \ffit/\partial \esymbol_{l}= \gul-\hul <0$.}
$\partial \ffit/\partial \esymbol_{l}<0$: expressing this enzyme would
decrease the fitness, which means that the enzyme {\price} $\hul$
exceeds the {\myvalue} $\gul$ (see SI Figure
\ref{fig:costbenefitcurves2}).  \co{JA! FN: what about density
  constraints? if the constraint is hit, all enzymes obtain an extra
  price, proportional to their size (size as defined in the density
  constraint!)}

\myparagraph{Flux burden and flux value balance} Enzymes are
costly. For each reaction, the enzyme investment per reaction flux
defines ``flux {\burden}''
$\hvl = \frac{\delta \hminus}{\delta v_{l}} = \frac{\delta
  \hminus}{\delta e_{l}}\,\frac{\delta e_{l}}{\delta v_{l}}$, an
effective overhead price of the flux.  Here we assume that enzymes are
reaction-specific and that each reaction is catalysed by a single
enzyme\footnote{\co{Woanders?}  To obtain a one-to-one mapping between
  reactions and enzymes in models, we may duplicate reactions that are
  catalysed by several enzymes, and also duplicate enzymes that
  catalyse several reactions. In models with such ``monoreactions''
  and ``monoenzymes'', the elasticity matrix $\Eunint$ is not diagonal
  and may be rank-deficient, i.e.~the Jacobian matrix
  $\NR\,\Eunint\,\Lmat$ may not be invertible. This has consequences
  for the calculations: instead of
  $\Eunu = \diag(\vv)\,\diag(\enzymev)\inv$, the enzyme elasticity
  matrix can be written as
  $\Eunu = \diag(\vv)\,\Escu\,\diag(\enzymev)\inv$, with a scaled
  enzyme elasticity matrix $\Escu$. This also works if our vector
  $\enzymev$ comprises variables other than enzyme levels,
  e.g.~temperature, with effects on several or all reactions. \co{REF
    to SI: say how flux burdens can be defined in this case}}. Under
this ``unique enzyme assumption'', we obtain a diagonal enzyme
elasticity matrix with elements
$\Eunul = \frac{\partial v_{l}}{\partial \esymbol_{l}} =
\frac{v_{l}}{\esymbol_{l}} = \ratelaw_{l}$. This matrix is invertible
unless $\ratelaw_{l}=0$ (e.g.~if reactions are in thermodynamic
equilibrium). With the help of this matrix and the enzyme {\price}
$\frac{\partial \hminus_{l}}{\partial \esymbol_{l}}=\hul$,
{\fluxburden}s can be computed as follows. Considering a variation of
an enzyme level and its \emph{direct} effect on reaction rates, the
flux {\burden} is defined\footnote{Flux {\burden}s provide a logical
  link between enzyme optimisation and Flux Cost Minimisation (FCM).
  In kinetic models, the minimal enzyme cost at which a given flux
  profile can be realised is called enzymatic flux cost. This cost, as
  a function of fluxes, can be used as a flux cost function in FCM.
  \co{logic requires envelope theorem!}  Moreover, flux {\burden}s $\hvl$
  from kinetic models can be used as coefficients defining linear flux
  cost functions for FCM \cite{lieb:14b,lieb:18fcm}.}
$\hvl = \frac{\hul}{\Eunul} = \frac{\hul}{\ratelaw_{l}} =
\frac{\hul\,\esymbol_{l}}{v_{l}}$. If the flux cost
$\fluxcost^{\rm enz}$ in an FBA model is given by an enzymatic flux
cost function (from an underlying kinetic model) \cite{lieb:18fcm},
the {\fluxburden} vector $\hvv$ (in the kinetic model) is equal to the
gradient $\nabla_{\vv}\fluxcost^{\rm enz}(\vv)$. Using these
definitions, our balance Eq.~(\ref{eq:valuebalanceeq}) for enzyme
values and prices can now be converted into a similar equation for
\emph{flux} values and prices: by multiplying with the ``enzyme
slowness'' $\esymbol_{v}/v_{l}$ (i.e.~dividing by the catalytic rate),
we obtain the flux value balance\footnote{In fact, by playing with
  these equations, the same optimality condition can be written in a
  multitude of ways, including
\begin{eqnarray}
 \label{eq:bla345differentForms} 
  \gul/\hul=1, \qquad
  \gvtotl/\hvl=1,  \qquad
  \gvtotl/\hul = 1/\ratelaw_l,  \qquad
  \gvtotl\,\ratelaw_l = \hul.
\end{eqnarray}
Each of the equations relates a point benefit (or value)  to a point cost (or
price)  and can be used to make sense of metabolic states.\co{explain this in more detail here; remove
  explanations further below!} }
\begin{eqnarray} 
 \label{eq:bla345} 
\underbrace{\gul\,\frac{e_{l}}{v_{l}}}_{\gvtotl}   = \underbrace{\hul\,\frac{\esymbol_{l}}{v_{l}}}_{\acostvl},
\end{eqnarray} 
 between flux value and flux burden in each reaction.
\co{JA! comment on minimal burden / price? kommt das irgendwo unten? CHECK! flux
  price: av = hu / r > humin / [kcat etamaxknown] = avmin; noch sagen:
  hu typischerweise ungefaehr konstant aber leicht steigend}

\section{\Summationconnectivitycondition}

\co{vielleicht wäre es doch besser, alle begriffe erst an der linearen
  kette (mit production objective, ohne metabolite cost) einzuführen
  und später zu verallgemeinern?}

\myparagraph{\ \\From local {\gain}s and {\price}s to global metabolic
  {\myvalue}s} The optimality conditions (\ref{eq:valuebalanceeq}) can
be used to check the results of a numerical optimisation, but it also
provides more general insights: it leads to general economic laws for
enzyme-optimal states, valid for any rate laws and cost function which
will be explored below. Below, starting from
Eq.~(\ref{eq:valuebalanceeq}), I derive the basic laws of metabolic
value theory and introduce the notion of economic potentials.  For
simplicity, we usually assume that flux benefit $\bbenefit(\vv)$ and
metabolite cost $\metcost(\cv)$ are linear functions, so {\fluxgain}s
and metabolite {\price}s are constant and known. \co{FN: With
  nonlinear cost and benefit functions, flux gains and metabolites
  prices vary but the form of the economic laws stays the same.}  A
simple production objective depends only on a single production rate;
in such cases, this single product has a production gain, while all
other production gains, flux gains, and metabolite prices vanish.
\todo{But this does not mean that other variables (in particular
  enzyme levels across the network) are not important. How can we
  infer the role of each enzyme, i.e.~the profile of enzyme
  {\myvalue}s across the network? And how are the two types of
  variables -- direct {\gain}s and {\price}s, and indirect enzyme
  {\myvalue}s -- related?}  If we perturb an enzyme, we may predict
the effect on value production by following causal chains in the
network. If we start from where production benefit is actually
realised, \todo{then to see how this is supported by enzymes elsewhere
  we need to follow causal chains in reverse, from effect to cause:
  step by step, value is ``acquired'' from variable to variable in
  backwards direction.}  \co{ref to a good figure! Wie fig 4, aber
  umgekehrt} How is this ``propagation of economic value'' shaped by
network structure and kinetics?  Remember, we are not interested in
anecdotical numerical results, but in general laws.  Thus, we will
ask: what can we learn from cost-benefit balance
Eq.~(\ref{eq:valuebalanceeq}) about optimal metabolic states?  Can we
learn something about possible flux profiles even without knowing the
rate laws?  At first sight, this may be surprising: the sensitivities
$\gul$, which must be matched by the enzyme prices, depend on enzyme
kinetics\footnote{The functions $\gplus(\u)$ and $\hminusfun(\u)$ are
  usually complicated and not explicitly known. If they are
  approximated by power laws or homogeneous functions, we can obtain
  simple economic laws.  \co{ref CBA labour
    \cite{lieb:cbalabour}}}. However, we can still learn about
metabolic fluxes by using metabolic control theory ({\MCA}).

\myparagraph{Enzyme {\myvalue}s reflect network-wide fitness effects}
Metabolic Control theory  describes how local parameter perturbations affect metabolic
 states. The effect of parameter perturbations on
steady-state fluxes and concentrations are quantified by sensitivities called metabolic response
coefficients. The metabolic response
coefficient $R^{z}_{\esymbol_{l}}=\partial z/\partial \esymbol_{l}$, between an enzyme level
$\esymbol_{l}$ and a state variable $z$, can be written as  a product
$R^{z}_{\esymbol_{l}}=C^{z}_{v_{l}}\,E^{v_{l}}_{e_{l}}$ where the
enzyme elasticity $E^{v_{l}}_{e_{l}}$ describes how an enzyme
perturbation perturbs the reaction rate (at constant metabolite
levels), and the control coefficient $C^{z}_{v_{l}}$ describes how
this rate perturbation changes our steady-state variable $z$.  To see
how MCT can help us characterise enzyme-optila fluxes, consider an enzyme variation
$\delta\esymbolv$, leading to a perturbed reaction rate. The metabolic
control coefficients describe the global effects of this perturbation
(see Figure \ref{fig:costbenefitcurves1}). By using these
coefficients, we can write the enzyme {\myvalue} as \co{show matrix
  form? JA! // $\gu = \bvtot\trans\,\Rvmat - \metcostc\Rcmat = (\bvtot\trans\,\Cvmat - \metcostc\Ccmat)\,\diag(\ratelawv)$}
\begin{eqnarray}
  \label{eq:enzymebalance0A}
\gul&=& \frac{\partial \gplus}{\partial \esymbol_{l}} 
  = \sum_{j} \bvtotj \, \Rv_{jl} - \sum_{i} \hci \, \Rc_{il}
  = \left(\sum_{j} \bvtotj \, \Cv_{jl} - \sum_{i} \hci \, \Cc_{il}\right)\, \frac{v_{l}}{\esymbol_{l}}.
\end{eqnarray}
The formula describes a chain of effects: an enzyme's direct effect on
a reaction rate (elasticity
$E^{v_{l}}_{\esymbol_{l}} = v_{l}/\esymbol_{l}$), the indirect influence of this rate on the stationary fluxes and concentrations
(metabolic control coefficients
$\Cv_{lr} = \frac{\partial \vsteady_{l}}{\partial
  \esymbol_{r}}\frac{v_{r}}{\esymbol_{r}}$ and
$\Cc_{ir} = \frac{\partial \csteady_{i}}{\partial
  \esymbol_{r}}\frac{v_{r}}{\esymbol_{r}}$), and their direct effects
on the metabolic objective (flux {\gain}s $\bvtotl$ and metabolite
{\price}s $\hci$). The difficult
terms in Eq.~(\ref{eq:enzymebalance0A}) are the metabolic control
coefficients $\Cv_{jl}$ and $\Cc_{il}$, which are nonlocal and
state-dependent (i.e.~to know them, we need to know the
solution to the optimality problem).  \co{WO? The formulae of
  metabolic value theory hide this complexity and describe enzyme
  economy entirely by the direct connections between network
  elements.}  However, Eqs (\ref{eq:valuebalanceeq}) and
(\ref{eq:enzymebalance0A})  yield a general rule:  the
balance condition (\ref{eq:valuebalanceeq}) requires that any  expressed enzyme must have a positive
value $\gul$ and therefore a non-zero catalytic rate $v_{l}/\esymbol_{l}$.  This means: if
an enzyme is  completely inhibited or if it  catalyses an equilibrium  reaction (and thus
$\ratelaw_{l}=v_{l}/\esymbol_{l}=0$), the enzyme must not be expressed
(``principle of dispensable enzyme'' \cite{lieb:18theory}).

\myparagraph{General conditions for enzyme-balanced {\flow}s} To do
this, let us formulate our optimality conditions in the language of
metabolic control theory.  By definition, an enzyme {\myvalue}
$\gul=\partial \gplus/\partial \esymbol_{l}$ is the metabolic response
coefficient between enzyme level $\esymbol_l$ and {\metabolicobjective}
$\gplus$, and the \emph{{\fluxvalue}}
$\gvtotl=\gul/\Eunul = \gul \frac{\esymbol_{l}}{v_{l}}$ is the
corresponding control coefficient (see SI
\ref{sec:proofenzymebenefitmetabolitevalue} for details).  With
control matrices $\Cvmat$ and $\Ccmat$, the flux values can be written
as (compare Fig.~\ref{fig:economicVariables})
\begin{eqnarray}
\label{eq:wvAndControlMatrices}
\gvtot\trans = \bvtot\trans\,\Cvmat - \metcostc\trans \Ccmat.
\end{eqnarray} 
Obviously, the {\fluxvalue}s $\gvtotl$ in
Eq.~(\ref{eq:wvAndControlMatrices}) cannot be inferred from network
structure alone: like other control coefficients, they depend on kinetics and on
the (optimal) metabolic state. So again, what can we learn about
metabolic values from network structure alone?

\myparagraph{{\Summationconnectivitycondition} for enzyme-balanced
  metabolic states} If the flux values $\gvtotl$ are  control
coefficients, they must
satisfy summation and connectivity theorems \cite{hesc:98}. By combining
Eqs  (\ref{eq:valuebalanceeq}) and (\ref{eq:enzymebalance0A}) and 
applying these theorems, we obtain the
{\summationconnectivitycondition} (Proposition \ref{th:theorem1} in
SI) 
\begin{eqnarray}
 \label{eq:fitnessbalance2x5}
 \bvtot \cdot \modevector &=& \frac{\hu\,\circ\esymbolv}{\vv} \cdot \modevector  \\
 \label{eq:fitnessbalance2x3}
 \hc \cdot \boldsymbol{\ell}  &=& \Eunint\trans\, \frac{\hu\circ\esymbolv}{\vv} \cdot \boldsymbol{\ell},
\end{eqnarray}
which must hold for all vectors $\modevector$ and $\boldsymbol{\ell}$
with the following properties.  The vector $\modevector$ in the flux
variation rule is a column (or linear combination of columns) of the
null space matrix $\Kmat$, i.e.~a stationary flux distribution
(satisfying $\Nint\,\modevector=0$). The vector $\boldsymbol{\ell}$ in
the concentration variation rule is a column (or linear combination of
columns) of the link matrix $\Lmat$, i.e.~a profile of internal
metabolic concentration variations that leave the conserved moieties
unchanged (satisfying $\Gmat\,\boldsymbol{\ell}=0$).  In short, both
vectors must describe valid, i.e.~constraint-respecting variations.
The dot $\cdot$ denotes the scalar
product, while multiplication $\circ$ and division of vectors apply
componentwise, and $\Eunint$ is the elasticitiy matrix in our
metabolic state. The variation rules relate flux {\gain}s $\bvtotl$
and concentration {\price}s $\hci$ to {{\enzymeinvestment}}s
$\hu \circ \esymbolv$ and inverse fluxes. With the flux {\burden}s
$\hvl=\hul\,\esymbol_{l}/v_{l}$, we can write them as
\begin{eqnarray}
 \label{eq:fitnessbalance2x0}
 (\bvtot -\acostv) \cdot \modevector&=&0 \\
 \label{eq:fitnessbalance3x0} 
 (\hc- \Eunint\trans\, \acostv) \cdot \boldsymbol{\ell} &=&0\;,
\end{eqnarray}
which must hold, again, for all valid \co{``test vectors'' - introduce
  and explain where?} vectors $\modevector\in \mbox{Span}(\Kmat)$ and
$\boldsymbol{\ell}\in \mbox{Span}(\Lmat)$. If we replace these vectors
by valid\footnote{A variation $(\delta\cv, \delta\esymbolv)$ is called
  {\valid} if it satisfies all model constraints, i.e.~if it leaves
  the conserved moieties unchanged (satisfying
  $\delta \cv_{\rm cm}=\Gmat\,\delta\cv=0$) and leads to a stationary
  flux variation $\delta\vv=\Eunc\,\delta\cv + \Eunu\,\delta\esymbolv$
  (satisfying $\Nint\delta\vv=0$).}  infinitesimal\co{FN: Variations
  indicated by $\delta$ are meant to be infinitesimal (defined
  mathematically by differential forms)} variations, we obtain the
{\summationconnectivitycondition} in differential form
\begin{eqnarray}
 \label{eq:fitnessbalance2x}
 (\bvtot -\acostv) \cdot \deltapar \vv&=&0 \\
 \label{eq:fitnessbalance3x} 
 (\hc- \Eunint\trans\, \acostv) \cdot \deltapar \cintv&=&0.
\end{eqnarray}
\co{use here deltapar, to indicate valid deltas? maybe different symbol?}
The {\summationcondition} (\ref{eq:fitnessbalance2x}) must hold for
all {\valid} flux variations $\deltapar \vv$ (i.e.~stationary
variations satisfying $\Nint\,\deltapar \vv=0$) and the
{\connectivitycondition} (\ref{eq:fitnessbalance3x}) must hold for all
{\valid} concentration variations $\deltapar \cintv$
(i.e.~moiety-conserving variations, satisfying
$\Gmat\,\deltapar \cintv=0$). The two rules refer to models with
active, enzyme-catalysed reactions and without metabolite
dilution. For models with inactive or non-enzymatic reactions, or for
models with metabolite dilution, the formulae must be modified (see SI
\ref{sec:proofSummationConnectivity}).

\co{gutes bild für economic  {\summationconnectivitycondition} (so wie fuer rules +
 bal eq) .. das waere eigentlich ein bild, in dem variationen gezeigt
 werden! existiert eigentlich schon, das nur klarer machen!
 elast-pfeile dicker!; bilder einzeln machen; oder doch noch ein
 zusaetzliches bild, in dem klar die variationen gezeigt werden!
 (variante des bildes der "compenated variation"; aber hier nicht
 kompensiert! sollte so aehnlich sein, dass man die beziehung der
 beiden leicht sieht (bzw spaeter mit diagrammen beides fast gleich
 aussieht! (wie kann man in den diagrammen die eigentliche variation
 zeigen (unabhaengig vom gewaehlten untersystem?)}

\begin{figure*}[t!]
\begin{center}
  \includegraphics[width=14.5cm]{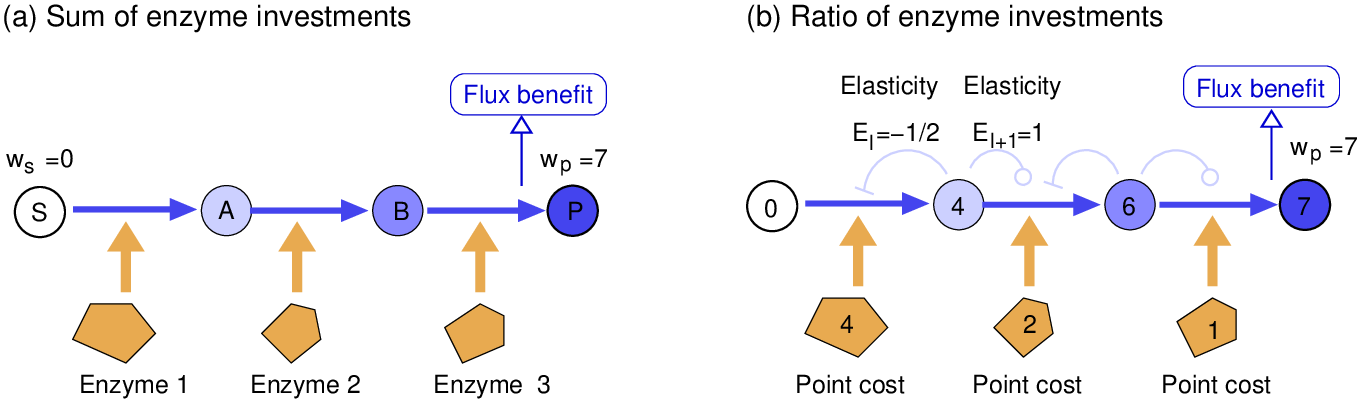}\\[4mm]
  \hspace{.8cm} \parbox{7.5cm}{\small
    {\Summationcondition} $\sum_l \bvtotl\,v_l = \sum_l \hudotl$:\\[1mm]
    Total {\fluxbenefit} = Total {\enzymeinvestment} }
  \parbox{7.5cm}{\small {\Connectivitycondition}
    $\frac{E_l}{E_{l+1}} =
    \frac{{\hudots}_{l+1}}{{\hudots}_l}$:\\[1mm] Enzyme investment
    ratio = Inverse elasticity ratio }
  \caption{Enzyme investments predicted by the
    {\summationconnectivitycondition}. (a) Metabolic pathway with
    production objective $\bbenefit(\vv) = w_{\rm P} \,v_{\rm P}$,
    where $w_{\rm P}$ and $v_{\rm P}$ are, respectively,
    the economic potential and the production rate of  product P. Aoccrding to the {\summationcondition}
    (\ref{eq:fitnessbalance2x}), in
    optimal states the sum of
    {{\enzymeinvestment}s} $\sum_{l} \hul\,\esymbol_{l}$ must be equal to the sum of
    {{\fluxbenefit}s}
    $\sum_{l} \Deltar \wintl \,v_{l}=w_{p}\,\prodrate_{p}$. The
    investments  accumulate along the chain, and by summing
    all {\enzymeinvestment}s upstream of a given metabolite, we obtain the
    investment embodied in this metabolite.  The ``embodied''
    investments in the product is equal to the total benefit. By
    dividing a metabolite's embodied investment by its production rate
    (i.e.~the pathway flux), we obtain the metabolite's economic
    potential (shades of blue). As expected, the economic potentials
    rise along the flux. If the initial substrate has an economic
    potential $w_{\rm S}>0$, this corresponds to a  substrate
    investment, which further  increases the economic potentials of all following metabolites. (b) The
    reaction elasticities determine the ratios of optimal enzyme
    investments.  The {\connectivitycondition}
    (\ref{eq:fitnessbalance3x}) states that {{\enzymeinvestment}}s
    for adjacent (producing and consuming) reactions are inversely
    proportional to the elasticities
    $\Eunvlci = \partial v_{l}/\partial c_{i}$. In the example
    (with flux $v=1$, substrate elasticities 1, product elasticities
    -1/2, and total benefit 7), we obtain  {\enzymeinvestment}s
    $\hudotl=(4, 2, 1)$. They decrease along the pathway, confirming the
    result from \cite{klhe:99}.}
 \label{fig:unbranchedfluxART}
\end{center}
\end{figure*}

\myparagraph{{\Summationconnectivitycondition} can be used to compute
  enzyme investments} The {\summationconnectivitycondition}
(\ref{eq:fitnessbalance2x}) and (\ref{eq:fitnessbalance3x}) for
enzyme-optimal states determine optimal {{\enzymeinvestment}}s.
Figure \ref{fig:unbranchedfluxART} shows an example, a linear pathway
with given {\fluxgain}s $\bvtot$, concentration {\price}s $\hc$, and
flux distribution $\vv$.  The {\summationcondition}
(\ref{eq:fitnessbalance2x}) shows that the scalar products
$\bvtot\cdot\modevector$ (``{\benefitshade}'') and
$\acostv\cdot\modevector$ (''{\costshade}'') must be equal for any
stationary flux variation $\modevector$ (given by ). If we use the
flux profile $\vv$ itself as a flux variation, the resulting equality
$\vv\cdot\,\bvtot = \vv \cdot (\hu \circ \esymbolv \oslashs \vv) =
\sum_l \hul\,\esymbol_{l}$ shows that the sum of enzyme investments is
determined by $\vv$ and $\bvtot$.  How will this investment be
distributed along a pathway? In the {\connectivitycondition}
(\ref{eq:fitnessbalance3x}), the ratio of {\enzymeinvestment}s around
a metabolite\footnote{The reason is simple: in models without
  conserved moieties, the link matrix $\Lmat$ in
  Eq.~(\ref{eq:fitnessbalance3x}) is given by an identity matrix
  $\Imat$. Without metabolite cost (metabolite prices $\hc=0$) and
  with equal fluxes in all reactions (stationary flux in linear
  chain), we obtain the condition
  $\Eunvlci\,\hudotl + E^{v_{l+1}}_{c_{i}}\,\hudotlplusone=0$ for each
  metabolite $i$. \co{explizit angeben!: and a similar equality for
    scaled elasticities} With a metabolite cost function
  $\metcost(\cv)$, the concentration {\price}s $\hc$ appear as an
  extra term.} depends on the reaction elasticities for this
metabolite.  Taken together, in a linear pathway with known
elasticities, the {\summationconnectivitycondition} determine all
{{\enzymeinvestment}}s completely.  With a linear enzyme cost
function, {\enzymeinvestment}s are proportional to enzyme abundance
and follow from proteomics data. But the
{\summationconnectivitycondition} can also be used in reverse: given
the {{\enzymeinvestment}}s, {\fluxgain}s, and flux directions, we may
predict the metabolic fluxes (proof in SI
\ref{sec:UniquenessProof}). Two simple examples (a linear pathway and
a branch point model) and an algorithm for larger networks are given
in SI \ref{sec:fluxesfromenzymecosts}.

\myparagraph{Optimality couples variables in different ways than
  simple kinetics} The {\summationcondition}
(\ref{eq:fitnessbalance2x}) relates fluxes and flux gains to enzyme
cost. If flux gains $\bvtotl$ and enzyme investments
$\hul\, \esymbol_l$ are known, we obtain linear constraints on the
inverse fluxes. In a linear pathway (or a network with only one flux
mode) this constraint can be used to scale our flux distribution. More
generally, the rule tells us -- given a change in some of the
variables -- how other variables must be adapted for the cell to
remain in an optimal state.  \co{etwas unklare beispiele .. edit! //
  etwas lang und unmotiviert} For example, a higher flux (at a
constant flux gain $\bvtotl$) leads to a higher flux benefit and
justifies a higher {\enzymeinvestment} $\hul\, \esymbol_l$. \co{nicht
  so relevant:?} In contrast, with lower flux gains $\bvtotl$ (and
constant investments $\hul\, \esymbol_l$), the flux must increase
(this requiring a higher catalytic rate).  And when enzyme {\price}s
$\hul$ increase and the  enzyme levels $\esymbol_l$ are fixed, the
fluxes must increase to maintain an optimal state. These links between
enzyme investments and fluxes are not due to kinetics alone, but to
our optimality postulate. By using kinetic relationships between
enzyme levels and fluxes, we can further limit the possible optimal
states. And even if $\hudot$ and $\Eunint$ are unknown, the simple
fact that the $\hudot$ must be positive puts constraints on the
fluxes. We will later come back to this point.

\section{Economic variables}
\label{sec:enzymebenefitmetabolitevalue} 

\myparagraph{\ \\Local economic laws} The
{\summationconnectivitycondition} characterise optimal states by
referring to {\valid} state variations, for instance stationary flux
variations that may concern the entire network. To consider such
variations, we need a global picture of the system considered. But can
optimal states also be characterised locally, by laws that describes a
single enzyme and its catalysed reaction, a single reaction and the
surrounding metabolites, or a single metabolite and the surrounding
reactions?  This seems unlikely because a local perturbation will have
effects elsewhere in the network: it will have indirect effects on
fitness through its action on other variables. We saw that optimality
conditions (e.g.~Eq.~(\ref{eq:valuebalanceeq})) depend on such
indirect effects. Hence, a local description may not suffice to
understand optimal states: instead, we need to consider network-wide,
indirect effects described by indirect economic values.  \co{gleich
  schon analogie mit ``preis'' einer ware fuer kosten, die woanders
  anfallen?}

\co{WO?  Analogy: In a market economy, a product has a value because
  it can be sold (i.e.~there is a demand). Materials or machines have
  a value because they can be used to make a product that can be
  sold. A machine that makes a machine has a value because it can make
  a machine that .. and so on.}

\myparagraph{Economic variables} In metabolic value theory, all
metabolites, reactions, and enzymes carry economic values. \co{We
  already know enzyme values .. WO flux values?}  \co{kurz def by
  control coefficients} Two other important types of economic values,
called economic potentials and economic loads, are assigned to
metabolites. An economic potential describes how a metabolite rate
contributes to the {\metabolicobjective} (i.e.~the (indirect) value of
metabolite production). To define it, we consider a virtual
  extra supply \co{explain! rephrase ``assume that''} of the
metabolite and ask how this would change the overall
{\metabolicobjective} by changing the system state.  The economic loads,
in contrast, describe the (indirect) value of metabolite
concentrations: they quantify how a virtual concentration change would
contribute to the benefit by changing the system state. \coout{Direct
  effects, arising if the {\metabolicobjective} function scores the
  metabolite concentration or production explicitly, are not
  included.}

\begin{figure*}[t!]
 \begin{center}
  \includegraphics[width=16.5cm]{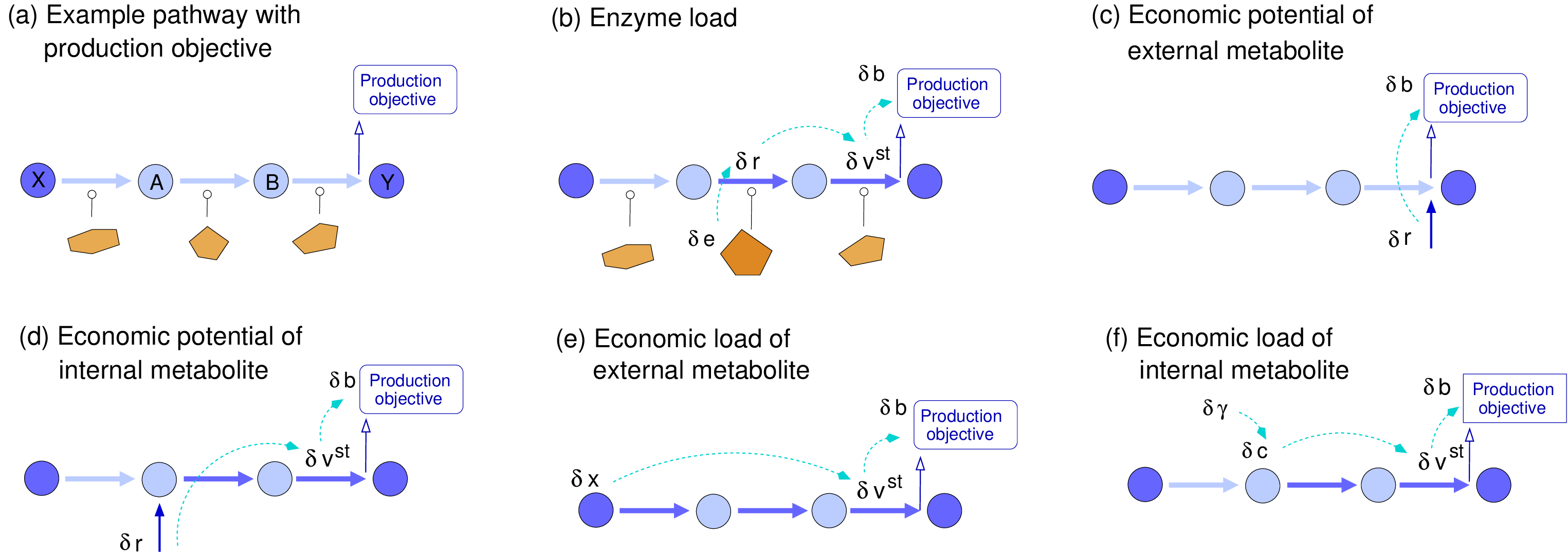}
 \end{center}
 \caption{Economic values in a linear pathway. (a) Example pathway
   with production objective (production of metabolite Y).  (b) Enzyme
   {\myvalue}s $\gus$ describe how enzyme level variations
   $\delta \esymbol$ change the {\metabolicobjective} in steady state.
   The indirect effect is mediated through a chain of effects
   $\delta \esymbolv \rightarrow \delta\vvsteady \rightarrow \delta
   \bbenefit$. Accordingly, the (indirect) enzyme value is
   obtained from a chain of derivatives (ratios of differentials
   $\gus = \frac{\delta \gplus}{\delta \esymbol} = \frac{\delta
     \gplus}{\delta v} \frac{\delta v}{\delta \ratelaw} \frac{\delta
     \ratelaw}{\delta \esymbol}$). The sensitivity
   $C^{\rm J} = \frac{\delta v}{\delta \ratelaw}$ (a flux control
   coefficient) describes the effect of a local rate perturbation
   ($\delta \ratelaw$) on the stationary flux ($\delta v$).Other
   economic variables are defined similarly. (c) Economic potentials
   $\wints = \delta \gplus/\delta \virtphi$ \co{int und external
     passen nicht, siehe auch (d)} of external metabolites describe
   the effect of a virtual production change
   $\delta \virtphi$\co{schoenere superscripts?}. \co{immer noch die
     CC-schreibweise dazu?} (d) Economic potentials
   $\wexts = \delta \gplus/\delta \virtphi$ of internal metabolites
   describe the effects of virtual supply fluxes $\delta \virtphi$.
   (e) Economic loads $\loads^{\rm x} = \delta \gplus / \delta x$ of
   an external metabolite $x$ describes the effects of virtual
   concentration changes $\delta x\co{^{virt}?}$. (f) Economic loads
   $\loads^{\rm c} = \delta \gplus / \delta \virtgamma$ of internal
   metabolites describe the effects of virtual concentration changes
   $\delta \virtgamma$.  \co{With a general metabolite objective
     $q(\vv,\cv)$, flux gains and metabolite prices can be defined
     similarly by derivatives!}}
  \label{fig:economicVariables} 
\end{figure*}

\co{The economic laws can be expressed in different form (``local'' vs
  ``extended'') and different types of scaling (``derivative'',
  differential'', ``difference'', ``point derivative'', or
  ``absolute'').  Both types of economic variables describe
  \emph{indirect} economic effects, that is, effects of a local change
  via changes in the global steady state (see Figure
  \ref{fig:economicVariables}).}

\myparagraph{Economic potentials} Let us first consider the economic
potentials (see Figures \ref{fig:economicVariables} (c) and (d)). Each
metabolite carries an economic potential, which assigns an indirect
value to the metabolite's production rate and which describes how a
steady extra supply of the metabolite would change the overall
fitness.  \co{bessere erklaerung. warum wlecher term wegfaellt} If it
increases the fitness, the economic potential $\wtoti$ (fitness change
$\delta \ffit$ per extra flux $\delta \prodrate_{i}$) is
positive. Generally, an economic potential consists of two terms: a
direct value (or ``production {\gain}'') and an indirect value (or
``production load''). For external metabolites $j$, the indirect value
vanishes and the potential is given by the direct value
$\wextj=\bpsij$. For internal metabolites $i$, the direct term
vanishes (because of the zero net rate) and only the indirect value
remains. This indirect value can be defined by control coefficients.
To define economic potentials mathematically, we imagine a virtual
supply flux $\virtphi_{i}$ that adds to the production of the
metabolite. To inspect its effects on the steady state and on
fitness\footnote{In our definition we require that the supply fluxes
  still allow for a steady state. This is not always the case.  First,
  if a supply flux contributes to a conserved moiety, this moiety
  cannot remain constant, thus excluding a steady state (structural
  instability). Second, a substrate-saturated enzyme limits the
  pathway flux and supply fluxes upstream of this enzyme will lead to
  unlimited substrate accumulatation (kinetic instability). In both
  cases, the system cannot buffer the supply flux and ends up in a
  non-steady state. Such variations are considered ``{\invalid}'' and
  are not allowed in the theory. For details, see SI section
  \ref{sec:importancereversible}}, we write the {\metabolicobjective} as
a function $\gplus(\enzymev,\cextv,\virtphiv)$ of enzyme levels
$\esymbol_{l}$, external levels $\cextj$, and virtual supply
fluxes\footnote{Our virtual exchange fluxes are only used as a
  mathematical tool and without a biological interpretation. However,
  in a thought experiment virtual fluxes may be realised by
  transporter proteins. In this case the economic potential of a
  metabolite would correspond to the ``fair'' {\price} $\hvl$ of the
  corresponding transporter, i.e.~the break-even point at which cost
  and benefit of the transporter cancel out.}  $\virtphi_{i}$ and
define the indirect production value of a metabolite $i$ by the
response coefficient\footnote{\co{das hier nur ganz kurz!
    hauptsaechlich (und auch den beweis) in CBA regulation!}  In this
  definition, the enzyme levels are meant to be
  constant. Alternatively, one could assume that, after applying the
  supply fluxes, enzyme levels are adapted to maximise fitness.
  However, the extra fitness increase would be a second-order effect
  and would not matter for the (first-order) economic potentials. This
  is why enzyme adaptation is ignored in our definition (see SI
  \ref{sec:proofadaptive}). \co{ref dort to envelope theorem?} }
$\gphialli=\frac{\partial \gplus}{\partial \virtphi_{i}}$.
\co{=$R^{q}_{r_{i}}=\bvtot\trans\,\Rv_{r_{i}} + \hc\trans\,\Rc_{r_{i}}
  = $ .. gleichung entsprechend (7))} We now set the economic
potential to $\winti = \gphialli$. In models with moiety conservation
(e.g.~if the sum [ATP]+[ADP] remains unchanged in all reactions),
additional supply fluxes (e.g.~a supply of ATP) may violate moiety
conservation (e.g.~the total concentration of ATP and ADP) and cause a
non-steady state. To avoid this, in the definition of economic
potentials \cite{rede:88} we describe supply fluxes by supply flux
vectors (describing simultaneous inflows and outflows of different
metabolites), which must allow for a steady state.  To construct such
vectors, we first consider a supply flux vector $\virtphiindv$ for
independent metabolites only, which can be chosen without constraints,
and then define the supply flux vector
$\virtphiv = \Lmat\,\virtphiindv$.  The economic potentials of
dependent metabolites are defined to be to zero by
convention\footnote{Instead of vanishing potentials, we may also
  assign arbitrary economic potentials to the conserved moieties. The
  change resembles a gauge transformation that changes the economic
  potentials themselves, but not the potential differences
  $\Delta \wtoti$, and therefore none of the measurable quantities
  (see SI \ref{sec:conservedmoieties}).}.  \coout{\co{formel wc = ?}}
\coout{Internal metabolite: total {\myvalue} = direct {\myvalue}
  (={\gain}) + indirect {\myvalue} (=load) = 0; therefore: load =
  -{\gain}. External metabolite: indirect {\myvalue} (=load) = total
  {\myvalue} = response coefficient.}

\co{In summary, a metabolite's economic potential denotes an indirect
  production rate {\myvalue} acquired through the metabolite's netw
  rate's effect on adjacent reactions (and therefore, on the entire
  metabolic state); for external metabolites, the potential is defined
  by the total production value, as defined through the flux benefit
  function. For internal metabolites, the net production vanishes.}

\co{\textbf{The difference trick}} \co{short paragraph based on CC
  formula!}  \co{For an enzyme, ``supporting the metabolic objective''
  can always be seen as ``producing metabolic value in the catalysed
  reaction'' (flux value multiplied by the flux).}  \co{Compound
  control coefficients and potential differences} \co{hier auch
  abschnitt zu delta w, allg splitting of control coeffs, meine
  ``metab'' control coeffs.}  \co{hier auch noch was zu den neuen
  kontrollkoeffizienten und ihren theoremen sagen (die ja auch
  klarmachen, wie lokale und globale groessen zusammenhaengen) (siehe
  SI)} \co{DEFS schon in SI? sonst in CBA local SI!}  \co{control
  coefficients (related to reactions) can be split into differences of
  metabolite properties (supply control coefficients)}

\myparagraph{Economic loads} We saw that economic potentials are
economic values associated with metabolite rates. Similarly, economic
loads are the economic values values associated with concentrations.
A metabolite concentration can influence the metabolic objective in
two ways: directly, by as described by its {\price}, and indirectly
via its effects on the steady state (see Figures
\ref{fig:economicVariables} (e) and (f)).  We see describe this by
considering virtual concentration changes. A change $\delta c_{i}$ of
metabolite $i$ changes the {\metabolicobjective}, and the concentration
{\myvalue} $\gci$ describes this effect.  A concentration {\myvalue}
$\gci = \loadi - \hci$\co{woher?}  consists of a direct {\myvalue}
(the negative metabolite {\price} $-\hci$) and an indirect
value\footnote{In metabolic value theory, ``load'' is a name for
  indirect values, but the term ``economic load'' is often used more
  specifically for concentration loads.}  $\loadi$, called
\emph{economic load}: the load describes how a virtual concentration
variation of our metabolite would affect the {\metabolicobjective}
indirectly, via changes of the network-wide metabolic state. External
metabolite concentrations $x$, as predefined variables, usually have
no direct value and their concentration {\myvalue}s are directly given
by their load $\loadj = \frac{\partial \gplus}{\partial
  \cextj}$. Internal metabolite concentrations, in contrast, can have
a {\price}, and the relationship between price and load depends on the
existence of conserved moieties\footnote{Metabolite values are closely
  related to the long-term effects of metabolite perturations. In
  models without moiety conservation, any perturbations of internal
  metabolite concentrations are cancelled by the metabolic
  dynamics. If perturbations of metabolite concentrations have no
  (steady-state) effect, the concentration {\myvalue}s $\gci$ of
  internal metabolites vanish, and since $\gci = \loadi - \hci$, the
  economic loads $\loadi$ and metabolite {\price}s $\hci$ must be
  equal.  In contrast, in models with moiety conservation, some
  variations cannot be cancelled by the dynamics (because they
  constantly change the conserved moiety concentrations). In this
  case, the concentration {\myvalue}s are given by
  $\gc = \Gmat\trans \,\gcm$, with concentration {\myvalue}s for
  conserved moieties in a vector $\gcm$, and the load vector reads
  $\loadint=\hc + \Gmat\trans\, \gcm$. In this general case, we obtain
  $\Lmat\trans\,\gc=0$, which implies the relationship
  $\Lmat\trans \,\loadint = \Lmat\trans\, \hc$.}. If metabolites in a
cell are diluted, their economic potentials contribute their
``effective economic load'' (see Section \ref{sec:EcBalEq}).  Finally,
also other model variables, such as growth rate, compartment volumes
or temperature, can be associated with economic loads.  \co{mention
  def of wv by virtual per variable. rel of yel, wvl, wci and wri to
  contol and response coefficients (list and explain them in a
  table?)}  In summary,  a metabolite load denotes an indirect concentration
  {\myvalue} acquired through the metabolite's concentrations effect on adjacent
  reactions (and therefore, on the entire metabolic state); for
  external metabolites, the total concentration value is directly given
  by the load, i.e.~by the response coefficient on the metabolic
  objective, and there is no concentration price. For internal
  metabolites, the total concentration {\myvalue} vanishes, and is
  given by the difference of load and concentration {\price}.

\co{\textbf{Overview of metabolic variables}} \co{an einer stelle die
  ganzen werte (inklusive shadow prices usw) aufdröseln - tabelle oder
  so?}  \co{schauen, was davon unten erklaert werden soll, nach
  oekonomischen regeln!}  \co{Metabolic
  value theory considers economic variables that complement the physical  variables. Each model variable (a concentration or
  flux), has an ``economic'' counterpart that is associated with the same
  network element. To define them, we consider variations (which may
  violate system constraints) and check how the fitness will be
  changed by these perturbations.}  \co{allgemein erst sagen, das all
  avriablen ec values haben, und zwar: enz values, flux values, conc
  values and econ pot?}  \co{JA!!!  change definition of potentials
  and loads, to correctly account for external loads usw?  oder
  stattdessen doch eher nur ``indirect .. value'' sagen?}  \co{alle
  variablen klar einfuehren! Mit definition! tabelle! // in dem
  abschnitt sollte fuer alle variablen klarwerden, aus welchen
  kontrollkoeffizienten sie entstehen (und schon wie die koeffizienten
  (durch multiplizieren mit N, Ec, Ec) zusammenhaengen. vorbereitung
  fuer rules!)}

 \co{``proxy
  variables'' (indirect effects, formally described as if they were
  direct)) // die logik ``indirekte werte'' sehr klar
  machen. ``abschneiden''; indirekte werte repraesentieren
  bestmoegliche adaptation des abgeschnittenen restsystems; mit einem
  trick repraesentieren wir diese aenderungen durch
  constraintverletzungen an der ``grenze'', beschriebene durch
  virtuelle variationen.}  \co{MERGE IN: Economic variables and rules:
  Bennennung nach physikalischen variablen und assoziation mit
  netzwerkelementen: jede constraint-gleichung ist mit einer variablen
  assoziert, zb mass balance constraint von metabolite i mit
  ``internal net rate phi-i''. der entsprechende Lagrange multiplier
  heisst dann ``production value'' und ist assoziiert mit dem element
  ``metabolite i''; entsprechend ``cost value'', ``catalytic rate
  value'', ``bound value'', ``total value''}

\co{JA nach Section 5! alle grundbegriffe aus diesem und den folgenden abschnitten
  sollten in einer abbildung klar gezeigt sein; dort sagen: arrows in
  fig 1 a! // eine abbildung mit definitionen und erklaerung aller
  oek var!! - direkt, proximal, und global; evtl was aus SI
  verwenden?
  
\includegraphics[width=5cm]{../../zeichnungen/CBA-dings3.jpg}

\co{JA! in paper noch ein kleines
  beispielbild mit direkten werten und eins daneben mit indirekten
  werten + control coeff (jutet bild mit den ganzen direkten gains +
  burdens!). EIN entsprechendes bild schon oben, aber mit DIREKTEN
  werten }

add a table with variables and corresponding derivatives + point
  derivatives: Variabl // Def // Flow; auch cost value: price/burden;
  ``benefit value'': gain/load
  
    \begin{tabular}{lll}
      Quantity & Positive term & Negative term \\\hline
      ``Absolute'' function & Benefit & Cost\\
      Derivative & Value (direct: ``gain''; indirect: ``load'') & Loss (direct: ``price''; indirect: ``burden'')\\
      Log-derivative & Point benefit (or ``usefulness'') & Point cost (or ``investment'') \\
    \end{tabular}
  }

\section{Local economic rules}

\begin{figure*}[t!] 
\begin{center}
  \includegraphics[width=15.5cm]{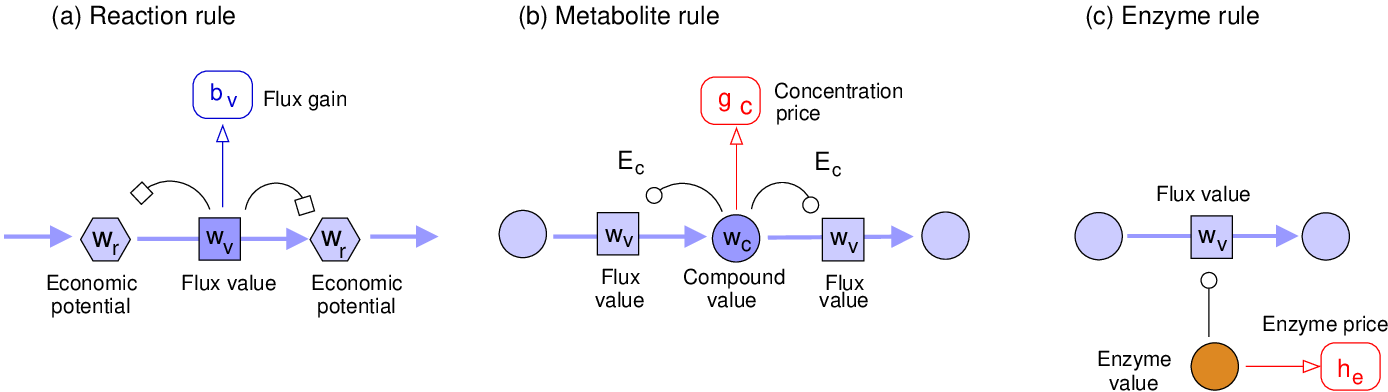}\\
\hspace{.8cm}  \parbox[t]{2cm}{} \parbox[t]{5cm}{$\gvtot
  = \bvdir + \underbrace{{\Ntot}\trans\,\wtot}_{\Deltar \wtot}$} \parbox[t]{5.2cm}{$\gc
  =  - \hc + \underbrace{\Eunint\trans\,\gvtot}_{\loadint}$} \parbox[t]{4cm}{$\fu
  =   - \hu + \underbrace{\Eunu\inv \gvtot}_{\gu}$} 
\end{center} 
\caption{Economic rules describe the economic values of neighbouring
  network elements. (a) Reaction rule. The flux {\myvalue} of a
  reaction consists of a direct {\fluxvalue} (the {\fluxgain}
  $\bvdirs$) and an indirect {\fluxvalue} (given by the economic
  potential difference $\Deltar w^{\rm c}$). (b) Metabolite rule. A
  concentration {\myvalue} $\gci$ consists of a direct {\myvalue} (the
  negative concentration {\price} $-\hcs$) and an indirect {\myvalue}
  (the economic load $\loadi$, acquired from the neighbouring flux
  values of adjacent reactions).  Other variables that impact reaction
  rates, like temperature, are associated with economic variables
  satisfying similar economic rules (not shown). (c) Enzyme rule. The
  total value (or ``economic stress'') of an enzyme is given by the
  enzyme's use {\myvalue} (or ``load'') $\gus = \gvtots \frac{v}{u}$
  minus the {{\enzymeprice}} $\hus$. Assuming that each reaction is
  specificylly catalysed by one enzyme, we can set
  $\Eunu\inv = \diag(\ratelaw)\inv =\diag(\enzymev)\diag(\vv)\inv = \diag(\tauv)$.  The
  ratio $\tau_{l} = e_{l}/v_{l}=1/k_{l}$ is also called ``enzyme
  slowness''. All these economic rules also hold for non-optimal
  states. In optimal states, the total values of enzymes must vanish,
  so enzyme use {\myvalue}s and {{\enzymeprice}}s must be equal.}
\label{fig:basicequations}
\end{figure*} 

\myparagraph{\ \\Local economic rules} If a models variable influences
the {\metabolicobjective} indirectly, this variable has a ``use value''
(quantified by an economic variable).  Computing this value may
require knowledge about the entire system. Formulae such as
Eq.~(\ref{eq:enzymebalance0A}) for individual enzymes, the control
coefficients refer to state variations in the entire network (caused
by a local variation, but extending over large parts of the system).
How can we describe metabolic variations and values locally, without
considering the entire system? To do so, we may consider ``invalid''
variations of a single reaction or a single metabolite (that violate
mass balances or moiety conservation and require compensation by
virtual perturbation variables). Describing these variables by
metabolic value theory, we obtain local economic rules that relate
economic variables between neighbour elements in the network.  \co{JA!
  (ref to proof in SI)} Each rule refers to a type of variable and
relates its economic value to the economic values of neighbouring
variables in the network (see Figure
\ref{fig:basicequations}). \co{die drei bloeck klarer und aehnlicher
  machen. EDIT! alles immer eins nach dem anderen}

\begin{enumerate}[leftmargin=5mm]
\item \textbf{Reaction rule} A flux value $\gvtotl$ describes the overall influence
  of a reaction flux $v_{l}$ on the
  metabolic objective.  The 
  \emph{reaction rule} (see
  Fig.~\ref{fig:basicequations} (a)) \co{proof wo!}
\begin{eqnarray} 
\label{eq:totalfluxvalue} 
\gvtotl = \underbrace{\bvdirl}_{\gvtotl^{\rm dir}} + \underbrace{\Deltar \wtotl}_{\gvindl},
\end{eqnarray}
describes it as a sum of two terms: a direct flux {\myvalue} (given by
the {\fluxgain} $\bvdirl$, \co{das expliczit in formeln zeigen, mit
  underbrace??}  plus a shadow value for fluxes that hit a
constraint), and an indirect {\fluxvalue}
$\gvindl=\Deltar \wtotl = \sum_{i} n_{il}\, \wtoti$ acquired from the
reactants and given by the difference of economic potentials along the
reaction (proof and explanations see SI
\ref{sec:SIderivationReactionRule} and
\ref{sec:proofenzymebenefitmetabolitevalue}). Thus, the economic
{\myvalue} of a flux -- a global systemic property! -- can formally be
attributed to the local conversion of metabolites of different values.

\item \textbf{Metabolite rule} A  concentration {\myvalue}
  $\gci$ describes the influence of a metabolite concentration
on  \co{define a fixed term for ``influencing''?}
  the metabolic objective.  According to the metabolite rule
(Fig.~\ref{fig:basicequations} (b)\co{proof wo?})
\begin{eqnarray}
 \label{eq:compoundrule} 
 \gci =  \underbrace{- \hci}_{\gci^{\rm dir}} + \underbrace{\sum_{l} \Eunvlci \,\gvtotl}_{\loadi},
\end{eqnarray}
it consists of an indirect and a direct {\myvalue}.  The direct value
is given by the negative concentration {\price} $-\hci$ (plus a shadow
value for metabolites that hit concentration bounds).  The indirect
{\myvalue} is called economic load and is given by $\loadi=\gci+\hci$.
The metabolite rule implies that economic loads are given by
$\loadi = \sum_{l} \Eunvlci \, \gvtotl$ (Figure
\ref{fig:basicequations} (b) and proof in SI \ref{sec:loadproof}).
What can we learn from this rule? Typically, external metabolites are
assigned a vanishing {\price}\footnote{Since external concentrations
  are given, their {\price}s do not matter and can be set to zero.  In
  contrast, if external concentrations themselves are choice
  variables, their {\price}s must be considered, for example to model
  ``ooportunity costs'' by which a higher (or lower) concentration
  would be beneficial for other systems outside the pathway modelled.
  Similarly, we do not score the metabolite rates of internal
  metabolites, because in our steady states, these rates vanish and
  their gains (direct economic values) do not matter. However, in
  models \emph{with} internal production rates (e.g.~rates balanced by
  dilution) non-zero production gains can be considered.}
$\hci=0$. In models with dilution, the sum in
Eq.~(\ref{eq:compoundrule}) contains an extra term $-\lambda\,\winti$
(see section \ref{sec:EcBalEq}), which describes a value loss due to
dilution. By incorporating the dilution term into the price $\gci$, we
obtain the effective price $\gci^{\rm eff}=\gci+\lambda\,\winti$. In
models without moiety conservation, \co{FN: This includes all metabolic
  models in which all metabolites are diluted} the metabolites'
concentration {\myvalue}s vanish ($\gci=0$), and so metabolite load
and metabolite price are balanced ($\loadi = \hci$). In models with
moiety conservation, we obtain the weaker condition
$\Lmat\trans \,\gc=0$, entailing a balance equation\footnote{Moiety
  conservation can be described by splitting the stoichiometric matrix
  into $\Nint = \Lmat\,\NR$ or by defining the left-nullspace matrix
  $\Gmat$, satisfying $\Gmat\,\Nint = 0$. The concentration
  {\myvalue}s in the vector $\gc = \Gmat\trans\,\gcm$, satisfy the
  optimality condition $\Lmat\trans\,\gc=0$. In models without
  conserved moieties (i.e.~$\Lmat=\Imat$), in optimal states $\gc$
  must vanish and so economic loads and concentration {\price}s must
  be equal, $\loadint = \hc$.  More generally, with conserved moieties
  we obtain the relationship $\Lmat\trans \,(\loadint - \hc)=0$.}
$\Lmat\trans \,\loadint= \Lmat\trans\,\hc$.

\item \textbf{Enzyme rule} The total value  (or ``stress'')
  $\ful = \partial \ffit/\partial \esymbol_{l}$  of an enzyme
is described by 
    the \emph{enzyme rule} (Fig.~\ref{fig:basicequations} (c)) \co{proof wo?}
\begin{eqnarray}
 \label{eq:enzymerule} 
 \ful =  \underbrace{- \hul}_{w_{e}^{\rm dir}} + \underbrace{E^{v_{l}}_{\esymbol_{l}}\; \gvtotl}_{\gul}.
\end{eqnarray}
The total enzyme value results from a direct {\price} (the enzyme
{\price} $\hul$) and an indirect {\myvalue} $\gul$ (or ``enzyme
load''), which represents the enzyme's influence on the metabolic
objective. The indirect value is acquired from the flux value
$\gvtotl$ of the catalysed reaction.  If a reaction is catalysed by a
(specific) enzyme, the enzyme elasticity is given by
$\Eunul = \esymbol_{l}/v_{l}$ and the enzyme load reads
$\loadv = \esymbolv / \vv \circ \gvtot =
\diag(\enzymev)\,\diag(\vv)\inv\,\gvtot$. \co{J! mit indizes
  schreiben} In optimal states, the total enzyme value must vanish
(because enzyme levels are control variables), unless the enzyme level
hits a bound (the bound $\esymbol \ge 0$ for positivity,
$\esymbol \le 0$ for some specified upper bound, or
$sum_{l}\esymbol_{l}=\esymbol_{\rm tot}$). In such constrained optimal
states, the total enzyme value must be balanced by a shadow value.  In
non-optimal states, the total value $\ful$ can be positive or negative
(implying, respectively, that the cell should increase or decrease the
enzyme level to reach an optimal state).
\end{enumerate}

Taken together, the economic rules are the basic laws for economic
potentials in metabolic networks, similar to  Kirchhoff's rules for
voltages and currents in electric circuits. \co{ref cba 
  fluxes?}

\myparagraph{Economic values represent direct and indirect effects}
All economic rules share the same  simple form: an economic value
consists of a direct and an indirect part, where the direct value is a gain
or price (i.e.~a direct fitness derivative, plus a possible shadow
values for upper and lower bounds), while the indirect value
(representing fitness effects via steady-state changes) is
``acquired'' from the neighbour network elements). 

\myparagraph{Shadow values as ``effective direct values''} In each of
the rules, the direct term represents a gain or price, which may
include a shadow gain or shadow price due to a bound on the physical
variable. For example, consider an enzyme level that hits the lower
bound $\esymbol=0$. In this case, a shadow value $\hul^{\rm bnd}$ is
subtracted\footnote{This makes sense: if an enzyme is idle, with no
  effect on the {\metabolicobjective} ($\gul$), then the enzyme price
  $\hul$ and the shadow value $\hul^{\rm bnd}$ cancel each other,
  leading to a zero enzyme stress
  $ \ful = - (\hul-\hul^{\rm bnd})=0$)} from $\hul$ in
Eq.~(\ref{eq:enzymerule}) \cite{lieb:18lagrange}. If we include this
term (as a direct term) into the enzyme price, we obtain the effective
enzyme price 
$\hul^{\rm eff}=\hul+\hul^{\rm bnd}$, and in an optimal state the effective
price will vanish.  Effective flux gains
$b_{v_{l}}^{\rm eff}=\bvdirl+\bvdirl^{\rm bnd}$ and metabolite
prices\footnote{Remember that, in our maximisation problems, a lower
  bound (keeping variables high) acts like a ``gain'', and an upper
  bound (keeping variables low) acts like a ``price''.}
$\hci^{\rm eff}=\hci+\hci^{\rm bnd}$ are defined similarly. For simplicity,
theis will be be explicitly mentioned below.

\myparagraph{Economic {\myvalue}s as projections of direct {\gain}s
  and {\price}s onto the metabolic network} The direct economic values
({\fluxgain}s $\bvtots$, metabolite {\price}s $\hcs$, and enzyme
prices $\hus$) arise from fitness derivatives and bounds on single
variables. While each enzyme (and possibly each metabolite) has a
direct price, we typically assume that only a few fluxes carry direct
values.  For example, with biomass production as the
{\metabolicobjective}, only the biomass-producing reaction has a direct
flux benefit.  But all reactions, metabolites, and enzymes may
contribute indirectly to this benefit and therefore carry a use value.

\co{For
example, each enzyme influences the overall metabolic state and
therefore the benefit function. This can be quantified by response
coefficients, our ``enzyme values'', and values of metabolic rates and
  concentrations are defined similarly. Relations between enzyme
values and direct gains and prices are established by the
{\summationconnectivitycondition}.  Each  value consists
  of a direct part (which may vanish) and an indirect part (which is
  come from  the values of neighbour elements). \co{FN: Note
    again (in their definition above, and agains below) that the
    shadow values are always included in bv, gc, hu}}

\co{Let us assume that the direct values are known. \co{FN: In a simple model without a metabolite cost, all metabolite prices vanish. Otherwise, we may assume that metabolite and enzyme cost represent occupied space. In this case, $\hci$ and $\hul$ represent effective molecule sizes}  How can we find the total values?}

The economic rules for the economic values of neighbour elements show
how variables ``acquire'' indirect use values from ``child
variables''. The direct values arise from the fitness function (and
bounds), and the indirect values arise from ``propagating'' these
direct values across the network. If the direct values in a network
are known, can we infer all other values from them? Assuming we know
the connection coefficients, this is fact possible: given the reaction
elasticities, flux {\gain}s, and metabolite {\price}s in an
enzyme-balanced state, the economic potentials and loads can be
directly determined. For generality, we consider models with moiety
conservation and dilution, see SI sections
\ref{sec:proofenzymebenefitmetabolitevalue} and
\ref{sec:SIproofMCAwithDilution}.  From the economic rules for such
models, we obtain a yield a formula for the internal economic
potentials \co{proof wo?}
\begin{eqnarray}
 \label{eq:solveForEcoPot} 
\wint &=& (\Lmat\,{\Mmat_{\lambda}}\inv\,\Lmatplus)\trans\,\underbrace{\hc -\Eunint\trans \,\bvtot}_{\hctot}.
\end{eqnarray} 
\co{FN? oder spaeter? we can read the ``forward tracing''
  Eq.~(\ref{eq:enzymebalance0A}) also as a ``backward tracing'', or
  ``acquiring'' of value (reaction acquires its value from
  {\metabolicobjective}, and enzyme acquires its value from reaction) //
  say that the tracing works similarly for other economic variables,
  too // das sind ja tatsaechliche modellparameter. unten:
  entsprechendes (tracing)-argument fuer virtuelle stoerungen} The
proximal concentration {\price} $\hctot$ consists of the direct price
and an indirect price acquired from adjacent reaction gains. In the
equation, $\Mmat_{\lambda} = \NR\,\Eunint\,\Lmat-\lambda\,\Imat$ is
the Jacobian matrix (for models with dilution), and the projection
matrix $\Lmatplus$ maps concentrations from internal metabolites to
independent internal metabolites (proof and example in SI
\ref{sec:SIProofsolveForEcoPot}). \co{auch geplant fuer CBA fluxes,
  CBA lagrange, CBA labour: from global to local: internal ec pot as
  ``abbild'' von ext production {{\myvalue}}s; allows for local
  understanding; extension of models} Equation
(\ref{eq:solveForEcoPot}) can be visualised geometrically: if the
{\fluxgain}s are described by a sparse vector in a high-dimensional
space, vector of economic potentials is a ``projection'' of this
vector (which concerns only one or a few reactions) \co{onto the null space if ..?}  onto the entire network. \co{rest des
  parag ausfuehrlicher! refer to SI on control coefficients} Similar
projections exist in {\MCA}: If a reaction rate is perturbed (e.g.~by
an enzyme inhibition), the perturbation can be described by a sparse,
non-stationary flux vector $\delta \vv$. The resulting flux change (a
stationary flux vector $\delta \vvsteady$) is obtained by
left-multiplying this vector with $\CJmat$, so $\CJmat$ acts as a
projector onto the subspace of stationary flux distributions. This
analogy between the two theories is not by chance: in one case, the
projection reflects the response of a mechanistic system (mediated
through chains of causal effects), in the other case it reflects
economic requirements (mediated through chains of incentives, in the
opposite direction). For details, see SI
\ref{sec:SIvirtualVariations}.

\co{load/potential immer als drittes! auch in anderen artikeln}

\begin{figure*}[t!]
 \begin{center}
  \includegraphics[width=15.5cm]{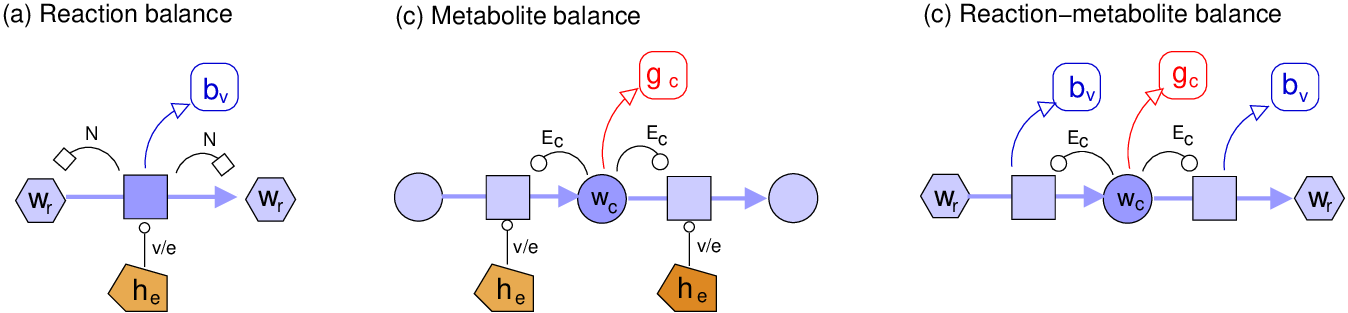}\\[3mm]
\hspace{1cm}
\parbox[t]{5cm}{$\underbrace{\Deltar \wtotl + \bvdirl}_{\gvtotl}=\underbrace{\hul\,\esymbol_l/v_{l}}_{\hvl}$}
\parbox[t]{4.cm}{$\loadi = \sum_{l}\Eunvlci\, \underbrace{\hul\,\esymbol_l/v_{l}}_{\hvl}$}
\parbox[t]{5.5cm}{$\loadi = \sum_{l} \underbrace{(\Deltar \wtotl + \bvdirl)}_{\gvtotl}\,\Eunvlci$}\\
\hspace{1cm}
\parbox[t]{5cm}{$(\Deltar \wtotl + \bvdirl)\,v_{l}=\hul\,\esymbol_l$}
\parbox[t]{4.cm}{$\loadi\,c_{i} = \sum_{l}\Escvlci\, \hul\,\esymbol_l$}
\parbox[t]{5.5cm}{$\loadi\,c_{i} = \sum_{l} (\Deltar \wtotl + \bvdirl)\,v_{l}\Escvlci$}
\caption{\todo{Economic balance equations.  Each equation can be
    written in ``value form'' (top) or ``value production form''
    (bottom).  (a) The {\fluxbenefitbalance} equation states that flux
    values and flux burdens of a reaction must be equal. It follows
    from the optimality condition for enzyme levels,
    $\ful=\gul-\hul=0$ \co{erklaeren!: wie ergibt sich das aus yel?}
    (see Figure (\ref{fig:basicequations})).  variables denote the
    economic potential difference $\Deltar \wtotl$, flux $v_{l}$,
    direct {\fluxgain} $\bvdirl$, price $\hul$, and flux burden
    $\acostvl=\hul\esymbol_{l}/v_{l}$. In reactions with a direct
    {\fluxgain} \todo{$b_{v}=0$}, the equation requires $\Delta \wintl$
    and $v_{l}$ to have the same signs and the economic potentials
    must increase along the flux.  Multiplying the equation with the
    flux $v_{l}$ yields the value production form (bottom). (b) The
    {\compoundbenefitbalance} equation (with economic load
    $\loadi = \hci$; flux cost weight $\hvl=\hudotl/v_{l}$; scaled
    elasticity $\Escvlci$) relates a metabolite's economic load
    $\loadi$ to the {{\enzymeinvestment}}s around the metabolite. In
    models without moiety conservation, optimality requires that load
    $\loadi$ and metabolite {\price} $\hci$ must be equal. In models
    with moiety conservation, we can augment $\hc$ to
    ${\hc}^{\rm eff} = \hc+\gc = \hc+\Gmat\trans\,\gcm$. \co{klar
      hier?} \coout{The balance equations are shown in the ``scaled''
      formulation (with benefit and cost, not {\myvalue} and {\price})
      (and in physical units of fitness)} (c) The {\loadbalance}
    relates the economic load of a metabolite (shaded) to the
    {\fluxvalue}s of neighbour reactions (and thus to economic
    potentials).}}
  \label{fig:balanceequations}
 \end{center}
\end{figure*}

\co{was besseres als ``neighbour'', etwas das child relationship anzeigt? `` subjacent? influenced?}

\section{Economic balance equations}
\label{sec:EcBalEq}

The economic rules (\ref{eq:totalfluxvalue}), (\ref{eq:compoundrule}),
and (\ref{eq:enzymerule}) explain a variable's economic value by a
direct value and by an indirect value acquired from the variable's
child variables (i.e.~variables directly influenced by it).  The rules
hold for all metabolic states, including non-optimal states and
non-enzymatic reactions. But we are particularly interested in enzyme
investments in optimal states.  If enzyme levels are choice
\co{``control'' statt ``choice'' variables uea in CBA!} variables (and
do not hit a constraint), their total value in optimal states must
vanish. By setting the enzyme stresses $\ful=0$ in
Eq.~(\ref{eq:enzymerule}), we obtain economic rules for optimal
states. Then, by combining the rules in pairs, we obtain balance
equations that relate enzyme investments to economic potentials,
loads, and costs in neighbouring reactions, metabolites, and enzymes
(see Figure \ref{fig:balanceequations}).

\begin{enumerate}[leftmargin=5mm]

\item \textbf{\Fluxbenefitbalance} \coout{use a consistent naming
    (conditions, rules, laws, balance equations, scaled and unscaled;
    local and global?)} By setting Eq.~(\ref{eq:enzymerule}) to zero
  (assuming expressed enzymes in an enzyme-optimal state) and
  inserting Eq.~(\ref{eq:totalfluxvalue}), we obtain the
  {\fluxbenefitbalance}  in ``enzyme value form''
\begin{eqnarray}
\label{eq:ReactionBalanceEnzymeValueForm}
(\Deltar \wtotl + \bvdirl)\,\Eunul = \hul
\end{eqnarray}
between {\fluxvalue} and enzyme {\price}, which must hold for all
active enzymatic reactions.  Assuming a one-to-one relation between
reactions and enzymes, the enzyme elasticity $\Eunul$ is given by the
catalytic rate $\ratelaw_l=v_l/\esymbol_l$.  Dividing
Eq.~(\ref{eq:ReactionBalanceEnzymeValueForm}) by $\ratelaw_{l}$ and
noting that $\hul/\ratelaw_{l}=\hvl$, we obtain the  equation in
 ``{\fluxvalue} form''
\begin{eqnarray}
\label{eq:FluxValueEq1}
 \Deltar \wtotl +  \bvdirl = \hvl
\end{eqnarray}
with the {\fluxburden}
$\hvl = \frac{\hul}{\ratelaw_{l}} = \frac{\hul\,\esymbol_l}{v_{l}}$
defined as above. Eq.~(\ref{eq:FluxValueEq1}) shows that the economic
potential (or ``use value'') $\wtoti$ of a metabolite is equal to an
``embodied value'': in reactions with out flux gains
(i.e.~$\bvdirl=0$) the potential increases from substrate to product
because of the flux burden $\hvl$, reflecting the enzyme investment in
the reaction. Therefore, along a pathway flux the metabolite values
will tend to increase and will embody all upstream enzyme (and
external substrate) investments.  The step from
Eq.~(\ref{eq:ReactionBalanceEnzymeValueForm}) to
Eq.~(\ref{eq:FluxValueEq1}) assumes that the $\esymbol_{l}$ are
actually enzyme levels (which appear as prefactors) in the rate laws,
that every reaction is enzyme-catalysed, and that enzymes are
reaction-specific. This ``unique enzyme assumption'' guarantees that
the enzyme elasticity matrix $\Eunu=\diag(\vv\oslashs\esymbolv)$ is
diagonal. If we further exclude zero enzyme elasticities, the matrix
will be invertible. Otherwise (e.g.~in the case of non-specific
enzymes other control variables $u_{l}$ such as temperature or membrane
potentials), the elasticity matrix will not be invertible and
Eq.~(\ref{eq:bla123}) must be replaced by modified
formulae\footnote{In this case, the elasticity matrix reads
  $\Eunu=\diag(\vv)\, \Escu\,\diag(\esymbolv)\inv$ with a
  non-invertible matrix $\Escu$.} (see appendix
\ref{sec:MethodsExtending}).

\coout{bild mit einfachen kriterien fuer economical {\flow}s?? (Gibts
 schon in ``structure of the theory''; einfacher? so wie in
 vortrag?)}

\coout{mention enzymatic vs nonenzymatic
 reactions in {\compoundbenefitbalance}}

\item \textbf{\Compoundbenefitbalance} Our second balance equation
  relates a metabolite's economic load  to the  {\fluxburden}s in 
  the adjacent reactions (reactions that  a  metabolite influences
  kinetically as a reactant, catalyst, or regulator). For internal
  metabolites (with concentations $c_{i}$ and loads $\loadi$) and
  external metabolites (with concentrations $\cextj$ and loads
  $\loadj$), the equalities read\footnote{To obtain  the equation, we assume
    an enzyme-balanced state and insert the {\fluxbenefitbalance}
    (\ref{eq:FluxValueEq1}) into the concentration-production balance
    (\ref{eq:loadfluxvalueequation}). \co{lieber: from (16):
      \co{$y_{c_{i}} = E^{v_{l}}_{c_{i}}\, w_{v_{l}}$}. dann aus (17)
    mit $t_{e_{l}}=0$: $h_{e_{l}}=v_{l}/e_{l}\,w_{v_{l}}$ so
    $w_{v_{l}}=\underbrace{a_{v_{l}}}_{h_{e_{l}}\,y/e_{l}}$ alles
    einbauen!}}  (proof see SI \ref{sec:proofinvestmentbalance})
\begin{eqnarray}
  \label{eq:concentrationbalanceequationext}
  \loadi = \sum_{l} \hvl \, \Eunvlci, \qquad
  \loadj = \sum_{l} \hvl\, \Eunextlj.
\end{eqnarray}
To derive the {\compoundbenefitbalance}
Eq.~(\ref{eq:concentrationbalanceequationext}), we assume that all
reactions are enzyme-catalysed.  A variants  of this equation can
include non-enzymatic reactions (see
Eq.~(\ref{eq:qloadinternal3}) in appendix).  If reaction rates depend
on variables other than metabolite concentrations (e.g.~temperature), these variables also have   economic loads satisfying
similar balance equations.
\item \textbf{\Loadbalance} The value of metabolite production and of
  metabolite concentrations are described, respectively, by economic
  potentials and loads. How are these  values related? In
  growing cells, with dilution fluxes $v_{i}^{\rm dil}=\lambda c_{i}$,
  metabolite concentrations and fluxes are coupled by
  $\cv = \frac{1}{\lambda}\Nint\,\vv$, on top of their coupling
  through rate laws.  Thus, concentration change affects the
  neighbouring reaction rates, which further affect metabolite net
  rates and eventually (in a steady growth state) their
  concentrations. How is all this reflected in value structure?
  The loads $\loadinti$ of internal metabolites are given by
  $\loadint=\hc - \Gmat\trans\, \gcm$. Eq.~(\ref{eq:compoundrule})
  relates a metabolic load to the {\fluxvalue}s $\gvtotl$ in adjacent
  reactions (with rates directly affected by the metabolite). By
  inserting the reaction balance (\ref{eq:FluxValueEq1}), we obtain
  the {\loadbalance}
\begin{eqnarray}
 \label{eq:loadfluxvalueequation} 
\loadi = \sum_{l} \underbrace{(\Deltar \wtotl + \bvdirl)}_{\gvtotl}\,\Eunvlci.
\end{eqnarray}
between the load of metabolite $i$, the flux {\gain}s $\bvdirl$ and
economic potentials $\wtoti$ in the adjacent reactions and the
elasticities $\Eunvlci$ between them. Similar equations exist for
external metabolites and  other variables that influence 
reaction rates (e.g.~temperature).
\end{enumerate}
As mentioned before, the direct value terms can contain shadow values
arising from bounds on the physical variables.  Interestingly,
reaction and metabolite balance resemble each other. The reason is
that enzymes can be seen as external metabolites: their concentrations
are constant, and they influence reaction rates kinetically.
Accordingly, the reaction balance
(\ref{eq:ReactionBalanceEnzymeValueForm}) resembles a metabolite
balance (\ref{eq:concentrationbalanceequationext}) with an enzyme
instead of an external metabolite (with elasticity
$\Eunextlj = \frac{v_{l}}{\esymbol_{l}}$ and a load $\loadj=\hul$
given by the {{\enzymeprice}}).

\myparagraph{Metabolic values in growing cells: the effective cost of
  dilution} The economic laws shown above hold for models without
dilution.  \co{sagen, ausnahme: dilution reactions, non-enzymatic
  lambda als eine art enzym? kosten entsprechen growth benefit????}
In growing cells, metabolite dilution can be described conveniently by
``degradation fluxes'' $v^{\rm dil}_{i}=\lambda \,c_{i}$ with the
growth rate $\lambda$ as a rate constant. For steady states, we obtain
a mass-balance equation $0=\Nint\,\vv-\lambda\,\cv$ that couples
concentrations directly to fluxes. This coupling has consequences for
metabolic economics. If cell growth is the objective, the dilution
rate of compounds, including metabolites and macromolecules, can be
treated as the fitness objective. The resulting balance equations
(with growth rate as a control variable and objective) are discussed in
\cite{lieb:18lagrange}. Here we consider a different problem: a
metabolic pathway with a given production objective, in which
metabolites are diluted at a given rate $\lambda$.  Dilution puts a
burden on metabolism, which reshapes the optimal enzyme investments.

For example, consider a linear pathway with a production objective
(scoring the last reaction flux). In growing cells, higher internal
metabolite concentrations will increase the dilution fluxes and the
waste of enzyme investment embodied in the metabolites.  \co{``loss'' of
  metabolite is costly (requiring reproduction!), and metabolites with
  a higher concentration have a higher loss, which is more costly
  (requiring more reproduction!)}  \co{effectively, that's just the
  same as if the metabolite had a higher price!} To keep the
metabolite concentrations low while maintaining the desired flux,
enzyme investments must be rearranged: upstream enzyme levels should
decrease and downstream enzyme levels increase.

\myparagraph{Dilution leads to effective metabolite prices} To model
this, we can describe dilution by ``dilution reactions'' with velocity
$v_{i}^{\rm dil} = \lambda\,c_{i}$, elasticity
$E^{\rm v^{\rm dil}_{i}}_{\rm c_{i}}=\lambda$, and flux values
$w_{v}=\Deltar \wints = - w_{r}$ (assuming there is no flux gain and
no ``product'' of the dilution reaction).  \co{FN: Losses similar to
  dilution can also be caused by non-enzymatic degradation reactions
  or membrane leakage. These can be described similarly to dilution,
  but with different rate constants for different metabolites.}  For
each metabolite, this reaction leads to an extra term
$-\lambda \,\winti$ on the right of the {\loadbalance}. We can see the
term $-\lambda\,\winti$ as a concentration {\price}, describing an
incentive to keep $c_{i}$ low: by including it into $\hci$, we obtain
the effective concentration prices $\hci' = \hci + \lambda\,\winti$,
which are higher than the ``real'' prices $\hci$ (or less negative,
for metabolites with a negative {\price}).  Alternatively, we can
bring the term $-\lambda\,\winti$ to the left and define the effective
economic load $\loadeffi = \loadi + \lambda \,\winti$.  In the
{\compoundbenefitbalance} equation, the extra term $\lambda\,\gvtotl$
on the left needs to be balanced by the sum on the right. To increase
this sum, investments are shifted from producing to consuming
reactions. This confirms our expectations: dilution favours enzyme
investments that keep \co{valuable} metabolite concentrations low.

\co{FN: JA! The load of compounds due to dilution provides an
  explanation for enzyme cost. If we think of a whoel-cell model,
  enzymes are not external ``knobs'', but are compounds like the
  metabolites.  To keep an enzyme at its favourable concentration, it
  be reproduced. The effort for reproducing the enzyme is one possible
  \todo{rationale} behind our enzyme cost function! see CBA local!}
\co{FN: In models with dilution, metabolite concentrations directly
  determine the steady reproduction rate. The coupling between
  potentials and loads directly reflects this coupling between
  metabolite rate and concentration. how 1/dilution acts as a sort of
  time derivative (and note how this impinges on laws for harmonic
  oscillations)}

\section{Balance equations for point cost and benefit}

\myparagraph{\ \\ Point variables and balance equations in
  {\pointbenefitform}} \co{punktgroessen. dann sagen, wir koennen das
  auch allgemeiner machen (mit log ableitungen)) bessere einleitung zu
  punktgroessen und den daruas folgenden gesetzen; evtl sogar hier
  ausgehend von wertstruktur ``differential'' variations and finite
  ``point'' groessen einfuehren, dann sagen, dass man dafuer auch
  gesetze machen kann.}  Economic values (such as {\gain}s, {\price}s,
potentials, or loads) are derivatives between fitness and physical
variables. If fitness is measured in units of Darwin (Dw), a
placeholder for the respective fitness unit used in a model, we obtain
the unit Dw/mM for economic loads (a fitness derivative for
concentrations) and Dw/(mM/s) for economic potentials (a fitness
derivates for metabolite rates), and possibly other units.  To make
all economic values comparable, we can define fitness derivatives with
respect to logarithmic variables,
$\ffit_{x}\partialder = \frac{\partial \ffit}{\partial \ln x} =
\frac{\partial \ffit}{\partial x} x=\ffit_{x}\,x$ (see Figure
\ref{fig:backpropagation}): all these ``point'' derivatives have units
of Dw, the unit of the fitness function itself! Note that the point
value of a variable is just the normal economic value, multiplied with
the variable's own numerical value. By writing economic laws with
these new derivatives, we obtain the laws in the so-called
``\pointbenefitform'' form (as opposed to our previous
``\valueform''). Now different processes can be directly compared: for
example, if a reaction substrate (characterised by a net consumption
rate) and and enzyme (characterised by a concentration) contribute to
the overall benefit, their benefit contributions (``point benefits'')
are directly comparable.

\co{Scaled derivatives make sense for multiplicative fitness functions! 

Consider $f = \frac{benefit(v)}{cost(e,c)}$

ln f = ln benefit - ln cost

Scaled derivative
$\frac{\partial \ln f}{\partial \ln v} = \frac{v}{f}\frac{\partial f}{\partial v} = \frac{v}{f} b_{v} = \hat b_{v}$\\
$\frac{\partial \ln f}{\partial \ln e} = \frac{e}{f}\frac{\partial f}{\partial e} = -\frac{e}{f} h_{e} = -\hat h_{e}$\\
$\frac{\partial \ln f}{\partial \ln c} = \frac{c}{f}\frac{\partial f}{\partial c} = -\frac{c}{f} g_{c} = -\hat g_{c}$}

\myparagraph{Balance between {\enzymebenefit} and enzyme investment}
Let us see an example. To rewrite the optimality condition
Eq.~(\ref{eq:valuebalanceeq}) in {\pointbenefitform}, we simply
multiply it by $\esymbol_{l}$ (see Figure \ref{fig:backpropagation}c).
The resulting equation
\begin{eqnarray}
 \label{eq:fitnessbalanceeq} 
  \underbrace{\gul\,\ul}_{\ubenel} = \underbrace{\hul\,\ul}_{\hudotl}
\end{eqnarray}
relates the enzyme point benefit (or ``value production'')
$\ubenel = \frac{\partial \gplus}{\partial \ln \esymbol_{l}} =
\gul\,\esymbol_{l}$ to the point cost (or \emph{{\enzymeinvestment}})
$\hudotl = \frac{\partial \hminusfun}{\partial \ln \esymbol_{l}} =
\hul \,\esymbol_{l}$. Except for the dots, the equation looks just
like Eq.~(\ref{eq:valuebalanceeq}). \co{FN: As mentioned before, with
  a linear cost function
  $\hminus(\esymbolv)=\sum_{l}\hul'\,\esymbol_{l}$, the enzyme
  investments is given by the enzyme cost $\hul'\,\esymbol_{l}$;
  details in CBA labour} \co{WO?  mention substrate investment more
  clearly, also in text!  evtl nicht hier in diesem beispiel, der
  einfachheit halber?}  \co{say: investments!}  An active enzyme
represents a positive investment: in an optimal state, it must also
have a positive point benefit!  By rewriting the left side
$\gul\,e_{l}=\gvtotl\,v_{l}$, we can express it as a
rate of value production.  The principle of local value
production\footnote{\co{NACH TEXT!}Metabolic states that satisfy the value production
  principle are called enzyme-economical.}, \co{define it!} a condition for
enzyme-optimal states, can be used as a constraint in flux balance
analysis. \co{(see section 9 and CBA fluxes)} Using flux values, it can
also be written as $\gvtotl\,v_{l}>0$.  Finally, by dividing
Eq.~(\ref{eq:fitnessbalanceeq}) by the flux $v_{l}$ and defining the
{\fluxburden} $\hvl = \frac{\hul\,\esymbol_{l}}{v_{l}}$, we reobtain
our balance equation (\ref{eq:bla345}).

\myparagraph{{\Fluxcostshade} and {\fluxbenefit}s} Let us now consider
economic variables in {\pointbenefitform} more generally.  if we
describe direct {\myvalue}s not by usual derivatives, but by
logarithmic derivatives, we obtain {\fluxbenefit}s
$\bvl = \partial \bbenefit/\partial \ln v_{l} = \bvtotl\,v_{l}$,
{\metabolitecost}s
$\metcostdotci = \partial \metcost/\partial \ln \cint_{i} =
\metcosti\,\cint_{i}$, and {\enzymecost}s (``{\enzymeinvestment}s'')
$\hudotl = \partial \hminusfun/\partial \ln \esymbol_{l} =
\hul\,\esymbol_{l}$.  A {\fluxbenefit} $\bvv = \bvtot \cdot\vv$
determines whether a {\flow} is beneficial ($\bvv>0$), {\futile}
($\bvv=0$, i.e.~satisfying
$\Nobj \, \vv = {\bvtot\trans \choose \Nint}\, \vv = 0$), or
{\wasteful} ($\bvv<0$). Futile or {\wasteful} {\flow}s are called
{\nonbeneficial}. We further introduce the {\fluxcostshade} (or ``flux
investment'') $\yvcostl=\hvl\, v_{l}$, which her, in kinetic models,
is equal to $\hul/\ratelaw_{l}$. \co{FN
  $= \frac{\hul}{\kcatl\, (1-\e^{-\theta_l})\,\eta^{\rm kin}(\cv)}$!}
In short, for our present models, we obtain $\yvcostl = \hudotl$,
i.e.~flux and {\enzymeinvestment}s are equal\footnote{We sometimes  search for an enzyme profile that realises a given  flux
  profile at a minimal enzyme cost (ECM problem).  The resulting total enzyme cost, written as a
  function of the flux profiles, is called enzymatic flux cost
  $\fluxcost^{\rm enz}(\vv)$ (see \cite{lieb:18fcm}). Its gradient
  $\acostv^{\rm enz}=\nabla_{\vv}\acostenz$ satisfies the
  relationship $\fluxcost^{\rm enz}_{\rm v}\,v =
  \hus\,\esymbol$.}. \co{practical aequivalenz von flux and enzyme
  cost; sagen. das steht fuer KINETISCHE flusskosten. WRITE!}
\co{tabelle mit allen namen und definitionen?}  \co{ausfuehrlicher;
  erwaehnen!}  ``Point'' versions of indirect values are defined
similarly.  Like in the example above, all economic laws can be
written in {\pointbenefitform} by by multiplying each economic
variable with the corresponding physical variable (i.e.~replacing
economic values by ``point'' values).\co{sometimes: just add the
  points, and the form of the law stays the same; sometimes need to
  replace unscaled elasticities by elasticities; more difficult with
  delta w ..}  \co{das mit dem punkt in den gleichungen unten
  tatsaechlich machen - also alle drei nochmal zusaetzlich (in einem
  eqnarray) in punktform schreiben} \co{hier econ rules in
  {\pointbenefitform}. frame them as conservation relations (very
  short)}
\begin{enumerate}[leftmargin=5mm]
\item \textbf{{\Fluxbenefitbalance}} The reaction balance in value production \co{uea statt ``point''}
  form relates value production to enzyme
  investment. In an optimal state, all active reactions must satisfy
  the value-price balance
  (\ref{eq:ReactionBalanceEnzymeValueForm}). By multiplying this
  balance with the enzyme level $\esymbol_{l}$, we obtain the reaction
  balance in {\pointbenefitform}\footnote{\co{kam schonmal: nach Eq
      23} Enzymes can also  be seen as ``external metabolites''. If
    we start from the {\loadbalance}
    (\ref{eq:loadfluxvalueequationPointCost}), replace the load
    $\loadi$ by the enzyme price $\hul$, and consider the scaled
    enzyme elasticities $E^{v_l}_{\esymbol_j}=\delta_{lj}$, we obtain
    the reaction balance.}
\begin{eqnarray}
 \label{eq:reactionbalanceeq}
 \underbrace{(\Deltar \wtotl + \bvdirl)\,v_{l}}_{\gul = \gtotlshade} = \underbrace{\hul\,\esymbol_{l}}_{\hudotl}.
\end{eqnarray}
It states that the  {{\fluxbenefit}}
$\gtotlshade= (\Deltar \wtotl + \bvdirl)\,v_{l}$ (the local value
production) must be equal to the {{\enzymeinvestment}}
$\hudotl = \hul \,\esymbol_{l}$, and therefore to the flux point cost
$\apointcostvl = \acostvl\, v_{l}$. Like the {\summationcondition}
(\ref{eq:fitnessbalance2x}), the {\fluxbenefitbalance}
(\ref{eq:reactionbalanceeq}) holds for any types of rate laws.  Here
are some practical consequences. Since active enzymes have positive
costs, {\fluxvalue} $\gvtotl=\Deltar \wtotl + \bvdirl$ and flux
$v_{l}$ must have equal signs, so in reactions without direct
{\fluxgain} ($\bvdirl=0$), the flux must lead from lower to higher
economic potentials. In reactions with direct {\fluxgain}s
($\bvdirl \ne 0$), fluxes may run in the orther direction if the
{\fluxgain} is sufficiently high\footnote{A model can always be
  rewritten without direct flux gains, by attributing all {\fluxgain}s
  $\bvdirl$ to the production of hypthetical external metabolites. In
  this reformulation, all fluxes follow the economic potential
  differences.}.  Turning this logic around, we can ask: given a flux
profile $\vv$, can there be internal economic potentials $\winti$ and
positive {{\enzymeinvestment}}s $\hudotl$ that satisfy the
\fluxbenefitbalance?  For economical {\flow}s $\vv$, the answer is yes
(Propositions \ref{th:theoremtestmode} and
\ref{th:theoremfutile}). For uneconomical {\flow}s -- e.g.~flux
profiles with futile cycles -- no consistent potentials $\wint$ can be
found. This closely resembles the role of chemical potentials in
thermodynamic flux analysis \cite{qibe:05}.

\item \textbf{\Compoundbenefitbalance} By multiplying the metabolite
  balance Eq.~(\ref{eq:concentrationbalanceequationext}) with the
  concentration $\cint_{i}$, we obtain the metabolite balance in point
  form\footnote{\co{diese herleitung schon oben vermerken; wie bei
      allen rules? // sagen, dass hier genauso wie oben
      gedankenexperimente verwendet werden koennen (siehe SI).}  In
    models without moiety conservation, the metabolite balance follows
    from a simple thought experiment.  In an optimal state, a
    concentration variation $\delta c_{i}$, has no fitness effect:
    $(\sum_{l}\gvtotl \Eunvlci - \hci) \,\delta c_{i}=0$.  Since this
    must hold for \emph{any} small variation, we can omit the term
    $\delta c_{i}$ and obtain the metabolite rule.}
  \footnote{Substrate and product elasticities have different signs,
    leading to positive and negative terms. Knowing the signs
    (assuming a positive flux, and therefore positive substrate and
    activator elasticities, and negative product and inhibitor
    elasticities) we can split the load into
    $\loadidot=\sum_{l\in \rm prod + act} \huldot \Escvlci -\sum_{l\in
      \rm sub + inh} \huldot |\Escvlci|$, with reactions $l$ in which
    the metabolite appears as a substrate, product, activator, or
    inhibitor. The sum terms themselves are all positive.}
 \begin{eqnarray}
  \label{eq:metabolitebalanceequation}
   \underbrace{\gci \,c_{i}}_{\loadidot} = \sum_{l} \underbrace{\hul \esymbol_{l}}_{\hudotl} \, \Escvlci
 \end{eqnarray}
 with the scaled
 elasticities $\Escvlci = \frac{c_{i}}{v_{l}}\Eunvlci$.
 \co{sort again the ideas of moiety conservation and relationship load
   - price:} The  equation relates a metabolite's {\point} load $\loadidot$ to
 the {{\enzymeinvestment}}s $\hudotl$ around the metabolite. On the right, we find
 a linear combination of enzyme investments with  scaled
 elasticities $\Escvlci$ as (positive or negative) prefactors. As we
 already know, in models without conserved moieties, $\loadidot$ is
 equal to the concentration {\price} $\hcidot$. If a metabolite load
 vanishes (e.g.~an internal metabolite without direct fitness
 effects that is not involved in moiety conservation), the sum on the
 right must vanish, so for each metabolite, we obtain a linear
 constraint on the enzyme levels. By defining ratios of enzyme levels,
 these constraints shape the proteome.

 What else can we learn from Eq.~(\ref{eq:metabolitebalanceequation})?
 Consider the pathway in Figure \ref{fig:unbranchedfluxART}.  If the
 fitness function contains no metabolite costs (and therefore,
 $\hc=0$), enzyme investments and reaction elasticities around a
 metabolite are inversely proportional:
 $\hudotl / \hudotlplusone = |E^{v_{l}}_{c_{l}} /
 E^{v_{l+1}}_{c_{l}}|$. We already know this from the
 {\connectivitycondition} (see Figure
 \ref{fig:unbranchedfluxART}). Typically, reaction substrates have
 larger scaled elasticities than reaction products. Hence, if
 ``production'' and ``consumption'' refer to flux directions (and not
 just to nominal reaction orientations) \cite{liuk:10}, producing
 reactions have higher {\fluxburden}s than consuming reactions, so
 {\fluxburden}s tend to decrease along the flux. \co{FN: in line with
   control coefficients ..} If our metabolite has a {\price}
 $\hci>0$, this {\price} appears in the balance equation and implies a
 positive load $\loadi>0$: in this case, consuming reactions must have
 higher elasticity-weighted {\enzymeinvestment}s than producing
 reactions\footnote{In unbranched metabolic pathways, this holds both
   for unscaled and scaled elasticities.}: this configuration makes
 intuitive sense because it keeps the metabolite concentration low.
 What about extracellular compounds?  Compounds with a positive
 influence on the {\metabolicobjective} (and with a positive
 concentration) have positive point loads, their import deserves an
 investment. In contrast, metabolites with a vanishing concentration
 or vanishing (or negative) load are not profitable for the cell:
 their transporters provide no benefit and should not be expressed.

\item \textbf{{\Loadbalance}} By multiplying
  Eq.~(\ref{eq:loadfluxvalueequation}) by $c_{i}$, we obtain the
  {\loadbalance} in {\pointbenefitform}
\begin{eqnarray}
 \label{eq:loadfluxvalueequationPointCost} 
\underbrace{\loadi\,c_{i}}_{\loadidot} =  \sum_{l} \underbrace{ (\Deltar \wtotl + \bvdirl)\,v_{l}}_{\gtotlshade} \,\Escvlci,
\end{eqnarray}
with the  point load $\loadidot$ and flux point
value $\gtotlshade$. We can  briefly write it as $\loadidot =  \sum_{l} \gtotlshade \,\Escvlci$.
\end{enumerate}

\myparagraph{Non-optimal states} \co{is there related text in the SI?}
When describing the value structure of metabolism, can we also
describe non-optimal states? The economic balance equations assume
optimal enzyme levels.  In reality, cells do not behave optimally, at
least not precisely, and certainly not for our simple optimality
criteria. Even without expression noise or leaky transcription, cells
would always be maladapted after perturbations such as gene
knock-downs.  \todo{Apparent non-optimality may arise from side
  objectives or from preemptive expression, and enzyme levels may not
  be optimal at all.  However, it may be practical to describe
  non-optimal states by using our optimality formalism.}  In fact,
metabolic value theory defines economic variables and rules for
\emph{any} metabolic state, not just optimal states. The only
difference is that, in non-optimal states, there are \emph{economic
  imbalances} (or ``stresses''\footnote{An economic stress can be seen
  as a force that pulls an enzyme towards its optimal expression
  level. If stresses could be sensed by the cell, they would be useful
  regulatory signals for steering the enzyme levels.  \co{(REF CBA
    regulation; briefly mention bones and trees here, rest in other
    artikel. DORT erwaehnen, dass stress der physikalischen spannung
    entspricht!}})
$\fudotl = \frac{\partial \ffit}{\partial e_{l}} =\gul-\hul$ that
describes a mismatch between the values and prices of enzymes.  Since
all stresses (of expressed enzymes) must vanish in optimal states,
they were not considered in the economic balance equations (meant to
describe optimal states). To describe non-optimal states, we can
include them as extra terms\footnote{\co{ans
    abschnittende!}\co{klar?}Note that the enzyme stress is different
  from the shadow value (i.e.~for an enzyme level that hits a lower or
  upper bound), but can have similar effects, turning the normal
  balance equation into an inequality.}, yielding the reaction
imbalance (see SI \ref{sec:violationsnonenzymatic})
\begin{eqnarray}
\label{eq:econForceEquation}
 (\Deltar \wtotl + \bvdirl) \, v_{l}&=& (\hul + \ful)\,\esymbol_{l}.
\end{eqnarray}
The  stress $\ful$ implies an \emph{imbalance}
between enzyme cost and benefit:  a positive stress (indicating that an enzyme
level is too low for an optimal state) yields the
economic imbalance
\begin{eqnarray}
\label{eq:econForceInequality}
 (\Deltar \wtotl + \bvdirl) \, v_{l}&>& \hul\,\esymbol_{l}.
\end{eqnarray}
In this case (i.e.~a flux stress with the same sign as the flux), the
enzyme's point benefit exceeds the point cost, and the cell would be
able to improve its fitness by increasing the enzyme level. Of course,
with a negative stress (i.e.~an enzyme level higher than required for
an optimal state), the inequality changes its sign, and the enzyme
level should be decreased.  If a non-optimal state is due to a
constraint (e.g.~a bound on a flux to model an enzyme knock-down), the
constraint will lead to a shadow value, and this shadow value can be
included into the flux {\gain} in brackets. \co{explain what that
  means} Non-zero stresses indicate a non-optimal state, and how the
cell can improve this state by changing the enzyme levels -- that is,
they hint at selection pressures.  Imagine that a cell cannot perform
some useful reaction because it has no enzyme for it. To quantify the
incentive for having this enzyme, we could start from the current
metabolic state and include the reaction into the network, but assume
that the system (with an enzyme price $\hul$) is not expressed.  The
flux value of the new, inactive reaction is given by
\begin{eqnarray}
\label{eq:econForceInequalityValueForm}
 \underbrace{\Deltar \wtotl + \bvdirl}_{\gvtotl}&=& (\hul + \ful)\,\underbrace{\esymbol_{l}/v_{l}}_{\ratelaw_{l}} = \acostvl + t_{v_{l}}.
\end{eqnarray}
A flux stress $t_{v_{l}}=t_{e_{l}}/k_{l} = \gvtotl-\acostvl$
describes an imbalance between flux value $w_{v_{l}}$ and flux
{\burden} $\acostvl$. A positive stress (or more precisely, a stress
with the same sign as the desired reaction rate) indicates that
evolving the enzyme would be profitable for the cell. \co{FN zu
  mutation cost, ref moderatel efficient enzyme supplement etc.}

\begin{figure*}[t!]
    \begin{center}
      \includegraphics[width=12.5cm]{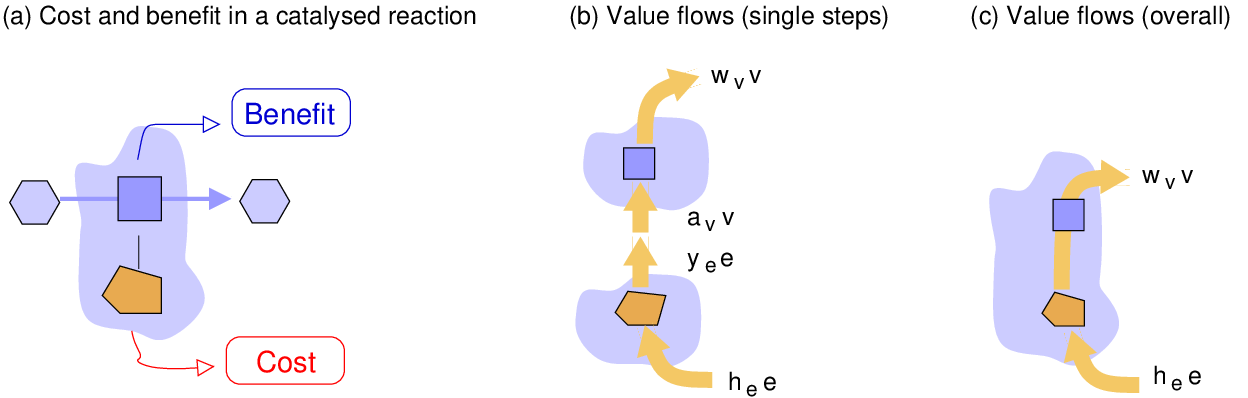}\\[3mm]
    \end{center}
    \caption{Conservation of economic value. (a) Example: a reaction
      (scored by a benefit function) and its catalysing enzyme (scored
      by a cost function). The subsystem considered is shown by a
      ``cloud''. From Figure \ref{fig:backpropagation} we know that
      economic variables can be defined in two ways.  Applying usual
      derivatives to benefit and cost functions yields values and
      {\price}s, while applying scaled derivatives (also called
      ``logarithmic'' or ``point derivatives'') leads to point
      benefits (also called ``utilities'' or ``importances'') and
      point costs (or ``investments''). Using these point derivatives,
      we can write all our economic laws in ``value production form'',
      simply equating ``inflowing'' and ``outflowing'' values. \co{REF
        to CBA labour=}\co{\cite{lieb:cbalabour}} (b) Value
      flows. Point costs and point benefits of enzyme (bottom) and
      reaction (top). Equation (\ref{eq:bla123}) tell us that all four
      quantities are equal (in optimal states). We can depict them as
      ``value flows'' entering and leaving our network elements. Value
      is ``conserved'' within each element, and also in between enzyme
      and reaction.  (c) The same value flows can also be used to
      describe the coupled subsystem of reaction and enzyme (and any
      larger subsystems of metabolic networks, not shown).}
 \label{fig:fourEqualities} 
\end{figure*}

\myparagraph{Value flows and equality between point cost and point
  benefit} Enzyme investments (point costs) and benefit contributions
(point benefits) have the same measurement units and satisfy
conservation relations, which suggests that they may be
interconvertible. We can see them as different forms of the same
``substance'', just like heat and work are different forms of a
``substance'' called energy.  \todo{More specifically, the value
  production equation can be seen as a conservation law for ``value
  flows'' (see Fig.~\ref{fig:fourEqualities}): in the ``value flow
  picture'', enzyme investments are values that flow into the system,
  are become benefit contributions, and eventually reach reactions and
  metabolites in which benefit is realised, and flow from there into
  the benefit function. The conversion between Enzyme investments and
  benefit contributions happens in every single reaction: coming from
  substrate and enzyme, value flows into the reaction \co{at rates
    given by point costs} and towards the product and into the direct
  flux benefit. \co{at rates given by point benefits}} As shown in
Fig.~\ref{fig:fourEqualities} (c), by rewriting enzyme {\myvalue}s as
$\gul=\gvtotl\,\frac{\esymbol_{l}}{v_{l}}$, the cost-benefit balance
Eq.~(\ref{eq:fitnessbalanceeq}) can be written as
\begin{eqnarray}
 \label{eq:bla123} 
 \underbrace{\gvtotl\,v_{l}}_{\gtotlshade} = \underbrace{\hvl\,v_{l}}_{\yvcostl} &=& 
 \underbrace{\gul\,\esymbol_{l}}_{\ubenel} = \underbrace{\hul\,\esymbol_{l}}_{\hudotl},
\end{eqnarray}
stating as equalities between {\fluxbenefit}, {\fluxcostshade}, enzyme
{\benefitshade} and {{\enzymeinvestment}} in all optimal states.
Thus, while value changes its form (being embodied in metabolite rates
or enzyme levels), in optimal states it is always conserved. In
non-optimal states, value is not conserved: it appears or disappears
wher value balances do not hold.

\section{The shape of economical metabolic {\flow}s}

\myparagraph{\ \\Economical {\flow}s} What flux distributions can we
expect to find in enzyme-optimal states?  Some general features, which
are independent of enzyme kinetics, follow from the variation
condition Eq.~(\ref{eq:fitnessbalance2x0}). A flux profile that
satisfies this equation with positive {{\enzymeinvestment}}s is called
\emph{economical}\footnote{This definition holds for {\complete}
  {\flow}s. Inactive reactions must be omitted from our model before
  the criterion can be applied.  Vanishing {\flow}s are defined to be
  uneconomical.}, and {\flow}s must be economical to appear in
enzyme-balanced states\footnote{\co{WO?}Finding economical flux modes
  or testing flux modes for being economical can be important in
  practice. For example, when computing enzyme investments by using
  the {\summationconnectivitycondition}, the assumed {\flow} must be
  economical. We need to be able to check this without knowing kinetic
  details.}! Importantly, economical {\flow}s are not just beneficial
($\bvtot \cdot \vv >0$), but must be \emph{locally beneficial}: each
enzyme must have a positive influence on the metabolic
benefit. Uneconomical {\flow}s entail a waste of enzyme, no matter
which underlying kinetics or enzyme cost functions are assumed. How
can we check in practice whether a flux profile is economical?

\begin{figure*}[t!]
\begin{center}
 \includegraphics[width=15.5cm]{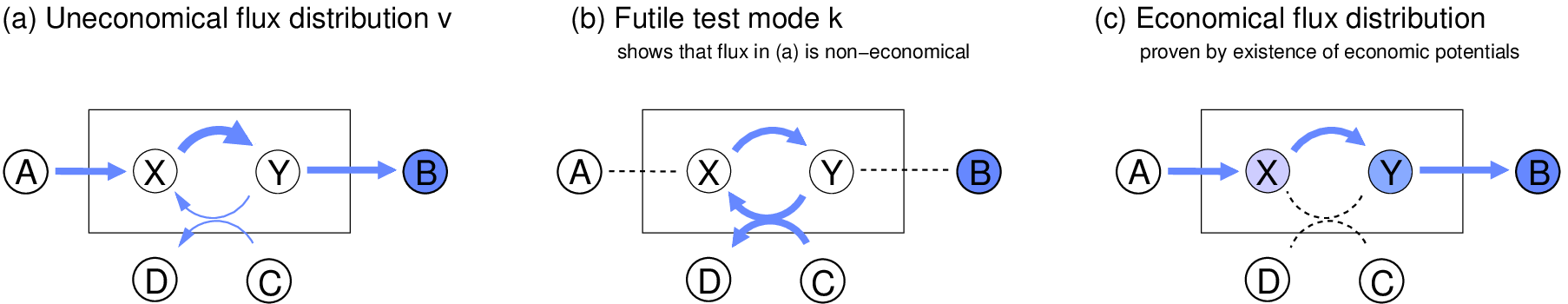}
\end{center}
\caption{Futilre submodes and economical {\flow}s.  A {\flow} is
  called economical if it can satisfy the {\summationcondition}
  (\ref{eq:fitnessbalance2x}) with positive enzyme investments. (a)
  Example pathway with production of metabolite B (blue circle) as the
  metabolic objective.  The {\flow} shown is uneconomical. We can see
  this by applying the submode criterion with the submode
  $\modevector$ shown in (b). An economical {\flow} cannot contain any
  futile submodes. Since this submode is futile (no B is produced) and
  conformal with the {\flow} $\vv$, $\vv$ must be uneconomical. We can
  see this from the {\summationcondition} (\ref{eq:fitnessbalance2x}):
  since $\bvtot$ scores the production rate of B, the right-hand side
  $\modevector \cdot \bvtot=0$ and the left-hand side
  $\modevector \cdot (\hudot \oslashs \vv) \ne 0$ cannot be equal. (c)
  Economical {\flow}. To show that this {\flow} is economical, we note
  that there are economic potentials (shades of blue) such that all
  fluxes lead from lower to higher potentials. In (a), for example,
  the flux cycle would make a consistent choice of economic potentials
  impossible. Note that the economic potentials of the external
  compounds A, B, C, and D are predetermined by the metabolic
  objective.  }
\label{fig:testmodetheorem} 
\end{figure*}

\myparagraph{Submode criterion for economical flux modes} \co{FN This
  rule, called submode criterion, can also be derived from the
  {\summationcondition} (\ref{eq:fitnessbalance2x0})\co{see SI ..?}.}
In the submode criterion for economic flux modes, we test for
non-beneficial flux motifs, i.e.~ local configurations of flux
directions that would exclude an optimal usage of enzymes. A motif (or
conformal submode\footnote{\co{redundant?} A conformal submode of a
  flux profile $\vv$ is a submode whose flux directions match the
  directions in $\vv$ (see appendix \ref{sec:appSubmodeCriterion}).})
in a flux profile $\vv$ is a set of active reactions that (by
themselves) can carry a stationary flux with the same flux
directions. Why do non-beneficial motifs make a flux profile
noneconomical? The explanation is simple: in an optimal state, any
flux variation must be fitness-neutral. Consider a flux variation
$\delta \vv$, caused by a change in enzyme levels that is itself a
submode of $\vv$.  Applying this flux variation will increase some of
the fluxes (but will not decrease any fluxes), so the enzyme demand
increases. Since (constraint-respecting) variations in enzyme-balanced
states must be fitness-neutral, the additional enzyme cost must be
balanced by an extra benefit. This means: in an enzyme-optimal state,
any conformal flux variations $\delta \vv$ must be beneficial, and
economical {\flow}s cannot contain non-beneficial submodes!  The
submode criterion refers only to flux directions (and not to flux
magnitudes) and \todo{can be used} by comparing the {\flow} to elementary
\co{proof??}  futile submodes. If futile motifs are present, it is
impossible to find economic potentials that satisfy the reaction
balance; conversely, if no futile motifs are present, this guarantees
that economic potentials can be found. \co{beweis?}  Figure
\ref{fig:testmodetheorem} shows an example: the {\flow} in (a)
contains, as a submode, the mode shown in (b). Since this submode is
futile, the {\flow} in (a) must be uneconomical and cannot be realised
by enzyme-optimal models. The {\flow} in (c), which does not contain
the submode, is economical. \co{be clear UEA about futile and
  wasteful!  distinguish!  analogy to thermodyn (in cba ii):
  endergonic {\flow}s are forbidden, but only cyclic submodes can be
  recognised without knowing chem pots!}

\myparagraph{Economical metabolic {\flow}s and flux cost minimisation}
Economical {\flow}s, as defined in kinetic models, are related to the
principle of minimal fluxes\footnote{\co{was ist mit corresspondence
    auch mit lag mult dabei?}  FBA considers fluxes and ignores
  concentrations and enzyme kinetics.  If pathways compete for enzyme
  resources is modelled by heuristic flux cost functions such as the
  sum of absolute luxes or weighted sums of fluxes as proxies for
  enzyme cost \cite{holz:04,holz:06}. Flux costs are less realistic
  than enzyme costs, but they can be computed without any kinetic
  information.} \cite{holz:04}, a heuristic rule in FBA. The principle
of minimal fluxes states that a {\flow} must satisfy the FBA
constraints (stationarity and flux bounds), realise a given
{\metabolicobjective} $\fluxbene'= \bvtotweight \cdot \vv$, and at the
same time minimise the sum of absolute fluxes. Flux Cost Minimisation
(FCM) \cite{lieb:18fcm} applies the same principle, but with general
cost functions\footnote{Flux cost functions must be differentiable and
  must increase with the flux, $\partial \acost/\partial |v|>0$, so
  flux prices $\partial \acost/\partial v_l$ and fluxes $v_l$ must
  have equal signs
  (i.e.~$\partial \acost/\partial \ln |v_l| = \partial \acost/\partial
  v_l\,v_l>0$ if $v_l\ne0$).}  $\acost(\vv)$ such as the sum of
absolute fluxes $\acost(\vv) = \sum |v_{l}|$ \cite{holz:04,holz:06}
and the weighted sums of absolute fluxes
$\acost(\vv) = \sum \acostvlweight |v_{l}|$ with cost weights
$\acostvlweight$ (or $\acost(\vv) = \sum \acostvlweight v_{l}$, if
fluxes are known to be positive).  An FCM problem is called
\emph{flux-enforcing} if its flux bounds exclude the equilibrium state
$\vv=0$ (by positive lower or negative upper flux bounds)\footnote{In
  FCM, flux bounds can be used to enforce a non-zero flux in an
  ATP-consuming maintenance reaction.  Similar flux bounds can be
  imposed in kinetic enzyme optimality problems. In both cases, the
  resulting shadow values can be included into the flux gain vector
  (see SI \ref{sec:inequality constraints}). The resulting effective
  flux gain vector $\bvtot$ changes the set of futile submodes, and
  previously uneconomical flux distributions become economical.}.  In
the corresponding FCM problems, shadow {\gain}s in the vector $\bvtot$
will lead to different flux solutions. FCM and metabolic value theory
lead to the same flux solutions: any non-flux-enforcing FCM
problem\footnote{This correspondence does not hold for flux-enforcing
  FCM problems because flux bounds could enforce futile submodes,
  which make the {\flow} uneconomical.} yields an {\flow} that is
economical, and any economical {\flow} can be obtained by some
non-flux-enforcing FCM problem (see Proposition \ref{th:theoremPMF1},
``Nocturno principle''). Starting from an enzyme-balanced state (with
{\fluxgain} vector $\bvtot$), we can construct many FCM problems with
flux objectives $\fluxbene(\vv) = \bvtot \cdot\,\vv$ and different
cost functions $\acost(\vv)$. All this holds both for general FCM
problems and for FCM problems with linear flux costs functions
$\fluxcost(\vv) = \sum_{l} \acostvl'\,v_{l}$ (for each given pattern
of flux directions). Thus, any economical {\flow}s can be predicted by
linear FCM by choosing appropriate flux cost weights! In theory, by
randomly choosing cost weights $\acostvl'$ and computing the flux
solutions, any economical {\flow} can be found.  We saw that
enzyme-optimal states correspond to solutions of FCM, a nonlinear
version of FBA.  In FCM, fluxes are optimised for a minimal cost,
which suppresses futile cycle fluxes.  In enzyme optimisation, it is
enzymes, not fluxes that are costly, but optimising the enzyme levels
leads again to economical fluxes.  So both methods restrict fluxes in
the very same way (see Proposition \ref{th:theoremPMF1} in SI), and
this even holds if the metabolic objective in enzyme optimisation
depends on metabolite levels: the reason is that the
{\fluxbenefitbalance} (\ref{eq:reactionbalanceeq}), our criterion for
economical {\flow}s, does not depend on concentration
{\price}s. Interestingly, FCM problems yield solutions that capture
metabolite costs! \co{reason: ``production economics'' and
  ``concentration economics'' act separately! see disc and appendix
  \ref{sec:AppFCMmetCost}} This also means that kinetic models can be
used to justify FCM: for any FCM solution $\vv$, there will be a
kinetic model that realises these fluxes by optimal enzyme
levels. Conversely, FCM (and even linear FCM) can be used to compute
economical {\flow}s to be realised in enzyme-optimal states.  \co{NEU
  SCHREIBEN! GANZ KURZ! REF to app A2! aber wenn die beiden
  vergleichbar sind, was entspricht dann den oekonomischen potentialen
  in fcm? und wie kann es sein, dass ME metabolitkosten zulaesst, die
  in FCM nicht vorkommen? das wird in CBA II diskutiert. The economic
  potentials are formally related to Lagrange multipliers in the FCM
  problems.}

\myparagraph{Economic potentials in flux balance analysis} Coming back
to FBA, how can we make sure that our flux solutions can also be
realised by kinetic models in enzyme-optimal states?  An FBA problem
assumes an objective $\bbenefit(\vv) = \bvtot \cdot \vv$ with a
constant vector $\bvtot$. Given the model, we can consider the set of
all possible kinetic models with the same network structure and
{\fluxgain} vector $\partial \fluxbene/\partial \vv=\bvtot$, and ask
about their enzyme-optimal states.  Can we restrict our FBA problem to
flux distributions that occur in one of these states? Since FBA does
not consider enzyme kinetics, this may seem difficult, bur in fact the
only thing we need to do is to exclude non-economical fluxes. However,
our criteria for economical fluxes -- submode criterion and existence
of compatible economic potentials and enzyme investments -- do not
depend on kinetics and can therefore be used in FBA. This is fairly
simple: in addition to stationary fluxes, we need to determine
economic potentials that satisfy the reaction balance
(\ref{eq:ReactionBalanceEnzymeValueForm}).  Like in thermodynamic FBA,
we obtain an extra constaint that restricts possible flux modes to a
number of segments in flux space (flux orthants or their
lower-dimensional surfaces).  In the resulting ``Value Balance
Analysis'', all flux solutions must be economical, i.e.~they must
satisfy a reaction balance with some choice of economic potentials.
The external economic potentials follow from the objective function,
while the internal potentials $\winti$ must be chosen to achieve a
positive value production $(\bvtotl+\Deltar \wintl)\, v_{l}>0$, which
means that flux values and fluxes must have the same
signs. Mathematically, the latter condition resembles the
thermodynamic constraint between chemical potentials and flux
directions \cite{hohh:07}, where economic and chemical potentials
correspond to each other.  In practice, economic FBA uses energetic
and economic constraints simultaneously: unphysical and uneconomical
metabolic {\flow}s are excluded at the same time.  \co{sagen,
  umgekehrt: shadow prices in FBA must satisfy the same laws (ref CBA
  lag + CBA fluxes)} Economic and thermodynamic constraints can shape
fluxes in similar ways.  For example, just like thermodynamics
excludes thermodynamically infeasible cycles, the value production
principle excludes futile cycles.  \coout{Here kinetic rate laws or
  elasticities do not appear, so the {\summationconnectivitycondition}
  (\ref{eq:fitnessbalance2x}) and (\ref{eq:fitnessbalance3x}) cannot
  be directly used. }

\section{Kinetic models in enzyme-optimal states}

\coout{Uebersicht: als uebrleitung zur rekonstruktion: nochmal
  zusammenhang von FBA, ME, potentialen, zyklen usw klarmachen. kurz
  klarmachen, dass alles auf der aequivalenz von fba und me (bzw der
  allgemeinheit von oekon fluessen) beruht}

\myparagraph{\ \\Economical {\flow}s can be realised by
  enzyme-balanced kinetic models} \co{ist alles noch ein bisschen
  bloed verteilt .. erst ein abschnitt hier, dann einer im supplement;
  dann die genaue erklaerung im ``proofs''-text ..} A search for
optimal metabolic states by numerical optimisation may lead to
irrelevant local optima, for instance a zero-flux state in which all
enzyme levels vanish and small expression increases would not pay
off. To obtain meaningful optima, can we use metabolic value theory to
construct kinetic models systematically in enzyme-balanced states? In
fact, metabolic value theory was initially developed from a simple
question: is there a systematic way to construct kinetic models in
enzyme-optimal states, as a starting point for assessing optimal
enzyme activity changes? \co{REF opt diff expr} We saw that
enzyme-optimal states must come with consistent economic potentials
and fluxes, which in turn requires the {\flow}s to be
economical. \co{so in theory, it may be possible to choose an
  economical flux profile and compatible UEA STATT CONSISTENT? economi
  potentials, and to construct a kinetic model around it.} But how to
obtain the economic potentials? If a given kinetic model with
metabolic state $(\vv,\cv,\esymbolv)$, the economic potentials can be
computed by taking derivatives. But can we turn this around?  Can we
choose an economical flux distribution and a set of economic
potentials, and construct a kinetic model with exactly these fluxes
and economic potentials? And will this model be enzyme-balanced, or
even be enzyme-optimal?

\co{\textbf{Proposition (economical fluxes and enzyme-balanced kinetic
    models):} Any economical metabolic state (defined by flux gains,
  economical fluxes, and compatible economic potentials) can be
  realised by an enzyme-balanced kinetic model with the same flux
  gains (and suitably chosen metabolite prices).}

\myparagraph{Systematic model construction} Due to this proposition,
if a metabolic network, a flux {\gain} vector $\bvtot$, and an
economical {\flow} $\vv$ are given, kinetic models with this \flow\
and with optimal enzyme levels can be constructed (see SI sections
\ref{sec:reconstruction} and SI \ref{sec:realisekineticmodels}). We
proceed in two steps. In the first step, the steady-state phase, we
choose thermodynamically feasible metabolite concentrations as well as
economic potentials satisfying the {\fluxbenefitbalance}.  To obtain  biologically plausible economic potentials, we can use
extra constraints\footnote{To put realistic constraints on the
  economic potentials, we may use Eq.~(\ref{eq:FluxValueEq1}) with
  flux {\price}s $\hvl = \hul/\ratelaw_{l}$, where
  $\ratelaw_{l}=v_l/\esymbol_l$ is the catalytic rate. For a positive
  flux $v_{l}$, we obtain the inequality
  $ \hul\,\esymbol_{l}/v_{l} \ge {\hul}^{\rm min}/k^{\rm cat} = h^{\rm
    v, min}_{l}$, where $k^{\rm cat}$ is the forward catalytic
  constant and \co{symbol?} ${\hul}^{\rm, min}$ is the minimum the
  {{\enzymeprice}}.  Enzyme {\price}s $\hul$ can be estimated using
  Eq.~(\ref{eq:investmentfunction}) from protein sizes and life times,
  normalised to a total investment
  $\sum_{l} \hul\,\esymbol_{l} = \sum_{l} \bvtotl \,v_{l}$, thus
  matching the total point benefit.}, heuristic assumptions
(e.g.~similar {{\costshade}}s for all enzymes \cite{lieb:14b}), or
data (e.g.~by fitting economic potentials to proteomics data as
proxies for enzyme costs).  Inactive reactions are omitted from the
model\footnote{In our model construction, vanishing fluxes can always
  be \co{in matlab: das am schluss nochmal pruefen!}  justified by
  assuming a large enzyme {\price} or a low catalytic constant.}.  In
the second step, the kinetic phase, we determine economic loads
$\loadi$ and reaction elasticities $\Eunvlci$ satisfying
$\gcint = \loadint - \hc$ and
Eq.~(\ref{eq:loadfluxvalueequation}). \coout{ in the form
  $0 = {\Escint} \trans\, \hudot - \loadint$, \co{as well as the
    general condition $\Lmat\trans \,\hc = \Lmat\trans \,\loadint$.}}
However, after our arbitrary choice of economic potentials  in step one,  there may be  no solution anymore  with our given 
metabolite price vector $\hc$. To obtain a solution anyway, we allow
for a (minimal) adjustment of $\hc$. We obtain a set of constraints
that define kinetically feasible elasticities and  choose
(e.g.~sample) the elasticities under  these constraints. From the elasticities, all kinetic
constants for the model can be constructed. \co{ref STM}  Aside from its practical
usage, this algorithm shows that any economical {\flow} can be
realised by some enzyme-balanced kinetic model.  Whether these models
are enzyme-optimal (i.e.~dynamically and economically stable) has to
be checked separately.

\begin{figure*}[t!]
\begin{center}
 \begin{tabular}{ll}
  (a) Economic potentials in respiration &
  (b) Economic potentials in fermentation \\ 
  \includegraphics[width=7.7cm]{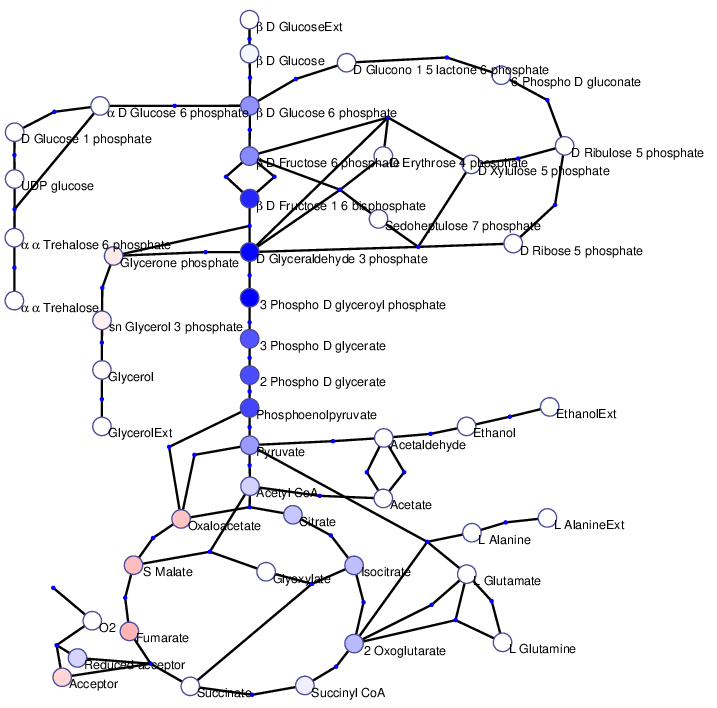} &
  \includegraphics[width=7.7cm]{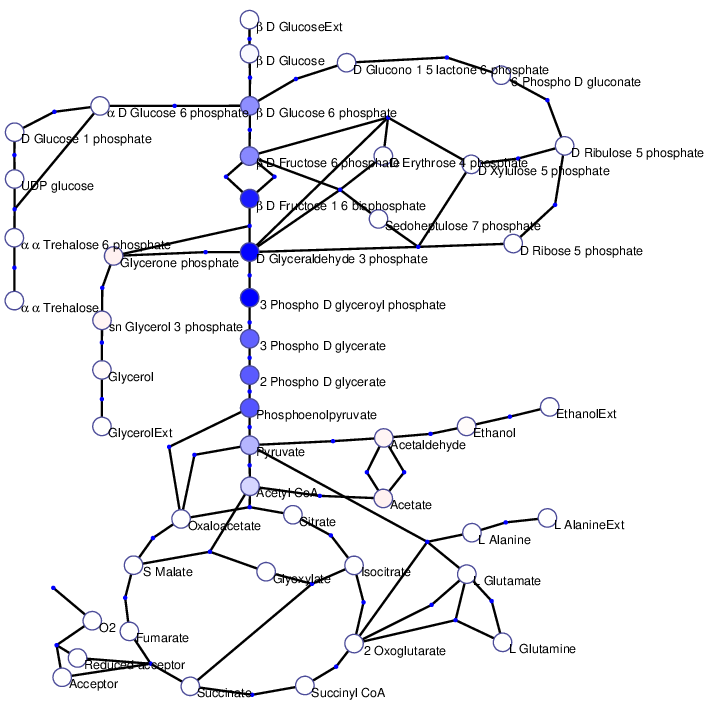}
 \end{tabular}
\end{center}
\caption{\co{add color legend; show fluxes} \co{SCHREIBEN!!} \co{fluesse zeigen;
    pfeile schoen; schriftgroesse; farben klarer, kreise groesser;
    schrift klarer; abkuerzungen Pfeile deutlicher! ADD COLOUR BARS}
  Economic potentials in the central metabolism of yeast, constructed
  from data and simple heuristic assumptions. (a) Economic potentials
  during respiration (usage of TCA cycle and oxidative
  phosphorylation). (b) Economic potentials during fermentation (with
  ethanol overflow). To reconstruct an enzyme-balanced state, I first
  determined thermodynamically feasible, economical fluxes by flux
  minimisation, where ATP production was used as the metabolic
  objective and some flux directions were predefined. Then, chemical
  and economic potentials were chosen within predefined ranges and in
  line with the known flux directions: the two types of
  potentials have  to decrease or increase, respectively,
  along the fluxes \co{(arrows)} (circle colours: pink: negative;
  white: zero; blue: positive; cofactors not shown).  Finally,
  economic loads and elasticities were chosen and the kinetic
  constants were computed. \co{note different color scales!} \co{genau
    klarmachen, welche infos in rekonstruierte potentiale geflossen
    sind und wie (ergibt das gesamtbild sinn?)} \co{sind
    proteindaten verwendet worden?  falls ja, sagen und auf proteomap
    in abb 1 verweisen}}
 \label{fig:reconstructed}
\end{figure*}

\myparagraph{Details of model construction} In our workflow for model construction we  can 
integrate various types of data including metabolite concentrations, fluxes,
kinetic constants, enzyme efficiencies, and protein cost.  For
example, we can choose economic potentials and {{\enzymeinvestment}}s
that comply with proteomics data and protein sizes, and then realise
the resulting economic state by a kinetic model.  \co{sagen, wie
  fluesse bestimmt wurden / bestimmt werden koennen}  Model
variables can  either be sampled, optimised, or chosen based on
experimental knowledge or heuristical rules. By sampling repeatedly,
we obtain an ensemble of kinetic models, each realising our {\flow}
under all kinetic, thermodynamic, and economic constraints.  Figure
\ref{fig:reconstructed} shows an example, a model of central
metabolism in yeast. To choose the economic potentials, I make the simple heuristic assumption that
the enzyme investments are similar between all enzymes.  Alternative assumptions would be that  known enzyme
levels are translated into {{\enzymeinvestment}}s to which
economic potentials could be fitted \cite{lieb:14b}, or flux
{\burden}s $\hvl=\hul/(v_{l}/\esymbol_l) \ge \hul/\kcatl$ are estimated from known $k_{\rm cat}$ values and enzyme sizes.
Using Eqs (\ref{eq:fitnessbalance2x}) and (\ref{eq:fitnessbalance3x}),
these burdens can be adjusted to satisfy all constraints of an
enzyme-optimal state.

\coout{Describe results for the yeast model in more detail. is it
 enzyme-optimal? depends on the cost hessian allowed; btw the cost
 hessian can be restricted by the cost gradient .. so there's only
 one free parameter in choosing the hessian .. does that fit with the
 construction? implement this ..}

\coout{heat can speed up reactions; but also perturb enzymes; include it
 in rate laws or fitness; bsp fuer alternative modellformulierung:
 auch : activitaet vs enz level; include ribosomes or not}

\coout{(like in Figure \ref{fig:metabolicEconomics})} \coout{Locality
 principle (i) puts a lot of constraints. WHICH ONES?.}
\coout{EXTRACT IMPORTNT EXAMPLES, DELETE THE REST If were microbes
 were optimised for fast growth under all circumstances, any gene
 knock-out in an experiment should impair growth, and there should be
 no buffering double knockouts. Many gene knockouts entail an
 adaptation of protein expression, but typically, this adaptation will
 not serve to maximise growth. Such non-optimal adaptations are the
 basis of MOMA (which predicts flux adjustments, assuming adjustments
 are \emph{small}) in comparison to FBA (which would assume that
 adjustments are \emph{optimal}). On the contrary, there are gene
 knock-outs that improve growth under laboratory conditions; under
 the optimality assumption, gene knock-outs should impair the fitness
 at least if knock-out and non-expression have the same fitness
 effects. Another example that rules out optimality is the
 fitness-buffering effects of some double knock-outs \cite{chck:07}.
 In all these cases, observed enzyme profiles are not optimised for
 growth under the conditions studied.}

\myparagraph{Dynamically and economically stable states} Will
enzyme-balanced states constructed as show above be enzyme-optimal?
While an enzyme balanced state satisfies the necessary optimality
conditions (ensuring stationary and enzyme-balanced states), an
enzyme-optimal state also needs to satisfy \emph{sufficient}
optimality conditions (ensuring dynamically and economically stable
states). In our construction, the necessary condition is satisfied via
the cost-benefit balance Eq.~(\ref{eq:fitnessbalanceeq}). To satisfy
the sufficient conditions, any small metabolic perturbations must be
buffered by the system dynamics, and any small enzyme variation must
lead to a (second-order) fitness decrease.  Thus, in a second-order
approximation, Jacobian matrix and fitness curvature matrix
$\Ffit_{\rm uu} = \partial^{2} \ffit/(\partial \esymbol_{l}\, \partial
\esymbol_{k})$ for active enzymatic reactions must be negative
definite\footnote{\co{ist das ein allgemeines problem? dann sollte es
    hier diskutiert werden! ansonsten, kann das nach cba synergies?
    oder zumindest nach SI?} The local shape of our fitness function
  $\ffit(\esymbolv)$ in enzyme space around an optimal state is
  described by the curvature matrix
  $(\partial^{2} \ffit/\partial \esymbol_{l}\,\partial \esymbol_{k})$.
  A vanishing eigenvalue shows that the fitness varies linearly (or
  remains constant) in some direction in enzyme space, so the optimal
  state is non-unique, zero, or does not exist (optimum at infinite
  enzyme levels).  This may happen, for example, in models with a
  linear cost function $\hminusfun(\enzymev)$ and a linear benefit
  function $\bbenefit(\vv)$: starting from any metabolic state, a
  proportional increase of all enzyme levels would change the fitness
  linearly, so the optimum is at $\enzymev=0$ or
  $\enzymev \rightarrow \infty$. To obtain a finite solution, we need
  to add a constraint (e.g.~an upper bound on enzyme
  cost). Alternatively, we may search for the optimal shape of an
  enzyme profile regardless of its absolute scaling.  To do so, we may
  maximise the {\metabolicobjective}/enzyme cost ratio or the return on
  investment ({\metabolicobjective} minus cost, divided by the
  cost). This function has a local optimum with a vanishing curvature
  and vanishing slope in the direction of an overall enzyme scaling
  (which we therefore need to constrain).  \co{bild?}}. The latter
criterion, called ``economic stability'', discards models with
dynamically and economically unstable states, that is, models in local
minima or saddle points of the fitness function.  To find such models,
we may construct enzyme-balanced models (as described above) and
select those with dynamically and economically stable states (for
details, see SI \ref{sec:reconstructionbalanced}).  Economic stability
may be formally ensured by strongly curved enzyme cost functions
(entailing positive curvatures in all directions in enzyme space), but
such cost functions may be biologically unrealistic.

\iftoggle{bookversion}
{\section{Conclusions}}
{\section{Discussion}}

\co{WO?
  hier logik: enzyme einzeln einstellen; auch moeglich: budget optimal
  verteilen (auch ``opportunity cost'') mention economic tenmperature
  / enzyme replacement kurz (lang in CBA labour)}

\begin{figure*}[t!]
\centerline{\includegraphics[width=15.5cm]{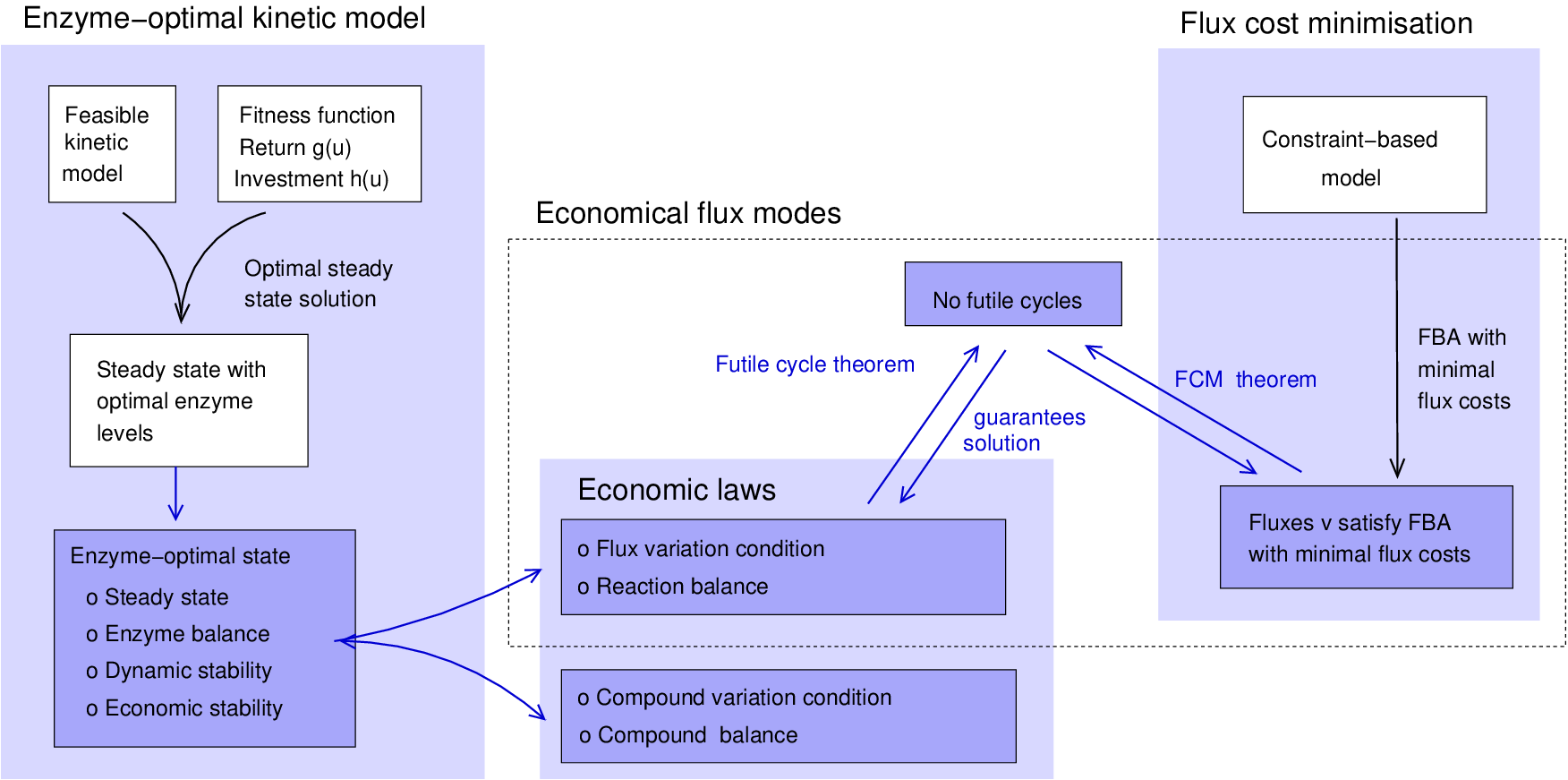}}
\caption{Economic laws for metabolic states. Conditions for kinetic models
  in enzyme-optimal states are shown on the left. For enzyme-balanced
  states, the necessary first-order conditions must be satisfied
  (stationary state and cost-benefit balance
  Eq.~(\ref{eq:fitnessbalanceeq}) describing a fitness extremum). For
  enzyme-optimal states, the sufficient second-order conditions must
  also be satisfied (dynamic stability, i.e.~a negative
  definite Jacobian; and economic stability, i.e.~a negative definite
  curvature matrix $\Fuu$, ensuring a fitness maximum). From the necessary
  conditions follows  a cost-benefit balance, entailing that active enzymes
  must have a positive control on the metabolic objective (``benefit
  principle''). This cost-benefit balance further  leads to the
  {\summationconnectivitycondition} (\ref{eq:fitnessbalance2x}) and
  (\ref{eq:fitnessbalance3x}) and to the economic balance equations
  (\ref{eq:reactionbalanceeq}) and
  (\ref{eq:metabolitebalanceequation}). Economical {\flow}s have 
  convenient properties (dashed box): they satisfy the
  {\summationcondition} and the reaction balance, are free of futile motifs, are solutions of
  FCM problems, and satisfy a {\fluxbenefitbalance} with positive
  enzyme investments (right box).} \label{fig:cbafbamsf}
\end{figure*}

\myparagraph{\ \\Metabolic value theory for kinetic models} Metabolic
value theory describes the value structure of metabolic states.  Here
we saw how economic variables and economic laws for kinetic metabolic
models can be derived from Metabolic Control Theory. Figure
\ref{fig:cbafbamsf} gives an overview.  In optimal states, all active
reactions must satisfy the cost-benefit balance
Eq.~(\ref{eq:fitnessbalanceeq}).  \co{We first obtained the
  \emph{cost-benefit principle}:} The {\myvalue} $\gul$ of an active
enzyme (describing its effect on the metabolic objective) must be
equal to the enzyme {\price} $\hul$ and must therefore be positive.
This principle leads to a number of other laws: using the summation
and connectivity theorems of {\MCA}, we obtain
{\summationconnectivitycondition} that relate {\fluxgain}s to enzyme
{\price}s along a flux mode, and metabolite {\price}s lead to the
enzyme prices around the metabolite.  In enzyme-optimal states, flux
modes must be economical and thus free of futile motifs\footnote{In
  models with flux bounds, the definition of futile motifs must be
  modified. A flux bound leads to a shadow values, which formally acts
  like a flux gain, and in the definition of futile motifs these extra
  flux gains must be taken into account.}.  Written in a local form,
the {\summationconnectivitycondition} yields economic rules and
balance equations. The economic variables in these laws can be defined
by metabolic control coefficients or shadow values
\cite{lieb:18lagrange} (proof in SI
\ref{sec:ProofsPotentialsAreIdentical}).  \co{mention again ``unique
  enzyme principle'' and ``principle of dispensable enzyme'' und wo
  sie herkommen bzw was passiert, wenn sie nicht gelten} Extensions of
the theory for models with other constraints or assumptions are
described in appendix \ref{sec:MethodsExtending}.  \co{models with
  isoenzymes, non-specific enzymes, non-enzymatic reactions, and
  constraints on model variables. let us consider here only one such
  exceptional case, cell models with dilution.}

\co{MERGE IN IN PARAG ABOVE: From this equation we obtain the
  (network-wide) {\summationconnectivitycondition}
  (\ref{eq:fitnessbalance2x}) and (\ref{eq:fitnessbalance3x}) and the
  (local) economic balance equations (\ref{eq:reactionbalanceeq}) and
  (\ref{eq:metabolitebalanceequation}). All of these laws are
  necessary conditions for enzyme-optimal states (proof in SI
  \ref{sec:proofequivalenceoptimalitytheorems}). Generally, states can
  be enzyme-optimal, economically stable, enzyme-balanced, or
  enzyme-economical, and there are states with economical fluxes (for
  explanations, see SI \ref{sec:SIdefinitionsAndTheorems}). The notion
  of economical fluxes (defined by the {\summationcondition}) is
  central to metabolic value theory and links it to flux cost
  minimisation. The balance equations apply both to single reactions
  and to entire pathways and the entire metabolic network.}

\myparagraph{What we can learn from Metabolic Value Theory} So what
did we learn about our initial questions?  \co{WO? nohcmal die
  ganzen arten von values erwaehnen, und dass sie und ihrer
  verbindungen die wertstruktur bilden} \co{In intro klare fragen
  stellen; hier klare antworten geben!}
  \begin{enumerate}[leftmargin=5mm]
  \item The  enzyme levels in optimal states depend on  network-wide fitness requirements. In metabolic value theory,
    these requirements are described by local economic
    variables. \co{sagen, dass bilanzgleichung auf zwei logischen
      schritten beruht: projizierte lokale groessen (geht fuer all
      outpurgroessen, auch ohne optimalitaetsgedanken; gleichsetzen
      mit enzymekosten (geht nur unter optimalitaetsannahme)} More generally, we saw
    that the values of metabolites, enzyme, and reactions (describing
    their ``network-wide'' fitness effects) satisfy local balance
    relations resembling Kirchhoff's rules for electric
    circuits. \co{This is largely enabled by the fact that the usual
      control coefficients (fr reaction perturbations) can be
      rewritten as
      ${C^{x}_{v}}\trans = {\Nint}\trans\,{C^{x}_{r}}\trans = \Delta
      {C^{x}_{r}}\trans$, which..} Using these rules, the indirect values and therefore the
    entire value structure can be inferred from direct gains and
    prices (describing direct effects of network elements on fitness)
    by projecting them onto the network. \co{if the optimal state is known}
  \item A metabolic pathway can be seen as a value chain: the enzymes
    (and substrates) invested lead to increasing metabolite values
    along the pathway flux.  \co{The substrate, enzyme, and cofactor
      investments accumulate along the chain, giving rise to
      ``embodied values''.} All these investments are defined as
    ``point costs'' and measured in fitness units, which makes them
    comparable. The embodied investment of a metabolite, divided by
    the metabolite's (absolute) production rate, defines its embodied
    value.  \co{WO? distinguish embodied value (defined along chain,
      sum he e / v) and ``value in use'' (normal ec pot
      definition). in optimal state the two are identical (bis auf
      previous substrate value!!  (das ueberall erwaehnen!)) // rules:
      value in use; set fu = 0, dann optimal embodiment!}
   \item Flux profiles  in  enzyme-optimal states
     must follow some simple algebraic rules and must be solutions of 
     FCM problems. Given an economical {\flow}, we can construct enzyme-optimal
     states by (i) optimising the enzyme levels by ECM or by (ii)
     finding possible economic potentials and building kinetic models
     around them.
  \end{enumerate}

  \coout{I think it's important here to "give answers" to the
    questions from where you've started; i.e.~to show that you've
    provided us with tools with which we can now answer these
    questions (without transforming them into other types of
    questions, e.g.~about specific objectives, etc.) // Discussion
    should meet al.~diegetic expectations: you would first show that
    you have provided tools to answer your original questions (maybe
    you can add a "tool question" to the series of questions in the
    introduction? -- Because I'm not sure those are in themselves
    questions to which one can give an answer in a paper?); then you
    would show why this may important even for people that don't
    particularly see why these questions matter, or why you claim that
    these questions are yet unanswered, because you would show that
    your theory also supplies us with stronger modeling approaches /
    tools / proofs (this should also indirectly convince the skeptical
    reader but it's also an independent aim).}

  \myparagraph{Economic state variables} Economic values provide a new
  perspective on metabolic states.  Kinetic models describe cells
  mechanistically by physical metabolic concentrations and fluxes.
  Thermodynamics adds a second layer of description, relating metabolite
  concentrations and fluxes by a notion of energies (in chemical
  potentials and thermodynamic forces). Metabolic value theory
  adds a third layer: a \emph{value structure} described by
  economic variables. The economic potentials describe the use value
  of metabolites, as defined through the fitness effects of
  hypothetical state variations\footnote{\co{WO?}To define metabolic values,
    metabolite variations were described by hypothetical influxes and
    their effects were assessed by response coefficients. However,
    instead of \emph{influxes}, we could also use \emph{degradation
      fluxes} (with a minus) to define economic potentials.
    Mathematically, this doe not make a difference, but in reality,
    degradation fluxes may be realised experimentally: e.g.~the
    economic potential of ATP (or, more precisely, the
    ATP-[ADP+phosphate] difference) can be measured by tuning a
    controllable ATPase and measuring the fitness effects. \co{kommt schon
    in abschnitt ueber experimentelle ueberpruefung in cba lagrange}}.
  The same logic applies  generally: by writing physical laws as
  constaints, all physical variables can be associated with dual
  economic variables.  The economic laws for these values allow us to
  study the value structure of metabolic states even if many model
  details are unknown. For example, we can explore the space of
  enzyme-balanced states, even without knowing the rate laws or enzyme
  cost functions. \co{kurz sagen: matching the dual variables between
    models allows for consistent model combination in optimal states
    (REF to CBA local\cite{lieb:cbalocal})} After constructing
  feasible economic states from the balance equations, we can realise
  them by enzyme-balanced kinetic models.

\co{WO? klar sagen, dass oek var und rules IMMER gelten,
    nicht nur im optimalen szustand.  dass im optimalen zustand
    zusaetzlich die enzyme rule = 0 gxesetzt wird, und damit die
    balance eqs (ohne zusaetzliche opportunity costs / fitness
    mismatch) gelten}

  \co{\textbf{Economic laws}} \co{find a word like ``form'' concerning
    scaling, another word concerning local vs global} \co{Both the
    form and the type have to do with the variations (implicitly)
    described by the laws}
  
\co{Principles for local value production, in different formulations (paragraph or box):

  [kommt nochmal unten:]Four ``scalings'':
  o Absolute cost and benefit (using homogeneous functions)\\
  o Point cost and benefit\\
  o Differential cost and benefit\\
  o Value and price
  
Two types of laws\\
  o quatitative balance \\
  o positivity or signs 

Principles with names:\\
o (Local)  Cost-benefit principle: reaction  balance: point cost = point benefit, or enzyme investment = enzyme usefulness (=a case of value flow conservation)\\
o derived from it: Principle of positive benefit: positive benefit!\\
o Local differential cost-benefit principle: same as before, for differentiyl quantity
o Local value / price principle: value production = burden in each reaction / enzyme\\
o derived from it: Principle of positive  value production: positive flux  value (in flux direction)!\\
o also mention: ``dispensable enzyme principle'' and ``unique enzyme principle''}

\co{WO? eventuelle hinweise auf oek temperatur nur ganz kurz! auf cba
  regulation verweisen}

\co{\textbf{Variations: from global variation rules to local economic
    rules} \co{The ``form'' of an economic law has to do with the
    extension of the variation described (local or extended in the
    network)} (erst intuitiv erklaeren (bezug zu thought experiments
  erwaehnen?) - danach, wie sie mathematisch auseinander
  hervorgehen).}  \co{ We can write the same laws in ``single-variable
  variation'' form (rules), ``local form'' (balances) and ``extended''
  form ( {\summationconnectivitycondition}) \co{The economic rate laws
    come in three types (explain for each types of variations (set of
    variables perturbed; including their adjacent constraints, or
    virtual variations to compensate them)

\co{WRITE!!
    
    1. Economic rules (``local'') concern single variable and direct
    neighbours; they refer to single-variable variations; replace
    surrounding effects by virtual compensation. rules jeweils named
    after the element or physical quantity: "metabolite net rate",
    "reaction rate", "metabolite concentration" "enzyme concentration"
    (or "compound concentration", for metabolite and enzyme together)
    
    2. Balance equations (``proximal''): metabolic elements and next
    neighbours (e.g. nearby adjacent enzyme costs); economic balance
    equations (referring to local variations)

    3. Variation conditions (``regional'' or ``network-wide''):
    extended; economic variation formulae (referring to real (legal)
    extended variations \co{clarify the relation ot summation and
      connectivity theorems: The {\summationcondition} and
      {\connectivitycondition} (and therefore, the reaction balance
      and the metabolite balance) are closely related to (and derived
      from) the summation theorem and the connectivity theorem,
      respectively}}
}

\co{WRITE!!}  \co{\textbf{From local laws to network-wide laws and
    back}} \co{hier nur kurz, ref to appendix
  \ref{sec:fromSumFormulaToLocal} // say that we can move back and
  forth by summing or by introduding economic potentials as proxy
  variables} \co{transition from extended to local, with economic
  variables as ``fixes'' for violated local dependence constraints;
  ``The concept of economic potentials can be directly derived from a
  notion of embodied enzyme investments and based on the flux
  variation condition (Section
  \ref{sec:FluxVariationConditionLeadsToEconomicPotentials})//
  entsprechend fuer metabolite variation condition and balance
  equation:
  \ref{sec:MetaboliteVariationConditionLeadsToMetaboliteBalance}} Note
that these variations are closely related to the
{\summationconnectivitycondition}; to better understand the ``virtual
compensation'', refer to SI (in which balance equations and economic
variables are directly derived).  while the balance equations refer to
single variables and their local neighbourhood, the
{\summationconnectivitycondition} can be seen as a ``regional'' form,
describing more extended (and proper!)  variations. \co{FN: COIN GOOD
  WORDS, something like ``integral'' and ``differential'' in field
  theories. explicitly refer to field theories + stokes' theorem. note
  that constraint violations (and virtual compensating variations) are
  something like the ``boundary integrals'' of a (local or regional)
  perturbation.} this has some practical applications.} \co{Note the
alternative derivation from the variation condition, yielding
e.g. Eq.~(\ref{eq:FluxBenefitBalanceFirst})} \co{HIER? Aside from
trick ... derive rules // see appendix. KOMMT DAS schon irgendwo?
briefly mention d'alembert principle + action principles as a wa to
formulate laws of motion. ``integral'' form. derivation of rules
(``direct form'') resembles derivation of Euler-Lagrange equation}
\co{But what is the meaning of these equations? briefly explain
  variations (d'Alembert) of flux modes, and then (derived from it)
  variations of single variables; differential \pointbenefitform
  .. daraus \pointbenefitform .. daraus value flows!}

\myparagraph{Scaled values: economic laws in different forms} The
elasticities and control coefficients in metabolic control theory
exist in scaled and unscaled form.  \co{FN. in MVT, quantities are
  usually only scaled ``from one side'', i.e. x is scaled, but not
  f. if we also scale f (which we can do without any problems and any
  major changes to our theory), this would exactly correspond to using
  scaled MCA sensitivities.} Likewise, the economic laws can be
written in different forms, referring to fitness derivatives
(``values'' $f_{x} = \partial f/\partial x$), fitness contributions
(``point derivatives''
$f\partialder=\partial f/\partial \ln x = \partial f/\partial x \cdot
x$), or differential fitness contributions (``variations''
$\delta f = \partial f/\partial x\,\cdot\delta x$). For example, the
{\costshade}s $\huldot$ (or ``enzyme investments'') are logarithmic
derivatives describing costly effects of (actual or virtual) enzyme
changes\footnote{\co{kam schon oben} In the case of a linear enzyme
  cost function $\hminus(\esymbolv)=\sum_{l}\hul'\,\esymbol_{l}$ with
  fixed coefficients $\hul'$, the enzyme {\costshade}s are given by
  $\hudotl=\hul'\,\esymbol_{l}$.}. \co{of course, not considering
  their enzymatic activity, which can have other beneficial or costly
  effects} This ``marginal'' definition of enzyme cost and benefit,
describing changes of a given cell state, resembles the empirical
definition in experimental studies
\cite{deal:05,szad:10}. Conveniently, point derivatives can be
obtained by multiplying each economic variable with the corresponding
physical variable. If we do the same to our economic laws, we obtain
economic laws in ``{\pointbenefitform}s''. In most cases, for going
from ``value form'' (with values and prices) to ``value production
form'' (with point benefits and point cost), we just need to put dots
on the economic variables and replace elasticities by scaled
elasticities. \co{WO?  say: {\pointbenefitform} is integral over
  \valueform (the integrand is the differential form)} \co{Will this
  work for all terms (also for variables related to bounds etc)?
  exception in delta wr: malnehmen mit v!!  ausserdem Eunscaled
  ersetzen durch Escaled.  is there a formal proof for this, and a
  rule about what physical and economic variables are dual to each
  other?} \co{does this hold for ALL economic laws? show also the
  variation formulae in \pointbenefitform? in SI? reaction variation
  formula: bezug zu both flux cost and enzyme cost (first: flux value
  and enzyme value)?}

    \co{\textbf{Values can also describe constraints and
      non-optimality}} \co{WD: Starting from a fitness objective --
    e.g.~maximising a production rate -- we \emph{go back} to the
    required fluxes, and further back to enzyme levels that support
    this objective, and figure out all their values.}  \co{[WO?
    stress can be seen as an opportunity cost!]}  \co{WO?
    Importantly, in economic laws, fitness effects and effects of
    constraints are treated the same. BEISPIEL!  That is, we can add
    them, and the two, and the effective {\gain} of a flux may arise
    from a fitness term, from a constraint, or from the sum of the
    two. The same holds for metabolite and enzyme {\price}s.}
  \co{``additional inputs'': This also works under additional side
    objectives, e.g.~the need to keep metabolite concentrations
    low. OTHER additional inputs, including non-optimality?}
  \co{bounds act like direct values, contribute to indirect
    values. DISCUSS non-optimality. rediscuss this in CBA labour for
    flows}

 \coout{If we consider a specific kinetic model and change its fitness
  function, an enzyme optimisation will yield different metabolic
  states with different economic variables, which are hard to
  compare. However, \co{if we compare different models whose optimal
   states are, dynamically, the same, ..... BLA BLA sharing the
   same vary $\bvtot$ (or $\hc$) separately,} keeping the metabolic
  state and the reaction elasticities constant, this will only change
  the economic potentials (or loads).} 
 \coout{If we focus on fluxes, we consider {\fluxgain}s and
  economic potentials as key variables and search for economical
  {\flow}s and consistent economic potentials.}

 \coout{Metabolic
  enzymes (marginal {\metabolicobjective} high) vs signaling molecules
  (absolute {\metabolicobjective} high). Discuss absolute, marginal, and
  scaled marginal} \coout{Notion of {\myvalue}s (unscaled derivatives
  of the {\metabolicobjective} function) and benefits (scaled derivatives)}

\myparagraph{Metabolic value as an inverse form of metabolic control}
Economic values are related to enzyme kinetics and metabolic
control. Like control coefficients, the indirect enzyme values are not
fixed molecule properties, but vary between metabolic states: they
depend on an enzyme's location in a metabolic network, on the fitness
function, on metabolic fluxes, and on resource allocation between
different enzymes. In optimal states, enzyme values must match enzyme
prices $\hul$, which may be constant (in models with linear enzyme
cost functions) or increase with increasing enzyme levels (assuming
convex cost functions).  In fact, metabolic value theory is an
inverted form of {\MCA}. While {\MCA} describes the \emph{forward
  effects} of enzyme changes (on the metabolic state), metabolic value
theory turns this logic around and quantifies the \emph{incentives}
for enzyme changes, given a desired effect on a metabolic
objective. In other words: to define metabolic values, we need to
start from the desired objective and go back to necessary enzyme and
metabolite changes. The {\fluxgain}s, concentration {\price}s and
{{\enzymeprice}}s describe how an objective is \emph{directly}
affected by fluxes and concentrations. In the network, they indicate
where cost and benefit are actually realised.  If we start from there
and follow the causal chains in reverse, we obtain the flux
{\myvalue}s, economic potentials, economic loads, and enzyme
{\myvalue}s that are \emph{indirectly} promoted by the fitness
function and acquired (in reverse direction) along causal chains.
Therefore, causality (describing forward effects)
and incentives (describing effects in reverse) are closely related,
and this is why MVT can be based on MCT.

\myparagraph{Production economics and concentration economics}
Metabolic dynamics arises from an interplay between metabolite
concentrations and rates, via mass balances and kinetics. \todo{In MCT,
  fluxes within a flux mode are described by summation theorems, while
  concentrations (e.g.~of a metabolite and the surrounding enzymes)
  are described by connectivity theorems. The two theorems hold
  independently. \co{If we compare MVT to {\MCA}, we note that the
    summation theorem underlies ``production economics'', while the
    connectivity theorem underlies ``concentration economics''.
    Similar to the two theorems in {\MCA}, in MVT there are two sets
    of equations that describe the same metabolic state, but from
    different angles.} Similar complementary laws exist also in
  metabolic value theory: there is an economics of metabolite
  production (described by {\fluxgain}s and economic potentials) and
  an economics of metabolite concentrations (described by
  concentration {\price}s and economic loads).}  \co{dann von
  summationstheorem ausgehend die ganze geschichte mit
  produktionswerten (mass balance!); von conn theorem ausgehend,
  konzentrationswerte (elasticities!); dann sagen, wie beide
  zusammenspielen (zB dilution) and warum die trennung pratisch ist
  fuer FBA} The economics of production concerns processes (including
metabolite conversion, fluxes and metabolite net rates) and describes
them by two sets of laws: from the {\summationcondition}, we obtain
the \fluxbenefitbalance. The economics of concentrations concerns
substance concentrations: the {\connectivitycondition} refers to
concentration {\price}s $\hc$ and gives rise to the metabolite
balance. \co{Their economic values are governed by separate laws:
  production {\myvalue}s ({\fluxgain}s and economic potentials) are
  described by {\summationcondition} and \fluxbenefitbalance, whereas
  concentration {\myvalue}s (metabolite {\price}s and economic loads)
  are described by {\connectivitycondition} and metabolite balance.}
The two sets of conditions hold simultaneously, but can be studied
separately.  In flux analysis, for example, we may consider the
{\fluxbenefitbalance} as a constraint, while ignoring metabolite
concentrations. This allows us to determine feasible flux patterns and
economic potentials without worrying about concentrations or specific
kinetics model in which these potentials are defined.

\co{ankuendigen: naechste abschnitte betreffen MVT and MCT}

\myparagraph{Enzyme investments and flux control} \co{translate the
  klipp+heinrich reasoning (for scaled and unscaled control
  cofficients) into the language of costs/benefits and prices/values.}
The definition of metabolic values by metabolic control can help us
make sense of the relation between control coefficients and enzyme
levels \cite{klhe:99}. For flux maximisation at a fixed enzyme budget,
Klipp and Heinrich proved two kinds of relationships that hold in
optimal states. \co{hier kurz fixed enz vs enz cost erklaeren}
\co{more generally, enzyme ``allocation'' vs optimal enzyme levels}
\co{sagen: hier opt enz levels one by one. an alternative reasoning:
  fixed enzym budget (biologically) - reallocation or optimal benefit
  at fixed cost (mathematically); equivalent, see CBA opt} First, the
unscaled flux response coefficients
$R^{\rm V}_{l}= \partial \vsteady/\partial \esymbol_{l}$ must be
equal. Second, the scaled flux control coefficients
$\hat{C}^{\rm V}_{l}= \partial \ln |\vsteady|/\partial \ln
\esymbol_{l}$ must be proportional to the enzyme levels $e_{l}$.
These relationships correspond exactly to our economic
laws\footnote{Here is a proof. For the first relationship, we note
  that maximising the flux (as a pathway objective) at a fixed total
  enzyme amount is equivalent to maximising the flux minus a linear
  enzyme cost function with equal enzyme weights $\hul'$.  By
  identifying $\partial V/\partial \esymbol_{l}$ with $\gul$, we
  obtain a {\price}-{\myvalue} balance $\gul = \hul'$, stating that
  all unscaled response coefficients must be equal.  For the second
  relationship, we note that scaled control coefficients are equal to
  scaled response coefficient and given by
  $\hat{C}^{\rm V}_{l} = \frac{e_{l}}{V}\frac{\partial V}{\partial
    \esymbol_{l}}= \frac{e_{l}}{V}\gul
  =\frac{1}{V}\gul\partialder$. Since $\gul=\hul'=\const$,
  $\gul\partialder = \gul\,\esymbol_{l}$ is proportional to
  $\esymbol_{l}$.}.  The first relationship reflects the
{\price}-{\myvalue} balance $\gul = \hul'$: if the cost function is
given by the sum of enzyme levels, all enzymes have equal prices and
must therefore have the same value. \co{which also strongly constrains
  the economic potentials.}  The second relationship reflects the
cost-benefit balance \co{FN Note that * }
$\gul\,\esymbol_l = \hul\,\esymbol_l$: if enzyme prices are fixed,
enzyme levels are proportional to enzyme investments and therefore to
enzyme point benefits (or ``value production''). Thus, metabolic value
theory confirms the relations from \cite{klhe:99}.

First, the equality $\gvtotl\,v_{l} = \gul\,\esymbol_{l}$ from
Eq.~(\ref{eq:bla123}) reflects the fact that the scaled enzyme
response coefficients and scaled control coefficients are equal
\cite{hesc:96}.  Second, there is an interesting relation between
{\fluxburden}s and control coefficients: in optimal states, the
{\fluxburden} vector $\hvv = \diag(\hu) \,\ratelawv\inv$ is equal to
the vector of flux {\myvalue}s $\gvtot$ and must therefore be a
nullvector of the matrix $(\Imat - {\CJmat})\trans$ (see SI
\ref{sec:SIproofsFluxPricesNullVector}). This is interesting news for
FCM, where $\hvv$ is the gradient of the flux cost function. If the
cost function is linear (as in FBA), (i.e.~if $\hvv$ is constant and
predefined), the predefined $\hvv$ puts constraints on the control
coefficients in the underlying kinetic model, assuming an optimal
system state.  \co{THIRD, REF + discuss briefly in disc (bei K+H): Ref
  "Enzyme cost reflects metabolic control" ! Noor 2016, SI 3.9 (proof
  in SI 7.4): The enzyme cost profile obtained by ECM is a linear
  combination of metabolic control profiles; das alles im buch in den
  haupttext?}

\co{say that costs/benefits have fitness units and are
  additive (``extensive''); while prices/values have different units
  (fitness divided by variables considered) and are not additive, and
  make the direct conection to enzyme efficiences (elasticities)) and
  economic potential differences.}  \co{nochmal sagen: verschiedene
  probleme. K+H: 1 target flux, fixed sum el.bei mir allgemeine
  objectives and enzyme cost functions. \co{remark on ``max benefit
    minus enzyme cost'' (hier) vs ``max flux at fixed enzyme''. isn't
    this a difference?}  difference yields fitness (with different
  prefactors) can be compared to optimality of one while havin the
  other one fixed, or to pareto optima.}  \co{say what simplifications
  were made. say whath this means for proteom.  - dann restlichen text
  von unten einbauen!}

\co{beim eingehen auf K+H, auch sagen was sich durch metabolite costs
  aendert! (welche kontrollkoeffizieneten zusaetzlich ins spiel
  kommen, und wie!)}

\co{refer to Fig 1:
  FIGURE part on Klipp + Heinrich (Case of flux maximisation): d ln
  g/du ln u = u/g dg/du = scaled flux control!  hu = const; same
  unscaled reponse coefficients scaled reponse coefficients
  proprotional to enzyme levels.}   \co{Naming: b(v,c) leads
  to ableitung ``use value'' (scaled derivative ``partial benefit'' or
  ``value flow'') // h(u) leads to ableitung ``embodied value''
  (related to exchange value) // hier und in CBA lagrange, CBA fluxes,
  CBA labour, CBA local: in optimal states, the embodied value and use
  value of a metabolite must be equal}

\co{Explain whether scaled
  control coefficients, generally, can be see as flows!
  \begin{figure*}[t!]
  \includegraphics[width=10.5cm]{/home/wolfram/projekte/cba/zeichnungen/q1.jpg}
\caption{\co{WO? WRITE ME}}
\end{figure*}
What about connectivity theorem?}

\myparagraph{Practical application} \co{WOHIN mit dem abschnitt?}  If
the enzyme {\price}s $\hul$ are known (e.g.~given by molecular
masses), the proteome (vector $\esymbolv$) defines an
investome. Simple cost-benefit principles tell us which enzymes should
be expressed, and at what relative levels.  Assuming an optimal state,
enzymes must satisfy the value-price balance
$\ratelaw_l\,\gvtotl=\hul$. If an enzyme efficiency $\ratelaw_l$ is
low (e.g.~because of a low substrate level or a reaction close to
chemical equilibrium), this balance cannot be satisfied and the enzyme
should not be expressed. Likewise, for reactions with low or negative
flux value $\gvtotl$ (control over the metabolic objective), the
condition cannot be satisfied. If enzymes are expressed, we may ask
about expression levels.  Any pair of enzymes must satisfy the
relation
$\frac{{\hus}_1}{{\hus}_2} = \frac{{\gus}_1}{{\gus}_2} =
\frac{{\gvtots}_1\,\ratelaw_1}{{\gvtots}_2\,\ratelaw_2}$: if two
enzymes share the same flux $v_1=v_2$ and if this flux is the
metabolic objective, we obtain the relation
$\frac{\CJ_1}{\CJ_2}\frac{\esymbol_2}{\esymbol_1} =
\frac{{\hus}_1}{{\hus}_2} = \const$ between the expression levels
$\esymbol_{1}$ and $\esymbol_{2}$ (because
$\ratelaw_l = v_l/\esymbol_l$ and ${\gus}_l = b_J\, \CJ_i$). Since
this holds for any pair of enzyme, in optimal states, enzyme levels
and flux control coefficients must be proportional, confirming the
result by Klipp and Heinrich \cite{klhe:99} for flux maximisation at a
given total enzyme amount.

\textbf{Metabolic Value Theory for general types of variables}
\co{Starting from Klipp + Heinrich, I generalise the link between
  control and economic values to all metabolic variables, including
  enzyme levels, steady-state concentrations and fluxes.} Also other
  model variables can be described by control coefficients, including
  peaks times in signaling system\co{REF}, the periods of metabolic
  oscillations\co{REF}, or the amplitudes of spatial modes in
  reaction-diffusion models\co{REF my preprint, to be written?}.  For
  a fitness-relevant target variable $x$, the control coefficients
  $\Cmat^{x}_{v}$ and $\Cmat^{x}_{r}$ can be used to define economic
  values, and if summation or connectivity theorems hold (e.g.~proven
  by time-scaling arguments), these theorems yield economic
  laws\footnote{\co{ISNT THERE already s section about this above?
      Reaction and metabolic control coefficients; splitting C = Cmat
      N; therefore delta w = ...  general explanation!}  \co{sort!
      erst CS, dann allgemein} If control coefficients can be written
    as differences (following the example
    $\Ccmat = -(\Nint \Eunc)\inv\,\Nint)$, where $\Nint$ acts like a
    difference operator\co{explain!}) we directly obtain a reaction
    balance equation.  \co{(gehört eigentlich nach CBA differential)
      (auch opt osc, ähnliche theoreme für periodic values?): mention
      theorems for differential expression!  (see diss verteidung
      vortrag, gegen ende)} This trick works for any objective
    function: not only actual biological objectives, but also
    functions that describe a goodness of fit to data. Thus, basing
    Metabolic Value Theory on {\MCA} is almost as general as basing it
    on Lagrange multipliers.}. \co{mention again difference trick and
    why it works quite generally!}

\myparagraph{Investome and metabolic control} Let us come back to our original question:
how can we  understand the  quantitative proteome  of a cell?  In the introduction, I suggested
that large protein investments must be justified
by a large ``importance''.  \co{sonst eher ``usefulness''. welches
  wort ist besser?}  of the protein. But  what do we mean by
``importance''? Transcription factors are  important for
cells, but their amounts are usually small. So if we claim that
``investment equals importance'', we need to define our terms more
precisely. We can paraphrase the result by Klipp and Heinrich by
saying: if ``investment'' stands for enzyme amount, and
``importance'' stands the scaled flux control coefficient, investment and importance are balanced.
``Enzyme investment'' and ``enzyme importance'' will increase with the enzyme amount, and by dividing by this amount,
we obtain  a
second equality, between ``enzyme price'' and ``enzyme value'', the
investment and importance \emph{per enzyme}. Also this second equality
follows from Klipp and Heinrich's results if we define ``enzyme
price'' as 1 (because all enzymes are weighted equally in the total
enzyme amount, which needs to be minimised) and ``enzyme value'' by
the unscaled response coefficient. Here we saw how these notions can
be generalised: in metabolic value theory, an ``enzyme price'' is the
derivative of an enzyme cost function, and ``enzyme values'' are
response coefficients between enzyme levels and the metabolic
objective. If we multiply price and value by the respective enzyme
level, we obtain ``investment'' and ``importance''. \co{explain again
  the distinction between normal and point derivatives (corresponding
  to unscaled and scaled response coefficients), and the resulting
  types of economic variables.}

\co{\textbf{FN: Enzyme amounts and enzyme cost: linear and
    homoegeneous cost functions}}\co{The economic variables are
  marginal quantities representing fitness derivatives.  How can we
  relate such variables to absolute quantities such as measured
  protein levels?  Here the ``point derivatives'' come into play. For
  example, aside from the enzyme price
  $\hul=-\partial \ffit/\partial \esymbol_{l}$, there is also the
  enzyme investment
  $\huldot=-\partial \ffit/\partial \ln \esymbol_{l} =
  \hul\,\esymbol_{l}$. By weighting a (measurable) enzyme level
  $\esymbol_{l}$ with the (objective-derived) enzyme price, we obtain
  the enzyme investment, which is ``marginal'', but resembles an
  absolute cost. In the case of a linear cost function,
  $\hminus(\esymbolv)=\sum_{l}\hul'\,\esymbol_{l}$, the relation is
  simple: the prices $\hul$ are constant and known (given by $\hul'$),
  and the investments are directly given by costs
  $\huldot=\hel'\,\esymbol_{l}$, and just a weighted version of the
  enzyme levels.} \co{briefly mention case of linear cost functions
  (constant prices) and homogeneous cost functions (extra termin
  balance eq); note extension to homogeneous functions. // hier nur
  kurz (genauer in CBA labour)!  (mit homogenen funktionen, regeln),
  dann erklaereung!  relation to control + wichtigkeit} \co{In models
  with linear enzyme cost functions $\hminus(\esymbolv)$\co{=sum h'el
    el}, \co{the enzyme price $\hul$ is given by the fixed preactor
    hel' and} the absolute cost $\hminus$ and {\costshade}
  $\hudot = \sum_{l}\hul\,\esymbol_{l}$ are equal.}

\co{Paragraph title?} \co{hier alles, was beziehung zwischen proteom
  und kontrollkoeffizienten betrifft; kurz absolute kosten /
  homogeneitaet erwaehnen (genaueres in CBA labour)} \co{TEILfrage:
  amount vs cost (absolut / marginal point cost): das unten an EINER
  stelle behandeln!  lieber erst alles mit linearer zielfunktion
  sagen. partial enzyme cost = ...; dann spaeter homogen; absolute und
  partial. faktor! verschwindet beim skalieren!} To apply these
notions to cell proteomes, we make three main assumptions: we assume,
first, that each protein has a price (derived from some cost function
to be specified, or maybe related to protein size); second, that
enzymes contribute (indirectly) to a metabolic objective or benefit
function (and that this contribution is described by control
coefficients); and third, that the proteome represents an optimal
state (where we can equate ``importance=investment'' and
``value=price'').  MVT yields critical insights about relation between
enzyme and metabolite values: enzyme values are not ``distributed at
random'', but acquired from flux values, which in turn reflect
economic potentials and flux gains. Thus, even if the enzyme values
are unknown, we know in principle how they emerge from the metabolic
network.  If the enzyme prices are constant and known\co{hier FN mit
  homogeneous functions usw, see below?}, we obtain strong constraints
on the possible metabolite values. And in the
``investment=importance'' equality, then enzyme importance can be
equated to a local value production (i.e.~flux value times flux, or
the balance value consumed and produced in reactions). Hence, the
higher an enzyme level (assuming fixed prices), the higher must be the
enzyme's value production. All this shows that the proteome, which may
appears structureless and ``arbitrary'' is in fact economically
structured. If the metabolic network structure is known, we can use it
to understand the ``value structure'' and make sense of the
proteome. \co{das waere ok als schluss - rest woanders hin?}

\co{lessons learned: (i) which enzymes are expressed? if enzyme prices
  (and flux gains) are constant, this strongly constrains the possible
  metabolite values, and therefore the possible (metabolic) control
  coefficients! (ii) how strongly are they expressed? (iii) how do
  metabolic potentials increase along a pathway?}  \co{Following H+K,
  for pathway flux as the target, optimality implies that investment =
  flux control. here clarified relation: amount - investment, by
  defining investment as log-marginal cost = exactly amount * price
  (marginal cost), where the price is assumed to be constant, and (eg
  related to enzyme mass)}

\section*{Acknowledgements}

I thank Mariapaola Gritti, Bernd Binder, Elad Noor, and Ron Milo for thinking with me. I am especially grateful to Bernd
who introduced me to labour value theory, the main inspiration for
this work. This work was funded by the German Research Foundation (Ll
1676/2-1 and Ll 1676/2-2).

\bibliographystyle{unsrt}
\bibliography{files/biology}

\clearpage

\begin{appendix}
  \section{Mathematical details}

\subsection{Conformity criterion and submode criterion}
\label{sec:appSubmodeCriterion}

\myparagraph{\ \\Beneficial, futile, and wasteful submodes} To
formulate the submode criterion for economical flux modes, we first
need to define submodes and flux motifs. A flux distribution $\modevector$ is
conformal with another flux distribution $\vv$ if all active reactions
in $\modevector$ are also active in $\vv$ and show the same flux
directions.  Given a flux mode $\vv$, a \emph{submode} of $\vv$ is a
stationary flux mode $\modevector$ on the subnetwork of active
enzyme-catalysed reactions in $\vv$ that is conformal with $\vv$.  The
sign pattern of a submode is calle flux motif. In a model with
{\fluxgain} vector\footnote{Note that the metabolite {\price} vector
  $\hc$ does not matter here.}  $\bvtot$, a submode is either
beneficial ($\bvtot\cdot \modevector>0$), {\futile}
($\bvtot\cdot \modevector=0$), or wasteful
($\bvtot\cdot \modevector<0$).
  
\myparagraph{Conformity criterion for economic {\flow}s} The
conformity criterion (Proposition \ref{th:theoremtestmode} in SI
\ref{sec:SIpropositions})  for beneficial, futile,
and wasteful submodes states the following: if $\vv$ is an economical {\flow},
then any beneficial submode $\modevector$ must contain a reaction with the same flux direction as $\vv$; any wasteful submode
$\modevector$ must contain a reaction with opposite flux
directions in $\modevector$ and $\vv$; and any futile submode
$\modevector$ must contain both sorts of reactions. The conformity
criterion follows from the flux variation condition
Eq.~(\ref{eq:fitnessbalance2x}), noting that $v_{l}$ and $v_{l}\inv$
have equal signs. Like the  condition itself, it holds only
for {\complete} {\flow}s $\vv$. If a {\flow} $\vv$ contains inactive
reactions, these reactions must be omitted: {\fluxgain}s
$\bvtotl$ and submodes $\modevector$ must be within   the active
subnetwork.

\myparagraph{Submode criterion for economic {\flow}s}  The
conformity criterion leads to the so-called  submode criterion: all submodes of an
economical {\flow} must be beneficial, implying that a {\flow} with
{\futile} or {\wasteful} submodes is uneconomical.  Also the opposite
holds: according to proposition \ref{th:theoremfutile} (``conditions
for economical {\flow}s''), all {\flow}s without non-beneficial
submodes are economical.  As a test for economical  {\flow}, we
can simply search for futile and wasteful submodes.  While the submode
criterion holds for all futile submodes, it is safe to consider  \emph{elementary} futile submodes
 for practical tests: if a  {\flow} contains a non-elementary futile submode, it
 must also contain an elementary futile submode. \co{proof wo?}  Therefore,
 if a {\flow} is free of elementary futile submodes, it is also  free
of any other futile submodes and is therefore economical.

\subsection{Flux balance models  and  metabolite cost}
\label{sec:AppFCMmetCost}

\co{aehnliches argument kommt in CBA II. hier lieber umschreiben und
  den zweiten teil betonen, die unabhaengigen gleichungen fuer raten
  und konzentrationen: ``production economics'' and ``concentration
  economics''. hier: related to summation theorems and connectivity
  theorems} How can flux cost minimisation models be reconciled with
kinetic models? In fact, how is this possible at all?  In FCM, costs
are assigned to fluxes while metabolite concentrations are not
described. In kinetic models, in contrast, the metabolic objective can
depend both an fluxes and on metabolite concentrations.  So how can
both methods predict the same fluxes? \todo{Here is an explanation.}
In FCM we can choose a flux cost function that effectively describes
enzyme and metabolite cost in a given kinetic model \cite{lieb:18fcm}.
A linearised version of this function will captures enzyme and
metabolite costs.  If we use this cost function in linear FCM, we
obtain shadow values (``economic potentials'') that refer to the sum
of both costs. In fact, we can also turn this around: with any
(feasible) set of economic potentials (e.g.~no matter if we choose them by sampling, by adjusting
them to proteomics data, or by assuming uniformly distributed enzyme
investments), there will always exist an enzyme-optimal kinetic model
that realises exactly these fluxes and economic potentials. \co{WD mit
  disc:}Seen from an abstract perspective, metabolic value theory
describes the economics of production rates (``production economics'')
and the economics of concentrations (``concentration economics''),
using separate variables (economic potentials versus loads) and
economic laws (e.g.~reaction balance versus metabolite balance, or
flux variation condition versus metabolite variation condition).  
In practice, they can be used separately.  Kinetic models contain two
types of direct values, a flux gain vector $\bvtot$ and a
concentration {{\price}} vector\footnote{\co{edit order - erst
    allgemein, dann ohne moiety cons?}  In models with predefined conserved moiety
  concentrations (e.g.~[ATP]+[ADP]),
 the term $\Gmat \trans \, \loadcm$ can  be included in an effective 
  $\hc$. In contrast,  if consered moiety concentrations  are treated as control variables, they also carry
  economic loads (in a vector $\loadcm$). This leads to a new term in
  the {\loadbalance},
  ${\Eunint}\trans\,[\Deltar \wtot+\bvdir] = \Gmat \trans \, \loadcm
  +\hc$, which couples concentration {{\price}}s and economic
  potentials. \co{linke seite is ye; der term
    $\Gmat \trans \, \loadcm$ fehlt sonst immer! in var rule, ec rul,
    met balance eq! (und load-pot eq?)}  Symbols denote the reaction
  elasticity matrix $\Eunint$, the left-kernel matrix $\Gmat$ of the
  stoichiometric matrix, and the conserved moiety load vector
  $\loadcm$ (describing the marginal benefits of conserved
  moieties). } $\hc$.  Focusing on fluxes (e.g.~in FCM problems), we
may consider the ``economics of conversion'' while ignoring the
``economics of concentrations'': we ignore metabolite concentrations
and metabolite {\price}s $\hci$ and describe fluxes and economic
potentials relying on the {\fluxbenefitbalance} alone.  The resulting
fluxes will be realisable, at least, by some (unknown) kinetic
model. Generally, there is a correspondence between models and states:
each kinetic model yields a (consistent) set of fluxes and economic
potentials, and any (consistent) fluxes and economic potentials
represent, implicitly, an underlying kinetic model (or in fact, many
of them) with specific enzyme kinetics and cost function.  By first
choosing consistent economic potentials and then reconstructing a
kinetic model (or a range of kinetic models) that realises this
profile, we can construct model ensembles.

\subsection{Detailed conditions for optimal states}
\label{sec:ConditionsForOptimalStates}

In our optimality problems for enzyme levels, metabolic objective
$\gplus(\vv,\cv)$ and enzyme cost $\hminus(\esymbolv)$ must be
continuous functions, and the cost function must be bounded. We
require that $\hminus(\enzymev)\le \hminus^{\rm max}$,
$\hminusfun(\enzymev) \ge \huweight \cdot \enzymev$, and for pathway
models we require that $\ffit(0)=0$, and $\ffit(\enzymev)>0$ for some
$\enzymev$. The existence of a local optimum can be proven easily:
$\ffit$ has a positive maximum in the enzyme polytope defined by
$\enzymev\ge 0, \husweight \cdot \enzymev \le \hminus^{\rm max}$.
However, there may be multiple optimum points, which may also form a
continuous manifold (e.g.~in models with kinetically identical
isoenzymes). Aside from physiologically plausible states, there may be
a ``locked state'' at $\enzymev=0$, describing an inactive system in
which fluxes and enzyme levels vanish. Despite its low fitness, this
locked state can be a locally optimal: an ``economic barrier'' for
enzyme levels must be overcome to reach more profitable states. Only
at higher enzyme levels, the system reaches the basin of attraction of
a more profitable enzyme-balanced state.  Finally, enzyme space may
contain regions with infeasible enzyme profiles, for which no steady
state exists (see SI \ref{sec:importancereversible})\footnote{\co{move
    this to SI?}  In some models (e.g.~with irreversible rate laws and
  completely saturated enzymes), certain enzyme profiles $\esymbolv$
  may not lead to a feasible steady state. The remaining feasible
  enzyme profiles form a subregion in enzyme space. If an optimal
  states is a boundary optimum in this region, the optimality
  condition contains an extra Lagrange term (which, however, does not
  allow from explicit model constraints, but from the fact that the
  model sometimes has no solution).  This complication can be avoided
  by excluding models with irreversible rate laws or completely
  saturated enzymes (see SI
  \ref{sec:importancereversible}). Alternatively, we may consider
  models with metabolite costs. In the scenario above, enzymes that
  are almost saturated will lead to a strong accumulation of
  substrate, which is penalised by the metabolite cost: at the
  subregion boundary, this cost would become very large, thus making a
  Lagrange multiplier on the boundary obsolete.}.

\section{Extending the theory}
\label{sec:MethodsExtending}

The models  above depict cells in a simplified
way: reactions are enzyme-catalysed and enzymes are
fully specific,  there are no isoenzymes,  enzyme profiles are
static and lead to stable steady states,  the metabolic objective
is a  function of fluxes and metabolite concentrations, and 
enzyme levels are the control variables to be optimised.  In reality,
things are more complicated: cells  show complex dynamics, are
subject to other objectives and constraints, and may  behave
non-optimally. For a  realistic picture of cells, our formalism
can be extended in various  ways (for details, see SI
\ref{sec:extensions}).

\textbf{(i) Isoenzymes and unspecific enzymes} Our premise ``one
reaction, one enzyme'' is not very realistic: biochemical reactions
may be catalysed by more than one enzyme (isoenzymes), and enzymes may
catalyse multiple reactions (unspecific or multifunctional enzymes).
In models this leads to a non-diagonal enzyme elasticity matrix, which
may not be invertible, and some formulae need to be modified.  To
convert $(\Deltar \wtotl + \bvdirl)\,\Eunul = \hul$ into
$(\Deltar \wtotl + \bvdirl)\,v_{l} = \hul\,\esymbol_{l}$, we assumed
that $\Eunu$ can be split into
$\Eunu = \diag(\vv)\diag(\esymbolv)\inv$. This allows us to move
$\diag(\esymbolv)$ to the right side, \co{JA!  derivations of Eqs
  (\ref{eq:FluxValueEq1}) and (\ref{eq:econForceEquation})} leading
to separate equation for each reaction.  If $\Eunu$ is not diagonal,
this is not possible, and all we can do is write
$\Eunu = \diag(\vv)\,\Escu\,\diag(\esymbolv)\inv$. Now , we obtain
$\sum _{l}(\Deltar \wtotl + \bvdirl),v_{l}\hat{E}^{v_{l}}_{e_{j}} =
h_{e_{j}}\,\esymbol_{j}$, still with $\diag(\vv)\Escu$ on the left,
and the reactions remain economically coupled.  In this new reaction
balance equation, the effects of an unspecific enzyme are captured by
summing over its catalysed reactions (see SI \ref{sec:isoenzymes}).
The metabolite balance remains unchanged, but the elasticities between
unspecific enzymes and their catalysed reaction are lower because an
enzyme needs to split its activity between several
reactions\footnote{As mentioned above, \co{WO? main text FN 7?} we can
  avoid these complications by writing our model with (redundant)
  ``monoenzymes'' and ``monoreactions'': a monoreaction is catalysed
  by a single monoenzyme, and a monoenzyme is an enzyme subpool
  catalysing a single monoreaction. However, this only shifts the
  problems: if a model contains reactions with the same
  stoichiometries, the Jacobian matrix may not be invertible, which
  complicates the definition of metabolic control coefficients (and
  therefore, of economic variables).}.

\co{\textbf{(ii) Coregulation} In metabolic steady states, mass
  balance conditions force some reaction fluxes (e.g.~the fluxes along
  an unbranched pathway) to be equal or proportional. Accordingly,
  some set of reactions must be either all on or all off (enzyme
  subsets), favouring a coexpression (on/off) or optimal coregulation
  (quantitative) of such groups of enzymes. In bacteria, this type of
  regulation may be implemented by operons. What do enzyme subsets
  mean for optimal enzyme profiles? \co{If two fluxes are
    stoichiometrically bound to vary proportionally
    ($v_{B}=\alpha\,v_{A}$) in all steady states, this means that,
    effectively, they can be replaced by one reaction with fluxes
    $v_{X}=v_{A}$ (by definition), stoichiometry
    $n_{iX}=n_{iA}+\alpha\,n_{iB}$, flux gain
    $w_{v_{X}}=w_{v_{A}}+\alpha\,w_{v_{b}}$, and linear enzyme cost
    $h(e_{X}) = h_{e_{A}}\,e_{A} + h_{e_{B}}\,e_{B} =
    [h_{e_{A}}+\frac{e_{A}}{e_{B}}h_{e_{B}}]e_{A}$. Entsprechend
    ganzer pathway: $v_{L}=v_{A}$ (reference flux),
    $n_{iL}=\sum_{l \in L} n_{il}$;
    $w_{v_{L}}=\sum_{l\in L} w_{v_{l}}$;
    $h_{e_{L}} = \sum _{l\in L}
    \alpha_{l}\,h_{e_{l}}\frac{e_{l}}{e_{A}}$} \co{insert a paragraph
    on dependent choice vars, enyzme subsets, operons etc.} was
  bedeuten die fur VBA?} \co{falls enzyme subset flussrichtungen
  umfasst, effectiv constraints zwischen economic potentials auf
  entfernung! in CBA kin appendix (oder auch wieder CBA fluxes?):
  falls vi = alpha vj, effektiv: betrachte nur eins davon, aber mit
  kosten fuer beide enzyme! effective problem within space of
  stationary solutions!}  \co{similar definition: reactions that can
  only vary proportionally, with fixed flux ratios and relative flux
  directions? effectively, one big reaction, with one complex enzyme,
  whose efficiency can change depending on ALL metabolite
  concentration involved in the original reactions}

\textbf{(ii) Economic balance equations for other cell variables} The
balance equations (\ref{eq:reactionbalanceeq}) and
(\ref{eq:metabolitebalanceequation}) refer to two basic motifs in
metabolic networks: a reaction together with its reactants, and a
metabolite together with the reactions it influences kinetically as a
substrate, product, or effector.  \co{FN: Interestingly, internal and
  external metabolites appear in the same way in the balance
  equations. the only difference is that the netw production of
  internal metabolites is constrained to zero by mass balance
  equations; while this generally shapes the possible metabolic
  states, the economic potentials assigned to internal and external
  metabolites are defined differently, in practice thay can be treated
  almost the same. this is good, because external metabolites in one
  model can be internal in another one (zoomable models).} But there
can also be other variables that affect reaction rates (such as
temperature, membrane potentials, etc), and the associated economic
values will satisfy balance equations.  A costly control variable $p$
that affects a single reaction (with {\price} $h_{p}$ and elasticity
$E^{\rm v}_{p}$) leads to the balance equation
$\gvtotl\,v_{l} = h_{p}\, p/\Escvlp$.  \coout{(proof idea:
  $\gvtotl \Eunvlpr = h^{\rm p}_{r}$, that is,
  $\gvtotl = h^{\rm p}_{r}/\Eunvlpr$) with variable value $p_{r}$,
  {\price} $h^{\rm p}_{r}$, and scaled elasticity $\Escvlpr$.}  In
general, several variables may affect a reaction, and several
reactions may be affected by one variable. In the first case, we can
sum the balance equations for the variables $p_n$ and obtain a balance
between {\fluxvalue} and weighted average {\price}
$\gvtotl = \langle h_{p_{n}}/\Eunvlpn \rangle_r$ (proof: from
$\sum_{n} \gvtotl\,v_{l} = \sum_{r} h_{p_{n}}\, p_{n}/\Escvlpr$
follows
$\gvtotl\,v_{l} = \frac{1}{N} \sum_{n} h_{p_{n}}\, p_{n}/\Escvlpn$).
In the second case, we can sum over the balance equations for
different reactions and obtain a balance between elasticity-weighted
average {\fluxvalue} and {\price},
$\sum_{l} \gvtotl\,\Eunvlpr = h^{\rm p}_{r}$ (proof:
$\sum_{l} \gvtotl\,v_{l}\,\Escvlpr = h^{\rm p}_{r}\, p_{r}$).

\textbf{(iii) Choice variables besides enzyme levels} In the models so far,
metabolite concentrations and fluxes were treated as state variables and
enzyme levels  as control variables.  But if our network
also contains macromolecule synthesis (protein production, mRNA
transcription, protein translation, or protein degradation), all
macromolecules (including enzymes) should be described as metabolites.
This is  possible: other variables, such as mRNA levels, cell
growth rate, temperature, concentrations in the extracellular medium,
or administered drug levels may be used as control variables.

\textbf{(iv) Constraints on state variables} Cell physiology puts
limits on enzyme concentrations, metabolite concentrations, and
fluxes.  Concentrations may be bounded by limited space or a cell may
require some minimum ATP level or minimum maintenance flux to
survive. In our optimality problems, such bounds can be treated as
constraints, and the resulting shadow values of shadow prices appear
as terms in the balance equations and can be included in $\bvtot$,
$\hc$, and $\hu$ (SI \ref{sec:inequality constraints}). For example,
consider the constraint $\esymbol_{l} \ge 0$ for positive enzyme
levels. If an enzyme is not expressed, this constraint is active, and
if we include the Lagrange multiplier as a negative enzyme shadow
{\price} $-{\hul^{\rm bnd}}$ into the cost-benefit imbalance
$\gul < \hul$, we obtain an equality $\gul = \hul - \hul^{\rm bnd}$.
Similarly, constraints on fluxes or metabolite concentrations lead to
shadow flux {\gain}s or {\price}s. For example, if an ATP-consuming
maintenance reaction is constrained to show some positive flux, this
leads to a (positive) shadow gain which adds to the {\fluxgain} and
contributes to the {\fluxvalue} of the reaction. In the flux benefit
balance, this shadow gain can justify a non-zero flux even if the
reaction consumes ATP without any benefit.  Similarly, upper bounds on
the enzyme abundance in cells, compartments or membranes
(e.g.~constraining the amount of photosystem complexes or ATP
synthases) lead to shadow values, which add to the enzyme prices.

\textbf{(v) Soft constraints on enzyme levels} Metabolic limitations
may either be described by constraints or by cost terms. For enzyme
levels, instead of putting a constraint $\esymbol_{l} \ge 0$, we may
add a cost that increases to infinity as $\esymbol_{l}$ goes to
zero. This cost may also have a biological meaning. For example, if
protein expression is leaky, and completely suppressing this is
energy-demanding, this can be described by a penalty on very small
enzyme levels.  The resulting negative price for small enzyme levels
replaces the shadow {\price} that we get from hard constraints. With
this cost term, enzymes will be expressed, but possibly at very low
expression levels (see SI \ref{sec:boundaryoptimaavoidedleaky}).

\textbf{(vi) Cost of unreliable enzyme expression} Due to stochastic
gene expression, protein molecule numbers follow random distributions
and optimal enzyme levels can never be realised precisely.  In a
finite cell volume and assuming a Poisson distribution for enzyme
molecule numbers $n$, the mean and standard deviation are given by
$\bar{n}$ and $\sigma_{n}=\sqrt{\bar n}$, and the coefficient of
variation $\sigma_{n}/\bar n = \sqrt{1/\bar n}$ decreases at higher
(mean) expression levels. To guarantee sufficient enzyme levels
despite this randomness, a cell may increase the mean enzyme level by
adding a ``safety margin'', for example $a$ times the standard
deviation (and then fine-tuning enzyme activities by small-molecule
regulation).  If $n$ is Poisson-distributed and $\bar n$ is the
desired expression value, the safe mean expression value would be
$n + a \sqrt{n}$. The safety margin leads to a modified cost
$\hminus(\esymbolv')$, where $\esymbolv'$ denotes a ``safe'' enzyme
level
$\esymbol'=\esymbol +\frac{a}{\sqrt{N_{\rm A}\,V_{\rm
      cell}}}\sqrt{\esymbol}$, with Avogadro constant $N_{\rm A}$ and
cell volume $V_{\rm cell}$. The relative increase due to the safety
margin is highest at low expression levels. Unlike the usual cost
functions (which are assumed to be linear or convex), the new cost
function can be concave\footnote{Moreover, the cost has an infinite
  price at $e_{l}=0$ -- however, around $e_{l}=0$, the assumption of
  continuous enzyme concentrations breaks down and a (stochastic)
  model with discrete enzyme molecule numbers is needed.}.

\textbf{(vii) Inactive enzymes} How does metabolic value theory
describe inactive reactions? With a convex enzyme cost function
(i.e.~a non-decreasing enzyme price), the enzyme value of any inactive
enzyme (with $v_{l} = 0$) must be below the minimum enzyme {\price}:
then, expressing even small amounts of this enzyme (instead of none)
would not pay off, but lead to a loss\footnote{Inactive enzymes with
  exactly balanced value and price are theoretically possible but very
  unlikely and can be ignored.}. Such reactions will remain inactive
even under infinitesimal perturbations (e.g.~of external metabolite
levels) and can be ignored in the model. When constructing a model
from given fluxes, zero fluxes can always be justified by assuming a
high enzyme {\price}.  However, it is important to note that some main
laws of metabolic value theory -- the
{\summationconnectivitycondition} (\ref{eq:fitnessbalance2x0}) and
(\ref{eq:fitnessbalance3x0}) and economic balance equations -- hold
only for {\complete} {\flow}s: before applying them to a given
{\flow}, all inactive reactions must be omitted, also the in
connection matrices such as $\Nint, \Kint, \Lmat, \Eunint$, and test
modes $\modevector$ always refer to the active part of the network.
Another way to model inactive reactions is to set enzyme levels to
zero by explicit constraints.  In the economic laws, the resulting
shadow price $\hul^{\rm bnd}$ leads to an effective enzyme {\price}
$\hul = \hul + \hul^{\rm bnd}$. Alternatively, we can require an
imbalance $\gul>\hul$ or a positive \emph{\stress} $\ful=\gul-\hul$
equal to $\hul^{\rm bnd}$. By inserting
$\gul=\gvtotl\, \frac{\partial v_{l}}{\partial \esymbol_{l}}$, we
obtain the economic imbalance for fluxes,
$\gvtotl < \hul (\frac{\partial v_{l}}{\partial \esymbol_{l}})\inv =
\acostvl$, that is, the flux value is below the flux {\burden}.

\begin{figure*}[t!]
\begin{center}
\includegraphics[width=7cm]{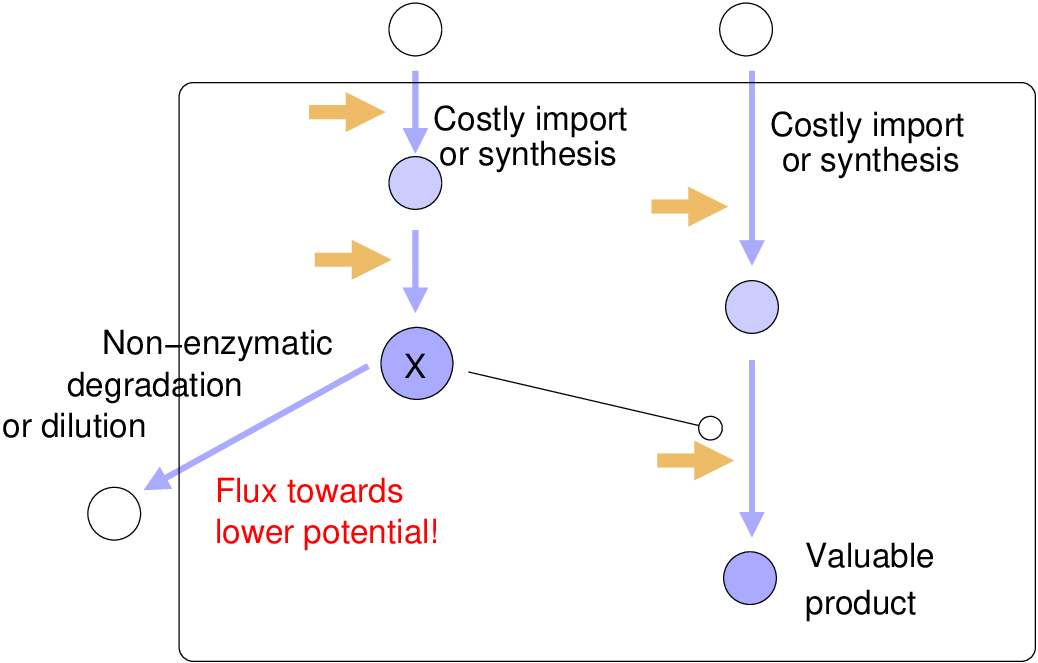}
\end{center}
\caption{Non-enzymatic reactions with a seemingly uneconomical flux
  from high to low economic potentials. In the example, metabolite X
  acts as a catalyst and needs to be kept at its optimal
  concentration. Since X is non-enzymatically degraded or diluted, it
  needs to be reproduced. In its synthesis pathway, the economic
  potentials first increase, but then drop in the non-enzymatic
  reaction. In an enzymatic reaction such a drop would indicate a
  non-optimal state, so if we wrongly assume that our reaction is
  enzymatic, the flux distribution appears futile.  However, in fact
  such pseudo-futile flux modes may be the best choice for a cell. To
  correctly classify them as economical, the flux variations in the
  flux variation rule (in the forms  (\ref{eq:fitnessbalance2x5}), ( \ref{eq:fitnessbalance2x0}), and (\ref{eq:fitnessbalance2x}))
 must be restricted to  enzymatic reactions.}
  \label{fig:nonEnzymatic} 
\end{figure*}

\textbf{(viii) Non-enzymatic reactions} In metabolic models,
non-enzymatic reactions are often ignored. However, but they can be
important: they may degrade valuable metabolites, produce toxic
compounds, or replace costly enzymatic reactions (e.g.~membrane
diffusion may replace costly transporters).  Even if a reaction is
uncatalysed, its flux can be indirectly controlled by surrounding
enzymatic reactions. Therefore, the presence of non-enzymatic
reactions creates new incentives that shape the optimal enzyme profile
(see Figure \ref{fig:nonEnzymatic} for an example).  In our theory, to
account for non-enzymatic reactions, some formulae need to be
modified. (i) In the flux variation rules
(\ref{eq:fitnessbalance2x5}), (\ref{eq:fitnessbalance2x0}), and
(\ref{eq:fitnessbalance2x}), flux variations must comprise enzymatic
reactions only.  (ii) In the {\summationconnectivitycondition}
(\ref{eq:fitnessbalance2} and (\ref{eq:fitnessbalance3}), extra terms
for non-enzymatic reactions need to be added (see SI
\ref{sec:proofnonenzymatic}).  (iii) In the reaction imbalance
Eq.~(\ref{eq:econForceEquation}), non-enzymatic reactions can be
described by hypothetical enzymes with non-optimal concentrations (see
SI \ref{sec:violationsnonenzymatic}) and by an imbalance term (or
``tension'') $\fudotl = \partial \ffit/\ln \esymbol_{l}$ on the right
side of the balance equation (see Eq.~(\ref{eq:econForceEquation}
))). In beneficial non-enzymatic reactions, an imbalance $\fudotl$
denotes the minimal enzyme {\myvalue} at which this reaction would be
profitable (if the enzyme existed). In contrast, a negative imbalance,
describes the loss caused by the non-enzymatic reaction.  (iv) In the
metabolite balance (\ref{eq:concentrationbalanceequationext}), a
non-enzymatic production, degradation, or dilution of the metabolite
leads to extra terms, which can be seen as effective loads or
concentration {\price}s. (v) The definitions of economic potentials
and loads as well as the \loadbalance\ remain unchanged. \coout{Where
  are the proofs?}

\textbf{(ix) Metabolite dilution} In growing cells, dilution tends to
decrease compound concentrations, as if all compounds were degraded by
reactions with a linear rate law $\lambda \,c_{i}$ (where $\lambda$ is
the cell growth rate).  Dilution has effects on stationary fluxes,
model dynamics, and cellular economics, including economic potentials
and enzyme investments.  In metabolic value theory, all results for
non-enzymatic reactions also apply to dilution reactions, leading to
simple formulae.  Dilution reactions with rate constant $\lambda$ lead
to an extra term $-\lambda\, \Imat$ in the Jacobian matrix, which
reappears in the summation and connectivity theorems for control
coefficients (SI \ref{sec:SIproofMCAwithDilution}), in the economic
laws, and in some of the formulae for economic variables. To account
for dilution in the {\compoundbenefitbalance}
(\ref{eq:concentrationbalanceequationext}) and in the {\loadbalance}
(\ref{eq:loadfluxvalueequation}), we consider an effective dilution
reaction with flux $v^{\rm dil}_i = \lambda\,c_i$, elasticity
$\lambda$, and a flux {\myvalue} $\gvtoti=-\winti$ (with a minus sign
because the metabolite is ``consumed''). We obtain the metabolite
balance $\loadi = \sum_{l} \hvl \, \Eunvlci -\lambda \,\winti $. The
new term $-\lambda\,\winti$ can be moved to the left and be included
into the metabolite load, yielding the effective load
$\loadeffi = \loadi + \lambda\,\winti$.  Effectively, adding dilution
\emph{increases} the metabolite {\price} by $\lambda \,\winti$. This
makes sense: when a metabolite is diluted, lower concentrations lead
to a lower value loss, thus creating an incentive for low
concentrations: this resembles an extra metabolite {\price}.  Also in
the {\connectivitycondition}, we can implement dilution by assuming an
effective metabolite {\price} $\hc^{\rm eff} = \hc + \lambda\,\wint$
(SI \ref{sec:demandDilution}). The {\summationcondition}, in contrast,
is not changed by dilution.  \coout{consider case where
  $F=\ln \lambda$, so $\partial F/\partial \lambda = 1/\lambda$, which
  gives another constraint on the $\wint$ {\myvalue}s.}


\coout{WEG?  consider the flux benefit balance in an unbranched
  pathway with production objective: if the internal metabolites are
  diluted, the fluxes $v_{l}$ decrease along the chain.  \coout{PRUEFEN
    ???} Since $\hc=0$, the sum rule $\sum_{l} \hudotl = \sum_{l}
  \bvtotl v_{l}$ must still hold. If we compare this model to a model
  without dilution and the same {{\fluxbenefit}}, the sum of marginal
  costs $\sum_{l} \hudotl = \sum_{l} \hul\, \esymbol_{l}$ must be higher in
  the presence of dilution.}  \coout{effektiv: in metabolite rule und
  balance: ignore dilution and assume effective load term}

\textbf{(x) Metabolic oscillations} Mathematically, metabolic states
with dilution are closely related to oscillating states
\cite{lieb:14c}.  This resemblance is useful if we model
\cite{lieb:2005,lieb:14c,lieb:18lagrange} metabolic oscillations in
which oscillating metabolite concentrations and fluxes are enforced by
oscillating enzyme activities.  In an optimality problem, all
oscillating variables are scored by a fitness function and our aim is
to optimise enzyme amplitudes and phase angles for a maximal fitness.
Oscillating metabolite concentrations and fluxes are approximated by
sine waves (of circular frequency $\omega = 2\,\pi \,f$), and their
amplitudes and phases are described by complex-valued vectors
$\cvt(\omega)$ and $\vvt(\omega)$.  Instead of the usual stationarity
condition $\Nint\,\vv=0$, we consider a mass balance equation
$i\,\omega\,\cvt = \Nint\,\vvt$.  Formally, the term on the left
resembles a dilution term (with steady-state condition
$\Nint\,\vv-\lambda \,\cv=0$), with an imaginary number
$\lambda = i\,\omega$ (with real-valued circular frequency $\omega$)
instead of a real-valued dilution rate $\lambda$.  \co{At the same
  time, a linear approximation of the rate laws yields the kinetic
  dependence
  $\tilde{\vv}=\Eunc\,\tilde{\cv}+ \Eune\,\tilde{\esymbolv}$. These
  dynamic equations, together with a metabolic control analysis of
  periodic perturbations [REF MCA periodic], define complex-values,
  frequency-dependent economic values of the amplitude variables
  $\tilde{\vv}, \tilde{\cv}$, and $\tilde{\esymbolv}$. For the
  metabolite amplitudes, these values are called periodic economic
  potentials [SEE CBA lag].}

\textbf{(xi) Multi-objective problems} Multi-objective problems
describe compromises between diferent objective functions. They can be
used to describe organisms in changing environments, organisms
anticipating uncertain challenges, and populations occupying several
ecological niches in which all objectives are important, but to
different extents in different niches.  A ``Pareto-optimal state'' is
a state in which no objective can be improved without compromising one
of the other objectives.  However, any Pareto-optimal state is also an
extremal point of some single-objective problem (in which one target
is optimised while constraining the others, or a convex combination of
the different targets is optimised) \cite{lieb:18theory}. Therefore,
metabolic value theory can be applied to Pareto-optimal states
\cite{lieb:14b}.

\end{appendix}

%
%
%
%
%
%
%
%
%

\end{document}